\documentclass[12pt]{article}
\usepackage{multirow,verbatim}
\usepackage{latexsym, amsbsy, amssymb}
\usepackage{natbib}
\usepackage{tikz}
\usepackage{graphicx}
\usepackage{epsfig}
\usepackage{amsmath}
\usepackage{amsthm}
\usepackage{hyperref}
\usepackage{txfonts}
\usepackage{dsfont}
\usepackage{physics}
\usepackage{mathrsfs}
\newenvironment{keywords}{\noindent\textbf{Keywords:} }{}

\usepackage{subcaption}
\usepackage{authblk}

\usepackage{listofitems} 
\usetikzlibrary{arrows.meta} 
\usepackage[outline]{contour} 
\contourlength{1.4pt}

\usepackage{listings}

\usepackage{xcolor}
\colorlet{myred}{red!80!black}
\colorlet{myblue}{blue!80!black}
\colorlet{mygreen}{green!60!black}
\colorlet{myorange}{orange!70!red!60!black}
\colorlet{mydarkred}{red!30!black}
\colorlet{mydarkblue}{blue!40!black}
\colorlet{mydarkgreen}{green!30!black}

\usepackage{algpseudocode}
\usepackage{multirow}
\usepackage{latexsym, amsbsy, amssymb}
\usepackage{verbatim}
\usepackage{color}
\def\boe{\begin{enumerate}}
\def\eoe{\end{enumerate}}

\newtheorem{proposition}{{\bf Proposition}}
\newtheorem{lemma}{{\bf Lemma}}
\newtheorem{corollary}{{\bf Corollary}}
\newtheorem{assumption}{{\bf Assumption}}
\newtheorem{definition}{{\bf Definition}}
\newtheorem{theorem}{{\bf Theorem}}

\newcommand\lo[1]{_{\nano{#1}}}

\def\E{\mathbb{E}}

\def\L{{\cal L}}

\def\var{\mathrm{var}}

\def\nano{\scriptscriptstyle}

\def\L2T{L \lo 2 (T)}
\def\L2TX{L \lo 2 (T\lo X)}
\def\L2TX{L \lo 2 (T\lo Y)}

\def\eod{\end{document}}

\newfont{\rsfsten}{rsfs10 scaled 1100}
\newfont{\rsfstena}{rsfs10 scaled 800}
\newfont{\rsfstenb}{rsfs10 scaled 500}

\def\var{{\mathrm{var}}}

\makeatletter
\renewenvironment{proof}[1][Proof]{%
  \par
  \pushQED{\qed}%
  \normalfont
  \topsep6\p@\@plus6\p@\relax
  \trivlist
  \item[\hskip\labelsep{\scshape #1\@addpunct{.}}]%
  \ignorespaces
}{%
  \popQED
  \endtrivlist
  \@endpefalse
}
\makeatother

\def\nano{\scriptscriptstyle}

\def\var{\mathrm{var}}

\def\ker{\mathrm{ker}}

\def\nano{\scriptscriptstyle}

\def\nano{\scriptscriptstyle}

\def\b1{\mathbf{1}}

\newcommand{\RN}[1]{%
  \textup{\uppercase\expandafter{\romannumeral#1}}%
}
\newcommand{\coord}[3]{%
{\vphantom{\left[#2\right]}}_{#1}\!\left[#2\right]_{#3}
}

\allowdisplaybreaks

\def\b{{\bf b}}

\def\L{{\bf L}}

\def\0{{\bf 0}}
\def\1{{\bf 1}}

\usepackage[dvipsnames]{xcolor}

\def\squarebox#1{\hbox to #1{\hfill\vbox to #1{\vfill}}}
\def\boxit#1{\vbox{\hrule\hbox{\vrule\kern6pt\vbox{\kern6pt#1\kern6pt}\kern6pt\vrule}\hrule}}

\definecolor{mygreen}{RGB}{34, 139, 34} 

\newtheorem{innercustomgeneric}{\customgenericname}
\providecommand{\customgenericname}{}
\newcommand{\newcustomtheorem}[2]{%
  \newenvironment{#1}[1]
  {%
   \renewcommand\customgenericname{#2}%
   \renewcommand\theinnercustomgeneric{##1}%
   \innercustomgeneric
  }
  {\endinnercustomgeneric}
}

\newcustomtheorem{customtheorem}{Theorem}
\newcustomtheorem{customlemma}{Lemma}
\newcustomtheorem{customproposition}{Proposition}
\newcustomtheorem{customcorollary}{Corollary}
\newcustomtheorem{customclaim}{Claim}

\title{A Kernel-Based Test for Multi-Way Conditional Independence}

\author[1]{Haoning Chen}
\author[2]{Yin Tang}
\author[1]{Bing Li}

\affil[1]{Department of Statistics, The Pennsylvania State University}
\affil[2]{Dr. Bing Zhang Department of Statistics, University of Kentucky}
\begin{document}
\maketitle

\begin{abstract}
Conditional independence involving more than two variables, or multi-way conditional independence,  is a fundamental probabilistic structure useful in many areas of statistical research, including latent-variable and graphical models. However, most existing works focus on pairwise conditional independence.  To address this gap, we propose a kernel-based test of multi-way conditional independence. Specifically, we characterize this hypothesis through the simultaneous vanishing of a vector of conditional cross-covariance operators and construct a test statistic based on the norm of the vector of their estimators. We derive the statistic's asymptotic distributions under the null hypothesis and local alternatives. As a related methodological contribution, we also develop a test of multi-way unconditional independence and provide a procedure for its implementation. An application to RNA data illustrates the method's use in genomics.

\end{abstract}


\section{Introduction}

Conditional independence is a fundamental concept in statistics and provides the structural basis for a wide range of methods, including graphical models \citep{lauritzen1996graphical}, causal inference \citep{pearl2009causality}, sufficient dimension reduction \citep{cook1998regression,li2018sufficient}, and latent-variable models \citep{bartholomew2011latent,allman2009identifiability}. However, most of the existing works on this problem are concerned with conditional independence between two variables given a third, and less attention has been given to conditional independence among several variables.

In latent-variable models, in particular, multivariate settings are common. The dependence among several observed variables is assumed to be fully explained by a latent state $Z$. For observed variables $X_1,\ldots,X_d$, this assumption is expressed as
\begin{align*}
P_{X_1,\ldots,X_d\mid Z} = \bigotimes_{j=1}^d P_{X_j\mid Z}
\qquad P_Z\text{-almost surely}.
\end{align*}
We refer to this property as multi-way conditional independence and write $X_1 \Perp \ldots \Perp X_d \mid Z$. This factorization is imposed as a key structural assumption in many latent-variable models. Under additional model-specific conditions, this factorization enables identification of the proportions and component distributions \citep{hall2003nonparametric,allman2009identifiability} and supports their nonparametric or moment-based estimation \citep{hall2003nonparametric,bonhomme2016nonparametric,anandkumar2014tensor}. It is therefore essential to test whether this condition actually holds. Residual dependence after conditioning on $Z$ may indicate omitted latent structure, direct interactions, or model misspecification, thus undermining subsequent inference. 

The question of whether a shared observed/latent state accounts for joint dependence is particularly relevant in genomics and neuroscience. Gene-expression latent-variable models represent dependence among genes through shared latent factors and observed covariates \citep{carvalho2008high,lopez2018deep}. Neural population models similarly use shared latent states or common inputs to explain correlated activity \citep{yu2009gaussian,macke2011common}, while cortical recordings reveal higher-order correlations that persist after accounting for stimulus-dependent activity \citep{koster2014modeling}. These examples motivate testing whether a given state $Z$ fully explains joint dependence through the conditional factorization above.

Although model-specific diagnostics have been developed for certain parametric latent-class models \citep{hagenaars1988latent,oberski2013montecarlo,lidonni2020latent}, a general nonparametric test for multi-way conditional independence has yet to be developed. In contrast, nonparametric testing of bivariate conditional independence has been extensively studied; see \citet{li2020nonparametric} for a comprehensive review. Representative procedures are based on nonparametric Hellinger distances \citep{su2008nonparametric}, maximal nonlinear conditional correlation \citep{huang2010testing}, more general metric characterizations \citep{huang2016flexible}, permutation-based reductions to two-sample test \citep{doran2014permutation}, regression residuals under structural restrictions \citep{zhang2017causal,shah2020hardness}, and distribution-free transformations \citep{cai2022distribution}.

Among these nonparametric bivariate conditional independence tests, the operator-based approach in reproducing kernel Hilbert spaces provides the foundation for our method. Under suitable conditions on the kernels, bivariate conditional independence can be characterized by the vanishing of the conditional covariance operator \citep{fukumizu2004dimensionality,fukumizu2008kernel}, which forms the basis of the kernel conditional independence test proposed by \citet{zhang2012kernel}. Subsequent work has developed related operator-based methods for nonparametric graphical modeling and high-dimensional conditioning \citep{lee2016additive,li2024sufficient}, as well as extensions to functional data \citep{li2018nonparametric,tang2026reproducing}. Alternative kernel formulations include kernel partial correlation and distance-based characterizations of conditional dependence \citep{huang2022kernel,sheng2023distance}.

Building on this operator characterization, we extend the bivariate framework to multi-way conditional independence. The key observation is simple. For any fixed ordering of the variables, multi-way conditional independence can be decomposed into a sequence of pairwise conditional independence conditions:
\begin{align*}
    X_1 \Perp \cdots \Perp X_d \mid Z
    \quad \Longleftrightarrow \quad
    X_j \Perp X_{\{j+1,\ldots,d\}} \mid Z,
    \quad \text{ for all } j=1,\ldots,d-1.
\end{align*}
Thus, the multivariate null hypothesis can be expressed through $d-1$ bivariate conditional-independence restrictions. To construct the test statistic, we apply the conditional covariance operator characterization to each restriction and combine the resulting operators into a single omnibus test statistic of multi-way conditional independence. We further develop the asymptotic distribution of our test statistic under the null and local alternative distributions. To the best of our knowledge, this is the first general nonparametric test for multi-way conditional independence.

In particular, the multi-way unconditional independence test can be regarded as a special case by taking $Z$ as constant, which also adds up to the existing literature. For example, joint independence based on the $d$-variable Hilbert--Schmidt independence criterion \citep{pfister2018kernel} and distance-based measures \citep{matteson2017independent,chakraborty2019distance}. We stay brief here as this unconditional degeneration is not the primary focus of the paper.

The remainder of the paper is organized as follows. Section~\ref{sec:backgrounds} reviews reproducing kernel Hilbert spaces and cross-covariance operators, which provide a framework for characterizing nonlinear dependence using linear operators. We construct the test statistics in Section~\ref{sec:test_statistics} and study their asymptotic properties in Section~\ref{sec:asy}.
The unconditional case then follows as a special case in Section~\ref{sec:unc}. Finally, Section~\ref{sec:num} describes the numerical implementation, and Section~\ref{sec:simulation} presents
simulation studies and applies our method to the OneK1K single-cell RNA-seq dataset.

\section{Background}\label{sec:backgrounds}

\subsection{Background on Reproducing Kernel Hilbert Spaces}

Let $\Omega_{X_j}=\mathbb{R}^{p_j}$ for $j=1,\ldots,d$, and $\Omega_{Z}=\mathbb{R}^{p_Z}$ \footnote{It can be generalized to second-countable locally compact Hausdorff spaces.}. Let $(\Omega,\mathcal{F},P)$ be a probability space, and let $X_j:\Omega\to \Omega_{X_j}$, $j=1,\ldots,d$, and $Z:\Omega\to\Omega_{Z}$ be random elements. For any subset $J \subset \{1,\dots,d\}$, let $X_J=(X_j)_{j\in J}$ denote the corresponding subvector. We use the pointwise product to define the kernel on the product space:
\begin{align*}
\kappa_{X_J}(x_J,x_J')
= \prod_{\ell\in J}
\kappa_{\ell}
(x_{\ell},x_{\ell}').
\end{align*}
Let $\mathcal H_{X_J}$ denote the reproducing kernel Hilbert space induced by $\kappa_{X_J}$.

We further define augmented variables $\ddot X_J = (X_J, Z)$. Here and throughout the paper, a double dot over a variable indicates that the variable has been augmented by adjoining $Z$. The associated kernel is then defined as
\begin{align*}
\kappa_{\ddot X_J}
\bigl((x_J,z),(x_J',z')\bigr)
= \kappa_{X_J}(x_J,x_J')\kappa_Z(z,z').
\end{align*}
Let $\mathcal H_{\ddot X_J}$ denote the reproducing kernel Hilbert space induced by
$\kappa_{\ddot X_J}$. Its inner product is also given by the product 
\begin{align*}
\left\langle f_1 g_1, f_2 g_2\right\rangle_{\mathcal H_{\ddot X_J}}
=
\left\langle f_1,f_2\right\rangle_{\mathcal H_{X_J}}
\left\langle g_1,g_2\right\rangle_{\mathcal H_Z}.
\end{align*}
where $f_1,f_2\in\mathcal H_{X_J}$, $g_1,g_2\in\mathcal H_Z$, and
$(fg)(x_J,z)=f(x_J)g(z)$. This definition extends by bilinearity to finite linear combinations and then by continuity to the whole space.
\begin{assumption}\label{asm:basic}
    The kernels $\kappa_{X_J}$ and $\kappa_Z$ are continuous, bounded, and positive-definite. The conditions hold for all $J \subset \{1, \ldots,d\}$.
\end{assumption}
By the boundedness of $\kappa_{X_J}$, the expectation functional $f\mapsto E[f(X_J)]$ is a bounded linear functional on $\mathcal H_{X_J}$. Applying Riesz representation theorem, there exists a unique $\mu_{X_J} \in \mathcal H_{X_J}$ such that
\begin{align*}
\langle f,\mu_{X_J}\rangle_{\mathcal H_{X_J}}
= \E[f(X_J)].
\end{align*}
We define $\mu_{X_J}$ to be the kernel mean embedding of $X_J$. Similarly, we define the kernel mean embeddings $\mu_Z\in\mathcal H_Z$ for $Z$ and $\mu_{\ddot X_J}\in\mathcal H_{\ddot X_J}$ for $\ddot X_J$, respectively. Then the corresponding empirical kernel mean embeddings are denoted by $\hat\mu_{X_J} = \E_n[\kappa_{X_J}(\cdot,X_J)]$, $\hat\mu_Z = \E_n[\kappa_Z(\cdot,Z)]$, and $\hat\mu_{\ddot X_J} = \E_n[\kappa_{\ddot X_J}(\cdot,\ddot X_J)]$, where $\E_n$ stands for the sample average based on i.i.d. data $\{(X_{1,i}, \dots, X_{d,i},Z_i); \ i=1,\dots,n\}$. 

We can then define the population-level and sample-level centralized kernels for $\kappa_{X_J}(\cdot,X_J)$ as $\tilde \kappa_{X_J}(\cdot, X_J) = \kappa_{X_J}(\cdot, X_J) - \mu_{X_J}$  and $\hat \kappa_{X_J} = \kappa_{X_J}(\cdot, X_J) - \E_n[\kappa_{X_J}(\cdot, X_J)]$, respectively.  Similarly define $\tilde \kappa_{\ddot X_J}(\cdot, \ddot X_J)$ and $\hat \kappa_{\ddot X_J} (\cdot, \ddot X_J)$, as well as $\tilde \kappa_Z(\cdot, Z)$ and $\hat \kappa_Z (\cdot, Z)$. We denote the space of continuous real-valued functions on $\Omega$ vanishing at infinity by $C_0(\Omega)$.

\begin{assumption}\label{asm:kernel_ex_uni} Assume the following two conditions hold, for any $J \subset \{1,\ldots, d\}$:
    \begin{enumerate}
        \item $\E [f(X_J)\mid Z =\cdot] \in \mathcal{H}_Z$ for all $f\in \mathcal{H}_{X_J}$,  and $\E [g(\ddot{X}_J)\mid Z =\cdot] \in \mathcal{H}_Z$ for all $g\in \mathcal{H}_{\ddot{X}_J}$;
        \item $\kappa_{X_J}$ and $\kappa_Z$ are $c_0$-universal, i.e., $\mathcal{H}_{X_J}$ and $\mathcal{H}_Z$ are dense in $C_{0}(\Omega_{X_J})$ and $C_{0}(\Omega_{Z})$ under the supremum norm. Then $\mathcal{H}_{\ddot{X}_J}$ is dense in $C_0(\Omega_{X_J} \times \Omega_{Z})$.
    \end{enumerate}
\end{assumption}

\subsection{Cross-Covariance Operators and Conditional Independence}

We adopt the following convention for defining cross-covariance operators. As a representative example, consider the pair $(X_J,Z)$ for some $J \subset \{1,\ldots,d\}$. The cross-covariance operator from $\mathcal{H}_Z$ to $\mathcal{H}_{X_J}$ is denoted by $\Sigma_{X_J Z}:\mathcal{H}_Z\to\mathcal{H}_{X_J}$ and is defined through the identity
\begin{align*}
    \langle f, \Sigma_{X_J Z} g \rangle_{\mathcal{H}_{X_J}}
    =
    \E[f(X_J)g(Z)]
    -
    \E[f(X_J)]\E[g(Z)],
\end{align*}
for all $f\in\mathcal{H}_{X_J}$ and $g\in\mathcal{H}_Z$, see \cite{conway2019course} and \cite{li2018linear}. By Riesz representation theorem, $\Sigma_{X_J Z}$ can be equivalently represented as
\begin{align*}
    \Sigma_{X_JZ} = \E \left[\tilde{\kappa}_{X_J}(\cdot,X) \otimes \tilde{\kappa}_{Z}(\cdot,Z)  \right].
\end{align*}
Here $f \otimes g$ stands for the linear operator defined by $(f\otimes g) h = \langle g, h \rangle f$. At the sample level, the cross-covariance operator is estimated by
\begin{align*}
\hat\Sigma_{X_JZ}
=  \E_n\left[ \hat\kappa_{X_J}(\cdot,X_J) \otimes \hat\kappa_Z(\cdot,Z) \right].
\end{align*}
The cross-covariance operators can then be defined for any other pair of random elements by replacing $X_J$ and $Z$ with the desired ones. For example, we define 
\begin{align}\label{eq:sigma-xddotz-zz}
\Sigma_{\ddot{X}_JZ} = \E \left[\tilde{\kappa}_{\ddot{X}_J}(\cdot, \ddot X_J) \otimes \tilde{\kappa}_{Z}(\cdot,Z)  \right]
\quad \text{and} \quad
\Sigma_{ZZ} = \E \left[\tilde{\kappa}_Z(\cdot,Z) \otimes \tilde{\kappa}_{Z}(\cdot,Z)  \right],
\end{align}
and the corresponding sample estimators
\begin{align*}
\hat\Sigma_{\ddot{X}_JZ} = \E_n \left[\hat{\kappa}_{\ddot{X}_J}(\cdot, \ddot X_J) \otimes \hat{\kappa}_{Z}(\cdot,Z)  \right]
\quad \text{and} \quad
\hat\Sigma_{ZZ} = \E_n \left[\hat{\kappa}_Z(\cdot,Z) \otimes \hat{\kappa}_{Z}(\cdot,Z)  \right].
\end{align*}
In particular, we give the equivalent representation of $\Sigma_{\ddot{X}_JZ}$ in \eqref{eq:sigma-xddotz-zz} as
\begin{align*}
\Sigma_{\ddot{X}_JZ} = \E \left[\big(\kappa_{X_J}(\cdot,X) \kappa_{Z}(\cdot,Z)\big) \otimes \tilde{\kappa}_{Z}(\cdot,Z)  \right],
\end{align*}
which will be repeatedly used in later sections.

Define the Moore--Penrose inverse $\Sigma_{ZZ}^{\dagger}$ on
$\operatorname{dom}(\Sigma_{ZZ}^{\dagger})
=\operatorname{ran}(\Sigma_{ZZ})\oplus\operatorname{ker}(\Sigma_{ZZ})$
as follows. For $g\in\operatorname{ran}(\Sigma_{ZZ})$, let $\Sigma_{ZZ}^{\dagger}g$ be the unique element $u\in\operatorname{ker}(\Sigma_{ZZ})^\perp$ satisfying $\Sigma_{ZZ}u=g$. For $g\in\operatorname{ker}(\Sigma_{ZZ})$, set $\Sigma_{ZZ}^{\dagger}g=0$. The definition is then extended by linearity to the entire domain. We then define the conditional cross-covariance operator following \cite{fukumizu2004dimensionality}. Let $J_1$ and $J_2$ be two disjoint subsets of $\{1,\ldots,d\}$, we define $\Sigma_{X_{J_1} X_{J_2}\mid Z}: \mathcal{H}_{X_{J_2}} \rightarrow \mathcal{H}_{X_{J_1}}$ as
\begin{align*}
    \Sigma_{X_{J_1} X_{J_2}\mid Z} = \Sigma_{X_{J_1} X_{J_2}} - \Sigma_{X_{J_1} Z} \Sigma_{ZZ}^{\dagger}  \Sigma_{Z X_{J_2}}.
\end{align*}
Similarly, we define the conditional cross-covariance operator for the augmented variable:
\begin{align*}
    \Sigma_{\ddot X_{J_1} X_{J_2}\mid Z} = \Sigma_{\ddot X_{J_1} X_{J_2}} - \Sigma_{\ddot X_{J_1} Z} \Sigma_{ZZ}^{\dagger}  \Sigma_{Z X_{J_2}}.
\end{align*}

\section{Construction of Test Statistic}\label{sec:test_statistics}
\subsection{Regression Operator}
We first express the conditional cross-covariance operators as cross-covariances of reproducing kernel Hilbert space regression residuals. These residuals are obtained by removing the components of the feature maps explained by $Z$. After mapping the variables into reproducing kernel Hilbert spaces, we transform nonlinear regression in the original variables into linear regression between their feature maps. To formalize this representation, we introduce the linear operators $B_{X_J\mid Z}$ and $B_{\ddot X_J\mid Z}$, which are the adjoints of the reproducing kernel Hilbert space regression operators considered in \cite{lee2016variable} and \cite{sang2026nonlinear}.

\begin{definition}\label{def:regression_op}
    For any $J \subset \{1,\ldots,d\}$, we define linear operators $B_{X_{J}\mid Z}: \mathcal{H}_{Z} \rightarrow \mathcal{H}_{X_{J}}$ and $B_{\ddot{X}_{J}\mid Z}: \mathcal{H}_{Z} \rightarrow \mathcal{H}_{\ddot X_{J}}$, respectively, as
    \begin{align*}
        B_{X_{J}\mid Z} = \Sigma_{X_{J} Z} \Sigma_{ZZ}^{\dagger},
    \qquad
        B_{\ddot{X}_{J}\mid Z} = \Sigma_{\ddot{X}_{J} Z} \Sigma_{ZZ}^{\dagger}.
    \end{align*}
\end{definition}
Note that both operators in Definition \ref{def:regression_op} should first be defined on $\operatorname{dom}(\Sigma_{ZZ}^\dagger) = \operatorname{ran}(\Sigma_{ZZ}) \oplus \ker(\Sigma_{ZZ})$. Under Assumption \ref{asm:kernel_ex_uni}, both operators are bounded, so their domains can be uniquely extended to the whole space $\mathcal{H}_Z$. Under Assumptions~\ref{asm:basic} and \ref{asm:kernel_ex_uni}, Corollary~1 of \cite{tang2026reproducing} shows that these operators represent the corresponding conditional mean embeddings:
\begin{align*}
\E\big[ \tilde\kappa_{X_J}(\cdot,X_J)\mid Z \big]
= & B_{X_J\mid Z}\tilde\kappa_Z(\cdot,Z),  \qquad
\E\big[ \tilde\kappa_{\ddot X_J}(\cdot,\ddot X_J)\mid Z \big]
=  B_{\ddot X_J\mid Z}\tilde\kappa_Z(\cdot,Z),
\end{align*}
which hold almost surely. Accordingly, we can define the corresponding feature-space regression residuals by
\begin{align*}
\xi_{X_J\mid Z}(\cdot,X_J;Z)
= & \tilde\kappa_{X_J}(\cdot,X_J) - B_{X_J\mid Z}\tilde\kappa_Z(\cdot,Z), \qquad
\xi_{\ddot X_J\mid Z}(\cdot,\ddot X_J;Z)
= \tilde\kappa_{\ddot X_J}(\cdot,\ddot X_J) - B_{\ddot X_J\mid Z}\tilde\kappa_Z(\cdot,Z).
\end{align*}
These are respectively $\mathcal H_{X_J}$-valued and $\mathcal H_{\ddot X_J}$-valued random elements, both measurable with respect to $\sigma(X_J,Z)$. By the preceding conditional mean identities, they are conditionally centered:
\begin{align}
\E\left[ \xi_{X_J\mid Z}(\cdot,X_J;Z) \mid Z \right] = & 0
\quad \text{and} \quad
\E\left[ \xi_{\ddot X_J\mid Z}(\cdot,\ddot X_J;Z) \mid Z \right] = 0
\qquad\text{almost surely}.
\label{eqn:mean_zero_reg_op}
\end{align}
For disjoint subsets $J_1$ and $J_2$, the corresponding conditional cross-covariance operator is the cross-covariance of the residual feature maps. That is,
\begin{align*}
    & \E \big[ \xi_{\ddot X_{J_1}\mid Z}(\cdot,X_J;Z) \otimes \xi_{X_{J_2}\mid Z}(\cdot,X_J;Z) \big] \\
    = & \E \big[ \big(\tilde\kappa_{\ddot X_{J_1}}(\cdot,\ddot X_{J_1}) - B_{\ddot X_{J_1}\mid Z}\tilde\kappa_Z(\cdot,Z)\big) \otimes \big(\tilde\kappa_{X_{J_2}}(\cdot,X_{J_2}) - B_{X_J\mid Z}\tilde\kappa_Z(\cdot,Z)\big) \big] \\
    = & \Sigma_{\ddot X_{J_1} X_{J_2}} - \Sigma_{\ddot{X}_{J_1} Z} \Sigma_{ZZ}^{\dagger} \Sigma_{Z X_{J_2}}
    - \Sigma_{\ddot{X}_{J_1} Z} \Sigma_{ZZ}^{\dagger} \Sigma_{Z X_{J_2}} + \Sigma_{\ddot{X}_{J_1} Z} \Sigma_{ZZ}^{\dagger} \Sigma_{ZZ} \Sigma_{ZZ}^{\dagger} \Sigma_{Z X_{J_2}} \\
    = & \Sigma_{\ddot X_{J_1} X_{J_2}\mid Z}.
\end{align*}
The Tikhonov-regularized empirical counterparts of the regression operators are
\begin{align*}
    \hat B_{X_J\mid Z}
    = & \hat\Sigma_{X_J Z}
    \big(\hat\Sigma_{ZZ}+\epsilon_n I\big)^{-1},
    \qquad
    \hat B_{\ddot X_J\mid Z}
    = \hat\Sigma_{\ddot X_J Z}
    \big(\hat\Sigma_{ZZ}+\epsilon_n I\big)^{-1},
\end{align*}
where $\epsilon_n>0$ is a regularization parameter satisfying
$\epsilon_n\to0$ as $n\to\infty$. Based on the residual representation, we define the empirical conditional cross-covariance operator as
\begin{align*}
    \hat\Sigma_{X_{J_1} X_{J_2}\mid Z}
    = & \E_n\big[
        \big(
            \hat\kappa_{X_{J_1}}(\cdot,X_{J_1})
            -\hat B_{X_{J_1}\mid Z}\hat\kappa_Z(\cdot,Z)
        \big)
        \otimes
        \big(
            \hat\kappa_{X_{J_2}}(\cdot,X_{J_2})
            -\hat B_{X_{J_2}\mid Z}\hat\kappa_Z(\cdot,Z)
        \big)
    \big] \\
    = & \hat\Sigma_{X_{J_1} X_{J_2}}
        -\hat B_{X_{J_1}\mid Z}\hat\Sigma_{Z X_{J_2}}
        -\hat\Sigma_{X_{J_1} Z}\hat B_{X_{J_2}\mid Z}^{*}
        +\hat B_{X_{J_1}\mid Z}\hat\Sigma_{ZZ}
            \hat B_{X_{J_2}\mid Z}^{*} \\
    = & \hat\Sigma_{X_{J_1} X_{J_2}}
    - \hat\Sigma_{X_{J_1} Z}
    \big(\hat\Sigma_{ZZ}+\epsilon_n I\big)^{-1}
    \hat\Sigma_{Z X_{J_2}} 
    - \epsilon_n \hat B_{X_{J_1}\mid Z}\hat B_{X_{J_2}\mid Z}^{*}.
\end{align*}
Similarly, for the augmented variable, we define
\begin{align*}
    \hat\Sigma_{\ddot X_{J_1} X_{J_2}\mid Z}
    = & \E_n\big[
        \big(
            \hat\kappa_{\ddot X_{J_1}}(\cdot,\ddot X_{J_1})
            -\hat B_{\ddot X_{J_1}\mid Z}\hat\kappa_Z(\cdot,Z)
        \big)
        \otimes
        \big(
            \hat\kappa_{X_{J_2}}(\cdot,X_{J_2})
            -\hat B_{X_{J_2}\mid Z}\hat\kappa_Z(\cdot,Z)
        \big)
    \big].
\end{align*}
The residual-based estimator defined above is closely related to the conventional conditional cross-covariance operator, given by
\begin{align*}
    \hat \Sigma_{X_{J_1} X_{J_2}} - \hat \Sigma_{X_{J_1} Z} \big(\hat \Sigma_{ZZ} + \epsilon_n I\big)^{-1} \hat \Sigma_{Z X_{J_2}}.
\end{align*}
Compared with our construction, as illustrated in the next subsection, this formulation includes an additional term, $\epsilon_n \hat B_{X_{J_1}\mid Z}\hat B_{X_{J_2}\mid Z}^{*}$. This term converges to zero in Hilbert--Schmidt norm, ensuring that the two constructions share the same consistency limit. However, under our choice of $\epsilon_n$, this term need not be negligible at the $n^{-1/2}$ scale. Moreover, the two constructions employ different coordinate mappings.

\subsection{Construction of Test Statistic}
We first state the theorem below to emphasize the key role played by (conditional) cross-covariance operators in discussing (conditional) independence.
\begin{theorem}\label{thm:cross_cor_op} Under Assumptions \ref{asm:basic}--\ref{asm:kernel_ex_uni}, we have
    \begin{enumerate}
        \item \cite[Theorem~4]{gretton2005measuring} $\Sigma_{X_{J_1} X_{J_2}} = 0 \Longleftrightarrow X_{J_1} \Perp X_{J_2}$;
        \item \cite[Corollary~9]{fukumizu2004dimensionality} $\Sigma_{\ddot X_{J_1} X_{J_2} \mid Z} = 0 \Longleftrightarrow X_{J_1} \Perp X_{J_2} \mid Z$.
    \end{enumerate}
    where $X_{J_1} \Perp X_{J_2} \mid Z$ means $P_{(X_{J_1},X_{J_2})\mid Z} = P_{X_{J_1}\mid Z} \otimes P_{X_{J_2}\mid Z}$, $P_Z$-almost surely.
\end{theorem}


The cross-covariance operators offer a convenient tool to characterize conditional independence. Another useful fact is that the mutual independence between $d$ variables can be equivalently derived by $d-1$ pairwise independence relations. For $j=1,\ldots,d-1$, write $j+ = \{j+1,\ldots,d\}$. Then
\begin{align}\label{eqn:indep_chain}
    X_1 \Perp X_2 \Perp \ldots \Perp X_d \mid Z \quad \Longleftrightarrow \quad 
    X_j\Perp X_{j+}\mid Z, \quad \text{for} \ j=1,\ldots,d-1.
\end{align} 
We combine the corresponding conditional cross-covariance operators through a Hilbert direct sum to obtain a unified characterization of multi-way conditional independence. Specifically, let
\begin{align*}
\mathbb H_S
=
\bigoplus_{j=1}^{d-1}
\mathcal{B}_2\big(
\mathcal H_{X_{j+}},
\mathcal H_{\ddot X_j}
\big),
\end{align*}
where $\mathcal B_2(\mathcal H,\mathcal G)$ denotes the space of Hilbert--Schmidt operators from $\mathcal H$ to $\mathcal G$. A member of $\mathbb H_S$ is a $(d-1)$-dimensional vector of linear operators in $\mathcal B_2(\mathcal H_{X_{1+}},\mathcal H_{\ddot X_1}) \times \dots \times \mathcal B_2(\mathcal H_{X_{(d-1)+}},\mathcal H_{\ddot X_{d-1}})$, with the inner product defined as follows: for any $U=(U_1,\ldots,U_{d-1})$ and $V=(V_1,\ldots,V_{d-1})$ in $\mathbb H_S$,
\begin{align*}
\langle U,V\rangle_{\mathbb H_S}
=
\sum_{j=1}^{d-1}
\langle U_j,V_j\rangle_{\mathrm{HS}}
=
\sum_{j=1}^{d-1}
\operatorname{tr}(U_j^*V_j),
\end{align*}
where $U_j^*$ denotes the adjoint of $U_j$. The induced norm satisfies $\|U\|_{\mathbb H_S}^{2} = \sum_{j=1}^{d-1}\|U_j\|_{\mathrm{HS}}^{2}$. Follwing the idea in \eqref{eqn:indep_chain}, we define $S\in\mathbb H_S$ as
\begin{align*}
S=
\big(
\Sigma_{\ddot X_1X_{1+}\mid Z},
\Sigma_{\ddot X_2X_{2+}\mid Z},
\ldots,
\Sigma_{\ddot X_{d-1}X_d\mid Z}
\big)^\top.
\end{align*}
Its squared norm is therefore
\begin{align*}
\|S\|_{\mathbb H_S}^2
=
\sum_{j=1}^{d-1}
\big\|
\Sigma_{\ddot X_jX_{j+}\mid Z}
\big\|_{\mathrm{HS}}^2.
\end{align*}
The following theorem establishes that $S$ is a well-defined element of $\mathbb H_S$ and characterizes multi-way conditional independence through its vanishing.
\begin{theorem}\label{thm:cond_indep} 
Under Assumptions \ref{asm:basic} and \ref{asm:kernel_ex_uni}, we have
\begin{align*}
    S = 0
    \quad \Longleftrightarrow \quad X_1 \Perp X_2 \Perp \ldots \Perp X_d  \mid Z.
\end{align*}
Furthermore, by boundedness of kernels, we have
\begin{align*}
    \|S\|_{\mathbb H_S} = 0 \quad \Longleftrightarrow \quad X_1 \Perp X_2 \Perp \ldots \Perp X_d  \mid Z.
\end{align*} 
\end{theorem}

This theorem provides the population foundation for the proposed test. It converts multi-way conditional independence into a single zero-norm condition in $\mathbb H_S$. Each component of $S$ characterizes conditional independence between one variable and the joint block of variables that follow it. The direct-sum norm aggregates these restrictions and is strictly positive whenever at least one of them fails. At the sample level, we collect the regularized empirical operators into
\[
\hat S=
\big(
\hat\Sigma_{\ddot X_1X_{1+}\mid Z},
\hat\Sigma_{\ddot X_2X_{2+}\mid Z},
\ldots,
\hat\Sigma_{\ddot X_{d-1}X_d\mid Z}
\big)^\top
\]
and define the test statistic as
\begin{align*}
T_n
= n\|\hat S\|_{\mathbb H_S}^2
=
n\sum_{j=1}^{d-1}
\big\|
\hat\Sigma_{\ddot X_jX_{j+}\mid Z}
\big\|_{\mathrm{HS}}^2.
\end{align*}
The value of the proposed statistic may depend on that ordering. If symmetry is desired, one may consider another construction by averaging the statistic over all permutations, but this can entail substantial additional computation. A more economical alternative is to compare each $X_j$ with the joint block $X_{\{1,\ldots,d\}\setminus\{j\}}$ and aggregate the corresponding squared Hilbert--Schmidt norms over $j=1,\ldots,d$. It uses $d$ components instead of $d-1$, and common computations involving $Z$ can be reused, so the additional computational burden is modest. Although we do not develop this alternative here, we could study it within the same framework and derive a similar characterization, adjusting the joint covariance operator and the resulting eigenvalue weights accordingly.

\section{Asymptotic Properties}\label{sec:asy}

\subsection{Convergence Rate of Regression Operators}

We first study the convergence of the empirical regression operators. Our analysis requires smoothness conditions on the relevant cross-moment operators and an eigenvalue-decay condition on $\Sigma_{ZZ}$. These conditions allow us to derive convergence rates and control the estimation errors that enter the subsequent asymptotic analysis. For the augmented operator $\hat B_{\ddot X_J\mid Z}$, we first decompose $\Sigma_{\ddot X_JZ}$ so that different moment components can be treated separately. To express this decomposition, we introduce the following
embedding operators.

For any subset $J \subset \{1,\ldots,d\}$, and any $\mu\in\mathcal H_{X_J}$, define the bounded linear operator $T_\mu: \mathcal H_Z \rightarrow \mathcal H_{\ddot X_J}$ as
\begin{align*}
    T_\mu f = \mu f, \qquad \text{for any} \ f \in \mathcal{H}_Z.
\end{align*}
Notice that $\norm{T_\mu}_\mathrm{op} = \norm{\mu}_{\mathcal{H}_{X_J}}$. Similarly, for any $\nu \in \mathcal{H}_Z$  we can define $R_\nu: \mathcal H_{X_J} \rightarrow \mathcal H_{\ddot X_J}$ as 
\begin{align*}
    R_\nu g = g \nu, \qquad \text{for any } g \in \mathcal{H}_{X_J},
\end{align*} 
with $\norm{R_\nu}_\mathrm{op} = \norm{\nu}_{\mathcal{H}_Z}$. Here, we slightly abuse notation by writing $T_\mu$ and $R_\nu$ without explicitly indicating their range/domain $\mathcal H_{\ddot X_J}/\mathcal H_{X_J}$, which varies with $J$ and will be clear from the context.

Since $\E[\tilde\kappa_Z(\cdot,Z)]=0$, we have
\begin{align}
    \Sigma_{\ddot X_J Z}
    = &
    \E\left[ \kappa_{X_J}(\cdot,X_J) \kappa_Z(\cdot,Z) \otimes \tilde\kappa_Z(\cdot,Z) \right] \notag \\
    = &
    \E\left[ \big( \mu_{X_J} + \tilde\kappa_{X_J}(\cdot,X_J) \big) \big( \mu_Z + \tilde\kappa_Z(\cdot,Z) \big) \otimes \tilde\kappa_Z(\cdot,Z) \right] \notag \\
    = &
    T_{\mu_{X_J}}\Sigma_{ZZ}
    +
    R_{\mu_Z}\Sigma_{X_J Z}
    +
    \E\left[ \tilde\kappa_{X_J}(\cdot,X_J) \tilde\kappa_Z(\cdot,Z) \otimes \tilde\kappa_Z(\cdot,Z) \right], \label{eqn:Sigma_Xddot}
\end{align}
where the third equality comes from the fact that for any $g \in \mathcal{H}_Z$,
\begin{align*}
    \E\left[ \left( \tilde \kappa_{X_J}(\cdot, X_J) \mu_Z \right) \otimes \tilde \kappa_{Z}(\cdot, Z) \right] g 
    = & \E\left[ \big( \tilde \kappa_{X_J}(\cdot, X_J) \mu_Z \big) \big(g(Z) - \E[g(Z)]\big) \right] \\
    = & \big(\Sigma_{X_J Z}  g \big)\mu_Z = R_{\mu_Z}\Sigma_{X_J Z} g.
\end{align*}

Therefore, $\Sigma_{\ddot X_J Z}$ consists of a term determined by the
marginal covariance operator $\Sigma_{ZZ}$, a term determined by the
cross-covariance operator $\Sigma_{X_J Z}$, and a genuinely centered
third-order term, denoted by
$\Lambda_{\ddot X_J Z} = \E\left[ \tilde\kappa_{X_J}(\cdot,X_J) \tilde\kappa_Z(\cdot,Z) \otimes \tilde\kappa_Z(\cdot,Z) \right].$ Similarly, we can derive the empirical version of the decomposition:
\begin{align*}
    \hat \Sigma_{\ddot X_J Z}
    = T_{\hat \mu_{X_J}}\hat \Sigma_{ZZ}
    +
    R_{\hat \mu_Z}\hat \Sigma_{X_J Z}
    +
    \E_n\left[ \hat \kappa_{X_J}(\cdot,X_J) \hat \kappa_Z(\cdot,Z) \otimes \hat \kappa_Z(\cdot,Z) \right].
\end{align*}

\begin{assumption}\label{asm:inverse_beta}
    There exists some $\beta >0$ such that for any $J \subset \{1,\ldots,d\}$ we have
    \begin{enumerate}
        \item $\Sigma_{X_{J} Z} = S_{X_{J} Z} \Sigma_{ZZ}^{1+\beta}$ for some bounded linear operator $S_{X_{J} Z}: \mathcal{H}_Z \rightarrow \mathcal{H}_{X_{J}}$;
        \item $\Lambda_{\ddot X_{J}Z} = S_{\ddot{X}_{J} Z} \Sigma_{ZZ}^{1+\beta}$ for some bounded linear operator $S_{\ddot{X}_{J} Z}: \mathcal{H}_Z \rightarrow \mathcal{H}_{\ddot X_{J}}$.
    \end{enumerate}
\end{assumption}

\begin{assumption}\label{asm:Sig_ZZ_eigen}
    Let $\gamma_1 \geq \gamma_2 \geq \ldots$ be ordered eigenvalues of $\Sigma_{ZZ}$. Assume $\gamma_k \asymp k^{-\alpha}$ for some $\alpha > 1$.
\end{assumption}

The preceding decomposition separates the component of
$\Sigma_{\ddot X_JZ}$ inherited from $\Sigma_{ZZ}$ from the remaining
components governed by the source condition. Together with the
eigenvalue-decay assumption, this decomposition yields convergence rates
for the empirical regression operator $\hat B_{X_J\mid Z}$ and the
corrected version of the augmented regression operator $\hat{B}_{\ddot X_{J}\mid Z}$, as stated in Theorem \ref{thm:rate_reg_operator} and Proposition \ref{prop:optimal_rate_reg_operator} below.
\begin{theorem}\label{thm:rate_reg_operator}
    Suppose Assumptions~\ref{asm:basic}--\ref{asm:Sig_ZZ_eigen} hold. If $\epsilon_n \rightarrow 0$ as $n \rightarrow \infty$, then
    \begin{enumerate}
        \item $\norm{\hat{B}_{X_{J}\mid Z} - B_{X_{J}\mid Z}}_\mathrm{op} = O_P \left(r_n \right)$, and
        \item $\big\|\big(\hat{B}_{\ddot X_{J}\mid Z} - T_{\hat \mu_{X_J}} \hat \Sigma_{ZZ} (\hat{\Sigma}_{ZZ} + \epsilon_n I)^{-1}\big) - \big(B_{\ddot X_{J}\mid Z} - T_{\mu_{X_J}} \Sigma_{ZZ} \Sigma_{ZZ}^\dagger \big)\big\|_\mathrm{op}  
        = O_P \left(r_n \right)$,
    \end{enumerate}
    where $r_n=n^{-1/2} \epsilon_{n}^{\beta \wedge 1 - 1} + \epsilon_{n}^{\beta \wedge 1} + n^{-1} \epsilon_{n}^{-\frac{3\alpha+1}{2\alpha}} + n^{-1/2} \epsilon_{n}^{-\frac{\alpha+1}{2\alpha}}$.
\end{theorem}
\begin{proof}
    See Appendix. For assertion $1$, the proof is the same as \cite{sang2026nonlinear}. 
\end{proof}


\begin{proposition}
\label{prop:optimal_rate_reg_operator}
Suppose the conditions of Theorem \ref{thm:rate_reg_operator} hold and
let $\epsilon_n \asymp n^{-\delta}$ for some $\delta>0$. The choice of $\delta$
that optimizes the convergence rate $r_n$ in Theorem
\ref{thm:rate_reg_operator} is as follows:
\begin{enumerate}
    \item If $\beta > \frac{\alpha-1}{2\alpha}$, then $\delta_{\mathrm{opt}} = \frac{\alpha} {2\alpha(\beta\wedge1)+\alpha+1}$. Under this choice,
    the optimal rate of $r_n$ is $O_P( n^{-\frac{\alpha(\beta\wedge1)}{2\alpha(\beta\wedge1)+\alpha+1}})$;
    \item If $0< \beta \leq \frac{\alpha-1}{2\alpha}$, then
    $\delta_{\mathrm{opt}}=\frac{1}{2}$. Under this
    choice, the optimal rate of $r_n$ is $O_P(n^{-\beta/2})$.
\end{enumerate}
\end{proposition}

\begin{proof}
    See Theorem~10 in \cite{sang2026nonlinear}.
\end{proof}

In the first regime, the exponent in the optimal rate is $-\frac{\alpha(\beta\wedge1)}{2\alpha(\beta\wedge1)+\alpha+1}$. For fixed $\alpha$, this exponent is decreasing in $\beta\wedge1$ and attains its smallest value when $\beta \geq 1$. In this case, the convergence rate becomes $O_P(n^{-\frac{\alpha}{3\alpha+1}})$. Since $\alpha>1$,
\begin{align*}
    \frac14 < \frac{\alpha}{3\alpha+1} < \frac13, \quad \frac{\alpha}{3\alpha+1} \rightarrow \frac13 \quad\text{as }\alpha\to\infty.
\end{align*}
Thus, the fastest rate allowed by $r_n$ can be arbitrarily close to $n^{-1/3}$. More generally, the rate is $o_P(n^{-1/4})$
whenever $\beta \wedge 1 > \frac{\alpha+1}{2\alpha}$.


Assertion 2 of Theorem \ref{thm:rate_reg_operator} controls the difference between the corrected regression operators, rather than $\hat B_{\ddot X_J\mid Z} - B_{\ddot X_J\mid Z}$ itself. To isolate the remaining projection component, fix a subset $J$ and define
\begin{align}
    R_n = T_{\hat\mu_{X_J}} \hat\Sigma_{ZZ} (\hat\Sigma_{ZZ}+\epsilon_nI)^{-1} - T_{\mu_{X_J}} \Sigma_{ZZ}\Sigma_{ZZ}^{\dagger}.  \label{eqn:def_Rn}
\end{align}
The dependence of $R_n$ on $J$ is suppressed in the notation.
Then
\begin{align*}
    \hat B_{\ddot X_J\mid Z}-B_{\ddot X_J\mid Z}
    = & \big[ \big( \hat B_{\ddot X_J\mid Z} - T_{\hat\mu_{X_J}} \hat\Sigma_{ZZ}(\hat\Sigma_{ZZ}+\epsilon_nI)^{-1} \big) 
    - \big( B_{\ddot X_J\mid Z} - T_{\mu_{X_J}} \Sigma_{ZZ}\Sigma_{ZZ}^{\dagger}
        \big)
    \big]
    +R_n.
\end{align*}
The first term on the right-hand side is controlled by Theorem~\ref{thm:rate_reg_operator}. In contrast, $R_n$ need not converge to
zero in operator norm, because the regularized empirical projection
$\hat\Sigma_{ZZ}(\hat\Sigma_{ZZ}+\epsilon_nI)^{-1}$ need not
converge to $\Sigma_{ZZ}\Sigma_{ZZ}^{\dagger}$ in operator norm.

Nevertheless, $R_n$ only appears after composition with a positive
power of $\Sigma_{ZZ}$ in the subsequent asymptotic expansion. The
following results show that this smoothing operation is sufficient to
make $R_n$ asymptotically negligible, which are parallel to \cite{tang2026reproducing}.

\begin{lemma}
\label{lem:smoothed_projection_error}
Suppose Assumptions \ref{asm:basic}--\ref{asm:Sig_ZZ_eigen} hold and
$\epsilon_n\to0$. Then, for every $\theta>0$,
\begin{align}
    \norm{ R_n\Sigma_{ZZ}^{\theta} }_{\mathrm{op}}
    = O_P\big( n^{-1/2}\epsilon_n^{\theta\wedge1-1}
        + \epsilon_n^{\theta\wedge1} \big).
    \label{eqn:smoothed_Rn_rate}
\end{align}
In particular, $\|R_n\Sigma_{ZZ}\|_{\mathrm{op}} = O_P(n^{-1/2}+\epsilon_n)$.
\end{lemma}

\begin{proposition}\label{prop:remainder_op1}
Suppose Assumptions \ref{asm:basic}--\ref{asm:Sig_ZZ_eigen} hold and $\epsilon_n = n^{-\alpha /(2\alpha(\beta\wedge1)+\alpha+1)}$, with $\beta \wedge 1 > \frac{\alpha + 1}{2 \alpha}$ and  $\alpha > 2$. For each $j=1,\ldots,d-1$, let $R_n$ denote the operator in \eqref{eqn:def_Rn} with $J=\{j\}$. Then
\begin{align}
    \sqrt n \norm{ R_n\Sigma_{ZZ} \big( \hat B_{X_{j+}\mid Z} - B_{X_{j+}\mid Z} \big)^*}_{\mathrm{HS}}
    = o_P(1).
    \label{eqn:mixed_projection_remainder}
\end{align}
\end{proposition}

\subsection{Asymptotic Null Distribution}

We now derive the asymptotic distribution of $\hat S$. Using the convergence results above, we show that the errors due to regression estimation and regularization contribute only an $o_P(n^{-1/2})$ remainder to $\hat S$ under the null hypothesis. We first establish asymptotic normality for each component and then use these results to derive the joint asymptotic distribution of $\hat S$.

\begin{lemma}\label{lem:asymptotic_normality}
Suppose Assumptions~\ref{asm:basic}--\ref{asm:Sig_ZZ_eigen}
hold. Choose $\epsilon_n = n^{-\frac{\alpha}{2\alpha(\beta\wedge1)+\alpha+1}}$ according to Proposition~\ref{prop:optimal_rate_reg_operator}, with $\beta \wedge 1 > \frac{\alpha + 1}{2 \alpha}$ and  $\alpha > 2$. For $j = 1, \dots, d-1$, define
\begin{align}\label{eqn:influence}
\psi(\ddot X_{j},X_{j+};Z)
=
\xi_{\ddot X_{j}\mid Z}(\cdot,\ddot X_{j};Z)
\otimes
\xi_{X_{j+}\mid Z}(\cdot,X_{j+};Z)
-
\Sigma_{\ddot X_{j}X_{j+}\mid Z}.
\end{align}
Then $\hat\Sigma_{\ddot X_{j}X_{j+}\mid Z}$ admits
the asymptotic linear representation
\begin{align}
\sqrt n\left(
\hat\Sigma_{\ddot X_{j}X_{j+}\mid Z}
-
\Sigma_{\ddot X_{j}X_{j+}\mid Z}
\right)
=
\frac1{\sqrt n}\sum_{i=1}^n
\psi(\ddot X_{j,i},X_{j+,i};Z_i)
+
o_P(1),
\label{eqn:component_asymptotic_linearity}
\end{align}
where the remainder is measured in the Hilbert--Schmidt norm.
Consequently,
\begin{align*}
\sqrt n\left(
\hat\Sigma_{\ddot X_{j}X_{j+}\mid Z}
-
\Sigma_{\ddot X_{j}X_{j+}\mid Z}
\right)
\xrightarrow{\mathcal D}
N\big(0,\Gamma_{\ddot X_{j}X_{j+}\mid Z}\big),
\end{align*}
where $\Gamma_{\ddot X_{j}X_{j+}\mid Z}:\mathcal H_{\ddot X_{j}}\otimes\mathcal H_{X_{j+}}
\to \mathcal H_{\ddot X_{j}}\otimes\mathcal H_{X_{j+}}$ is the covariance operator
\begin{align}\label{eqn:cov_influence}
\Gamma_{\ddot X_{j}X_{j+}\mid Z}
=
\E\left[
\psi(\ddot X_{j},X_{j+};Z)^{\otimes2}
\right].
\end{align}
\end{lemma}
Note that \eqref{eqn:component_asymptotic_linearity} indicates that $\psi(\ddot X_{j},X_{j+};Z)$ is the influence function of $\hat\Sigma_{\ddot X_{j}X_{j+}\mid Z}$.
Once these influence functions have been identified, the asymptotic distribution of $\hat S$ follows directly from its asymptotically linear
representation. Combining the asymptotic linear representations of the individual
components yields the following joint Gaussian limit:
\begin{theorem}\label{thm:S_asymptotic}
    Under the same assumptions as Lemma~\ref{lem:asymptotic_normality}, we have
    \begin{align}\label{eqn:test_asym}
        \sqrt{n} \big(\hat{S} - S\big) \xrightarrow{\mathcal{D}} N\left(0, \Gamma_{S\mid Z}\right)
    \end{align}
    where $\Gamma_{S \mid Z} :  \mathbb H_S \rightarrow \mathbb H_S$ is
    \begin{align*}
        \left(\Gamma_{S\mid Z}\right)_{(j,k)} = \E \left[ \psi (\ddot{X}_{j}, X_{j+}; Z) \otimes \psi (\ddot{X}_{k}, X_{k+}; Z) \right]
        = \E \left[\psi_{j} \otimes \psi_{k}\right].
    \end{align*}
\end{theorem}

The joint Gaussian limit determines the asymptotic null distribution of the test statistic. Under multi-way conditional independence $S=0$, the continuous mapping theorem yields the following weighted chi-square limit for $n\|\hat S\|_{\mathbb H_S}^2$.
\begin{corollary}
     Assume the null hypothesis $S=0$ holds. Then under the same assumptions of Theorem~\ref{thm:S_asymptotic}, we have
    \begin{align}\label{eqn:asym_dist}
        n \|\hat{S}\|^{2}_{\mathbb H_S}
        \xrightarrow{\mathcal{D}} \sum_{k=1}^{\infty} \lambda_{k} Z_{k}^{2}
    \end{align}
    where $\lambda_{k}$ are the eigenvalues of $\Gamma_{S\mid Z}$ in \eqref{eqn:test_asym} with ${\Sigma}_{\ddot{X}_{j} X_{j+} \mid Z} = 0$, and $Z_k$ are i.i.d. standard Gaussian variables.
\end{corollary}

\subsection{Asymptotic Distribution under Local Alternatives}

We next study the behavior of the proposed test under alternatives that
converge to the null hypothesis as the sample size increases. Let $P_n$ denote the distribution of each observation in row $n$, and attach a subscript $n$ to population quantities computed under $P_n$.  In particular, write $S_n$ for the vector of conditional cross-covariance operators and $\Sigma_{X_J Z,n}$, $\Sigma_{ZZ,n}$ for the cross-covariance operator under $P_n$; $\Lambda_{\ddot X_J Z,n}$ for the third-order term.

\begin{assumption} \label{asm:uniform_local}
For every $n$, assume Assumptions \ref{asm:basic}--\ref{asm:Sig_ZZ_eigen}
hold under $P_n$ with the same $\alpha$ and $\beta$. The kernels and their associated reproducing kernel Hilbert spaces do not depend on $n$. Moreover, for every relevant subset $J \subset \{1,\ldots,d\}$, there exist uniformly bounded linear operators $\{S_{X_J Z,n}\}_n$ and $\{S_{\ddot X_J Z,n}\}_n$, with
\begin{align*}
\sup_{n\geq1}\norm{S_{X_JZ,n}}_{\mathrm{op}}<\infty,
\qquad
\sup_{n\geq1}\norm{S_{\ddot X_JZ,n}}_{\mathrm{op}}<\infty,
\end{align*}
such that
\begin{align*}
\Sigma_{X_JZ,n}
=S_{X_JZ,n}\Sigma_{ZZ,n}^{1+\beta},\quad
\Lambda_{\ddot X_JZ,n}
=S_{\ddot X_JZ,n}\Sigma_{ZZ,n}^{1+\beta}.
\end{align*}
Let $\{\gamma_{k,n}\}_{k\geq1}$ be the ordered eigenvalues of $\Sigma_{ZZ,n}$. We assume that $\gamma_{k,n}\asymp k^{-\alpha}$
uniformly in $n$, with both the exponent $\alpha$ and the implicit
comparison constants independent of $n$.
\end{assumption}

Under $P_n$, define
\begin{align*}
\psi_{j,n}
=  
\xi_{\ddot X_j\mid Z,n}(\cdot,\ddot X_j;Z)\otimes \xi_{X_{j+}\mid Z,n} (\cdot,X_{j+};Z)
- \Sigma_{\ddot X_jX_{j+}\mid Z,n}, \quad j=1,\ldots,d-1,
\end{align*}
and let $\psi_n=(\psi_{1,n},\ldots,\psi_{d-1,n})^\top$, which serves as the influence function under the local alternative distribution. Also define $\Gamma_{S\mid Z,n} = \E_{P_n}[\psi_n\otimes\psi_n].$ Indeed $\Gamma_{S\mid Z,n}$ is a trace-class \footnote{For a bounded linear operator $A$ on $\mathbb H_S$, define
$\|A\|_{\mathrm{tr}}
=
\operatorname{tr} \left((A^*A)^{1/2}\right)
$. The operator $A$ is called trace-class if $\|A\|_{\mathrm{tr}}<\infty$.} operator on $\mathbb H_S$. The bounds on the kernels and regression operators imply that $\psi_n$ is bounded, then
\begin{align}
\|\Gamma_{S\mid Z,n}\|_{\mathrm{tr}}
&\le
\E_{P_n}\|\psi_n\otimes\psi_n\|_{\mathrm{tr}}
=
\E_{P_n}\|\psi_n\|_{\mathbb H_S}^{2}
<\infty. \label{eqn:gamma_trace_class}
\end{align}
Therefore $\Gamma_{S\mid Z,n}$ is a trace-class operator on
$\mathbb H_S$ by definition.

\begin{theorem}
\label{thm:local_alternative}
Suppose Assumption~\ref{asm:uniform_local} holds with $\alpha > 2$ and
$\beta \wedge 1 > \frac{\alpha+1}{2\alpha}$. Set $\epsilon_n = n^{-\frac{\alpha}{2\alpha(\beta\wedge1)+\alpha+1}}$, as in Proposition~\ref{prop:optimal_rate_reg_operator}. Assume further that
\begin{align}
\norm{
\Gamma_{S\mid Z,n}
-
\Gamma_{S\mid Z,0}
}_{\mathrm{tr}}
\to 0,
\label{eqn:local_covariance_convergence}
\end{align}
where $\Gamma_{S\mid Z,0}$ is a trace-class operator. Consider local alternatives satisfying
\begin{align}
S_n=n^{-1/2}S_h
\label{eqn:local_alternative}
\end{align}
for some fixed $S_h\in \mathbb H_S$. Then
\begin{align}
\sqrt n(\hat S_n - S_n) \xrightarrow{\mathcal D} G,
\quad G\sim N(0,\Gamma_{S\mid Z,0}),
\label{eqn:local_clt}
\end{align}
and
\begin{align}
n\norm{\hat S_n}_{\mathbb H_S}^2
\xrightarrow{\mathcal D}
\norm{G+S_h}_{\mathbb H_S}^2.
\label{eqn:local_norm_limit}
\end{align}
Suppose additionally that
$S_h\in\left(\ker(\Gamma_{S\mid Z,0})\right)^\perp$. If we adopt the decomposition
\begin{align*}
    \Gamma_{S\mid Z,0} =  \sum_{k=1}^{\infty}\lambda_k(v_k\otimes v_k), \quad
    S_h = \sum_{k=1}^{\infty}\sigma_k v_k,
\end{align*}
where $\lambda_k>0$ are the eigenvalues and $v_k$ are the corresponding orthonormal eigenvectors, then
\begin{align}
n\norm{\hat S_n}_{\mathbb H_S}^2
\xrightarrow{\mathcal D}
\sum_{k=1}^{\infty}
\big(\sqrt{\lambda_k}Z_k+\sigma_k\big)^2,
\label{eqn:local_weighted_chisq}
\end{align}
where $Z_1,Z_2,\ldots$ are independent standard normal random
variables.
\end{theorem}

\section{Unconditional Independence}\label{sec:unc}

The unconditional case is a mathematical subproblem of the preceding conditional framework, by taking $Z$ to be a degenerate random variable. Its asymptotic results are established separately below and do not require Assumptions \ref{asm:kernel_ex_uni}--\ref{asm:Sig_ZZ_eigen} concerning regression on $Z$. In the unconditional setting, we define $\mathbb H_S^{\mathrm{uc}}$
\begin{align*}
\mathbb H_S^{\mathrm{uc}}
=
\bigoplus_{j=1}^{d-1}
\mathcal B_2\big(\mathcal H_{X_{j+}},\mathcal H_{X_j}\big).
\end{align*}
In that case, the conditional cross-covariance operator degenerates to the unconditional one. Therefore, we define
\begin{align*}
    S^{\mathrm{uc}}
    =
    \big(
    \Sigma_{X_1X_{1+}},
    \Sigma_{X_2X_{2+}},
    \ldots,
    \Sigma_{X_{d-1}X_d}
    \big)^\top
\end{align*}
and its sample counterpart
\begin{align*}
    \hat S^{\mathrm{uc}}
    =
    \big(
    \hat\Sigma_{X_1X_{1+}},
    \hat\Sigma_{X_2X_{2+}},
    \ldots,
    \hat\Sigma_{X_{d-1}X_d}
    \big)^\top.
\end{align*}
For $j=1,\ldots,d-1$, we can similarly show that
\begin{align*}
\psi_j^{\mathrm{uc}}(X_j,X_{j+}) = \tilde\kappa_{X_j}(\cdot,X_j) \otimes \tilde\kappa_{X_{j+}} (\cdot,X_{j+}) - \Sigma_{X_jX_{j+}}
\end{align*}
is the influence function of $\hat\Sigma_{X_jX_{j+}}$, and define
\begin{align*}
    \psi^{\mathrm{uc}}
    =
    \left(
    \psi_1^{\mathrm{uc}},
    \ldots,
    \psi_{d-1}^{\mathrm{uc}}
    \right)^\top,
    \quad
    \Gamma_S^{\mathrm{uc}}
    =
    \E\left[
    \psi^{\mathrm{uc}}\otimes\psi^{\mathrm{uc}}
    \right].
\end{align*}
The corresponding results are proposed in the theorem below.
\begin{theorem}\label{thm:unconditional}
Suppose for $j=1,\dots,d-1$, kernels $\kappa_{X_j}$ and $\kappa_{X_{j+}}$  are continuous,
bounded, positive definite, and $c_0$-universal. Then the following statements hold.
\begin{enumerate}
    \item The random variables $X_1,\ldots,X_d$ are mutually independent
    if and only if $S^{\mathrm{uc}}=0$; or equivalently $\Sigma_{X_1X_{1+}} = 0, \Sigma_{X_2X_{2+}} = 0, \cdots, \Sigma_{X_{d-1}X_d} =0.$
    \item We have
    \begin{align*}
        \sqrt n
        \big(
        \hat S^{\mathrm{uc}} - S^{\mathrm{uc}}
        \big)
        \xrightarrow{\mathcal D}
        N\big(0,\Gamma_S^{\mathrm{uc}}\big).
    \end{align*}

    \item In particular, if $X_1,\ldots,X_d$ are mutually independent,
    then
    \begin{align*}
        n\left\|\hat S^{\mathrm{uc}}\right\|_{\mathbb H_S^{\mathrm{uc}}}^2
        \xrightarrow{\mathcal D}
        \sum_{k=1}^{\infty}\lambda_kZ_k^2,
    \end{align*}
    where $\{\lambda_k\}_{k\geq1}$ are the eigenvalues of
    $\Gamma_S^{\mathrm{uc}}$ and
    $Z_1,Z_2,\ldots$ are independent standard normal random variables.
\end{enumerate}
\end{theorem}
We next consider local alternatives for the unconditional test.
Let $P_n$ denote the distribution of each observation in row $n$, and attach a subscript $n$ to population quantities computed under $P_n$. In particular, define $\tilde\kappa_{X_J,n}(\cdot,X_J) = \kappa_{X_J}(\cdot,X_J)-\mu_{X_J,n}$. Then for $j=1,\ldots,d-1$, let
\begin{align}
\psi_{j,n}^{\mathrm{uc}}
= &
\tilde\kappa_{X_j,n}(\cdot,X_j)
\otimes
\tilde\kappa_{X_{j+},n}(\cdot,X_{j+})
-
\Sigma_{X_jX_{j+},n}, \label{eqn:def_psi_uc_j}\\
\psi_n^{\mathrm{uc}}
= &
\big(
\psi_{1,n}^{\mathrm{uc}},\ldots,
\psi_{d-1,n}^{\mathrm{uc}}
\big)^\top,
\quad
\Gamma_{S,n}^{\mathrm{uc}}
=
\E_{P_n}\left[
\psi_n^{\mathrm{uc}}\otimes\psi_n^{\mathrm{uc}}
\right] \in \mathbb H_S^{\mathrm{uc}}. \label{eqn:def_psi_uc}
\end{align}
By a similar argument as in \eqref{eqn:gamma_trace_class}, we know $\Gamma_{S,n}^{\mathrm{uc}}$ is a trace-class operator.
\begin{theorem}
\label{thm:unconditional_local_alternative}
Suppose for $j=1,\dots,d-1$, kernels $\kappa_{X_j}$ and $\kappa_{X_{j+}}$  are continuous,
bounded, positive definite, and $c_0$-universal. The kernels and their associated reproducing kernel Hilbert spaces are fixed and do not depend
on $n$. Assume that
\begin{align}
\big\|
\Gamma_{S,n}^{\mathrm{uc}}
-
\Gamma_{S,0}^{\mathrm{uc}}
\big\|_{\mathrm{tr}}
\to 0,
\label{eqn:uc_local_covariance_convergence}
\end{align}
where $\Gamma_{S,0}^{\mathrm{uc}}$ is a trace-class operator on $\mathbb H_S^{\mathrm{uc}}$. Consider local alternatives satisfying
\begin{align}
S_n^{\mathrm{uc}}
=
n^{-1/2}S_h^{\mathrm{uc}}
\label{eqn:uc_local_alternative}
\end{align}
for some fixed $S_h^{\mathrm{uc}}\in\mathbb H_S^{\mathrm{uc}}$.
Then
\begin{align}
\sqrt n\big(
\hat S_n^{\mathrm{uc}}-S_n^{\mathrm{uc}}
\big)
\xrightarrow{\mathcal D}G,
\quad
G\sim N\big(0,\Gamma_{S,0}^{\mathrm{uc}}\big),
\label{eqn:uc_local_clt}
\end{align}
and
\begin{align}
n\left\|\hat S_n^{\mathrm{uc}}\right\|_{\mathbb H_S^{\mathrm{uc}}}^{2}
\xrightarrow{\mathcal D}
\left\|G+S_h^{\mathrm{uc}}\right\|_{\mathbb H_S^{\mathrm{uc}}}^{2}.
\label{eqn:uc_local_norm_limit}
\end{align}
Suppose additionally that $S_h^{\mathrm{uc}}\in \ker(\Gamma_{S,0}^{\mathrm{uc}})^\perp$. If we adopt the decomposition
\begin{align*}
\Gamma_{S,0}^{\mathrm{uc}} =  \sum_{k=1}^{\infty}\lambda_k(v_k\otimes v_k),
\quad
S_h^{\mathrm{uc}} = \sum_{k=1}^{\infty}\sigma_kv_k.
\end{align*}
where $\lambda_k>0$ is the eigenvalue and $v_k$ is the corresponding orthonormal eigenvectors, then
\begin{align}
n\big\|\hat S_n^{\mathrm{uc}}\big\|_{\mathbb H_S^{\mathrm{uc}}}^{2}
\xrightarrow{\mathcal D}
\sum_{k=1}^{\infty} \big(\sqrt{\lambda_k}Z_k+\sigma_k\big)^2,
\label{eqn:uc_local_weighted_chisq}
\end{align}
where the $Z_k$ are independent standard normal random variables.
\end{theorem}

\section{Numerical Implementation}\label{sec:num}
We now describe the numerical implementation of the proposed test. We first derive the coordinate mapping for the test statistics, then obtain a Gamma approximation to its null distribution, and finally discuss the selection of kernel bandwidths and the regularization parameter.
\subsection{Coordinate Mapping for the Test Statistic}
For $j= 1,\ldots,d-1$, Let $K_{\ddot X_j}\in\mathbb R^{n\times n}$ denote the Gram matrix with entries $(K_{\ddot X_j})_{ik}=\kappa_{\ddot X_j}(\ddot X_{j,i},\ddot X_{j,k})$. Define $H=I-n^{-1}1_n1_n^\top$ and $\hat K_{\ddot X_j}=HK_{\ddot X_j}H$. Similarly define $\hat K_{X_{j+}}$. Using the empirically centered feature maps, consider the spanning systems
\begin{align*}
    \mathscr B_{\ddot X_j}
    =
    \{\hat\kappa_{\ddot X_j}(\cdot,\ddot X_{j,i}):i=1,\ldots,n\}, \quad
    \mathscr B_{X_{j+}}
    =
    \{\hat\kappa_{X_{j+}}(\cdot, X_{j+,i}):i=1,\ldots,n\}.
\end{align*}
Analogous notation is used for $Z$. Following \citet[Section~12.3]{li2018sufficient}, the empirical covariance operators have the matrix representations
\begin{align*}
    \coord{\mathscr B_{\ddot X_{j}}}
    {\hat\Sigma_{\ddot X_{j} Z}}
    {\mathscr B_{Z}}
    =n^{-1}\hat K_{Z},\quad
    \coord{\mathscr B_{Z}}
    {\hat\Sigma_{ZZ}}
    {\mathscr B_{Z}}
    =n^{-1}\hat K_{Z}, \quad
    \coord{\mathscr B_{Z}}
    {\hat\Sigma_{Z X_{j+}}}
    {\mathscr B_{X_{j+}}}
    =n^{-1}\hat K_{X_{j+}}.
\end{align*}
The same rules apply when either variable block is replaced by $Z$ or an augmented block. Then the regularized regression operators therefore satisfy
\begin{align*}
   \coord{\mathscr{B}_{\ddot X_{j}}}{\hat{B}_{\ddot X_{j}\mid Z}}{\mathscr{B}_{Z}}
    = & \coord{\mathscr{B}_{\ddot X_{j}}}{\hat{\Sigma}_{\ddot X_{j} Z}}{\mathscr{B}_{Z}}
   \coord{\mathscr{B}_{Z}}{\big(\hat{\Sigma}_{ZZ} + \epsilon_{n} I\big)^{-1}}{\mathscr{B}_{Z}} = \frac{1}{n} \hat{K}_{Z} \big( \frac{1}{n} \hat{K}_{Z} + \epsilon_{n}  I\big)^{-1} = \hat{K}_{Z} \big( \hat{K}_{Z} + n \epsilon_{n} I\big)^{-1}.
\end{align*}
Similarly we get $\coord{\mathscr{B}_{Z}}{\hat{B}_{X_{j+}\mid Z}^{*}}{\mathscr{B}_{X_{j+}}} = \left( \hat{K}_{Z} + n \epsilon_{n}  I\right)^{-1} \hat{K}_{X_{j+}}$. Applying them to the residual covariance formula gives
\begin{align*}
   \coord{\mathscr{B}_{\ddot{X}_{j}}}{\hat{\Sigma}_{\ddot{X}_{j} X_{j+}\mid Z}}{\mathscr{B}_{X_{j+}}} = &\coord{\mathscr{B}_{\ddot{X}_{j}}}{\hat{\Sigma}_{\ddot{X}_{j} X_{j+}} - \hat{B}_{\ddot{X}_{j}\mid Z} \hat{\Sigma}_{Z X_{j+}} - \hat{\Sigma}_{\ddot{X}_{j} Z} \hat{B}_{X_{j+}\mid Z}^{*} + \hat{B}_{\ddot{X}_{j}\mid Z} \hat{\Sigma}_{Z Z} \hat{B}_{X_{j+}\mid Z}^{*}}{\mathscr{B}_{X_{j+}}} \\
    = & \frac{1}{n} \big[ ( I - \hat{K}_{Z} (\hat{K}_{Z} + n\epsilon_{n} I)^{-1}) \big]^{2} \hat{K}_{X_{j+}} = \frac{1}{n} R_{Z}^{2} \hat{K}_{X_{j+}},
\end{align*}
where $R_Z = I - \hat{K}_{Z}\left(\hat{K}_{Z} + n\epsilon_{n}I\right)^{-1}
    = n \epsilon_n \big(\hat{K}_{Z} + n\epsilon_{n}I\big)^{-1}$.
Finally we obtain
\begin{align*}
    \big\|\hat{\Sigma}_{\ddot{X}_{j} X_{j+}\mid Z}\big\|_{\mathrm{HS}}^{2}
    = & \frac{1}{n^{2}} \trace \big( \hat{K}_{\ddot{X}_{j}}^{\frac{1}{2}} R_{Z}^{2} \hat{K}_{X_{j+}}^{\frac{1}{2}} \cdot \hat{K}_{X_{j+}}^{\frac{1}{2}} R_{Z}^{2} \hat{K}_{\ddot{X}_{j}}^{\frac{1}{2}} \big)
    = \frac{1}{n^{2}} \trace \big( R_{Z} \hat{K}_{\ddot{X}_{j}} R_{Z}  \cdot R_{Z} \hat{K}_{X_{j+}} R_{Z}\big).
\end{align*}
We define $\hat{K}_{\ddot{X}_j\mid Z} = R_{Z} \hat{K}_{\ddot{X}_{j}} R_{Z}$ and $\hat{K}_{X_{j+}\mid Z} = R_{Z} \hat{K}_{X_{j+}} R_{Z}$. Then we can conclude that
\begin{align*}
    n\left\| \hat S \right\|_{\mathbb H_S}^2 = n \sum \limits_{j=1}^{d-1} \big\|\hat{\Sigma}_{\ddot{X}_{j} X_{j+}\mid Z}\big\|_{\mathrm{HS}}^{2} = \frac{1}{n} \sum \limits_{j=1}^{d-1} \trace \big( \hat{K}_{\ddot{X}_j\mid Z} \cdot \hat{K}_{X_{j+}\mid Z}\big).
\end{align*}
\subsection{Gamma Approximation to the Null Distribution}\label{subsec:gamma}

We now describe how to determine its rejection threshold under the null hypothesis. Under the null hypothesis, $S=0$, the asymptotic null distribution of the test
statistic is
\begin{align*}
    \hat T = n\|\hat S\|_{\mathbb H_S}^2
    \xrightarrow{\mathcal D}
    T
    =
    \sum_{i=1}^{\infty}\lambda_i Z_i^2.
\end{align*}
Although this representation characterizes the null distribution, it requires estimating the eigenvalues $\{\lambda_i\}_{i\geq1}$ and then numerically evaluating the related weighted chi-square distribution. To obtain a simpler and computationally more efficient calibration, we follow \citet[Section~3.4]{zhang2012kernel} and use a two-parameter Gamma distribution that matches the first two moments of $T$; see also the equivalent Welch--Satterthwaite approximation in \citet[Sections~4.2 and~6.3]{tang2026reproducing}. These moments only require $\trace(\Gamma_{S\mid Z})$ and $\trace(\Gamma_{S\mid Z}^2)$. To see this,
\begin{align}\label{eq:ET_varT}
    \E[T] = \trace(\Gamma_{S\mid Z}), \quad
    \var(T) =  2\trace(\Gamma_{S\mid Z}^{2}).
\end{align}
Then we can use a Gamma distribution $Gamma(\alpha,\beta)$ with shape parameter $\alpha = \frac{\E[T]^2}{\var(T)}$ and rate parameter $\beta = \frac{\E[T]}{\var(T)}$ to approximate the weighted sum of independent $\chi^{2}$ random variables as in \cite{zhang2012kernel}. 

The influence function $\psi$ and its covariance operator $\Gamma_{S\mid Z}$ are given in \eqref{eqn:influence} and \eqref{eqn:cov_influence}. Their empirical counterparts under the null hypothesis are
\begin{align*}
\hat\psi_{j,i} = & \hat\xi_{\ddot X_j\mid Z}(\cdot,X_{j,i}; Z_i) \otimes \hat\xi_{X_{j+} \mid Z}(\cdot,X_{j+,i};Z_i), \quad
\hat\psi_i
=
\big(
\hat\psi_{1,i},\ldots,
\hat\psi_{d-1,i}
\big)^\top; \\
\hat\Gamma_{S\mid Z}
= &
\E_n \big[
\hat\psi\otimes\hat\psi
\big]
=
\frac1n\sum_{i=1}^n
\hat\psi_i \otimes \hat\psi_i.
\end{align*}
Indeed, $\hat\Gamma_{S\mid Z}$ is a consistent estimator. We state this as a lemma below.
\begin{lemma}\label{lem:gamma_consistent}
    Under the same assumptions as in Lemma~$2$, we have
    \begin{align*}
    \|\hat\Gamma_{S\mid Z}-\Gamma_{S\mid Z}\|_{\mathrm{tr}} =o_P(1).
    \end{align*}
\end{lemma}
Consequently, we obtain consistent estimates of the first two moments of $T$ by replacing $\Gamma_{S\mid Z}$ with $\hat\Gamma_{S\mid Z}$. The following proposition gives the corresponding matrix expressions, which are needed for Gamma approximation.

\begin{proposition}\label{prop:gamma_plugin}
Under the null hypothesis $S=0$, the plug-in estimates of mean $\E[T]$ and variance $\var(T)$ in \eqref{eq:ET_varT} are
\begin{align*}
\trace(\hat\Gamma_{S\mid Z}) = & \frac{1}{n}\sum_{j=1}^{d-1}
\trace\big(
\hat K_{\ddot X_j\mid Z}
\circ
\hat K_{X_{j+}\mid Z}
\big),\\
2\trace(\hat\Gamma_{S\mid Z}^{2}) = & \frac{2}{n^2}
\sum_{j=1}^{d-1}\sum_{k=1}^{d-1}
\trace\big[
\big(
\hat K_{\ddot X_j\mid Z}
\circ
\hat K_{X_{j+}\mid Z}
\big)
\big(
\hat K_{\ddot X_k\mid Z}
\circ
\hat K_{X_{k+}\mid Z}
\big)
\big].
\end{align*}
\end{proposition}
Then the asymptotic distribution of $\hat T$ can be approximated by a $Gamma(\hat\alpha,\hat\beta)$ distribution, where $\hat \alpha = \frac{\trace(\hat\Gamma_{S\mid Z})^2}{2\trace(\hat\Gamma_{S\mid Z}^{2})}$, $\hat \beta = \frac{\trace(\hat\Gamma_{S\mid Z})}{2\trace(\hat\Gamma_{S\mid Z}^{2})}$.

\subsection{Tuning} \label{sec:tunning}
We use Gaussian radial basis functions as kernels, equipped with a bandwidth parameter $h$. For each $X_J$ with $J \subset \{1,\ldots,d\}$ and $Z$, we choose the bandwidth parameter according to the median pairwise distances among the observations. Specifically, letting $\{X_{J,i}\}_{i=1}^n$ denote the sample of a generic variable $X_J$, the bandwidth is chosen as
\begin{align*}
h_{X_J}
=
\operatorname{median} \left\{ \|X_{J,i}-X_{J,j}\|:\; 1\le i<j\le n \right\},
\end{align*}
and the Gaussian kernel is
\begin{align*}
\kappa_{X_J}(x,x')
=
\exp\Big(
-\frac{\|x-x'\|^2}{2h_{X_J}^2}
\Big).
\end{align*}
As an analogy to \cite{li2018nonparametric}, we propose a GCV criterion based on the unaugmented feature maps of $X_j$ and $X_{j+}$ for the covariance operator $\Sigma_{X_j X_{j+}\mid Z}$, for $j=1,\ldots,d-1$, as below. Denote the smoothing matrix
\begin{align*}
    P_Z^{\epsilon_n} = \hat K_Z \big( \hat K_Z+ n \epsilon_n I \big)^{-1}.
\end{align*}
Using this matrix, we define
\begin{align}\label{eqn:gcv}
\mathrm{GCV}_{X_{j} X_{j+}}(\epsilon_n) 
= \frac{ \norm{\hat K_{X_{j}} \circ \hat K_{X_{j+}} - (P_Z^{\epsilon_n} \hat K_{X_{j}}) \circ (P_Z^{\epsilon_n} \hat K_{X_{j+}}) }_F^2 }
{ \big( 1 - n^{-1} \trace [ ( \hat K_Z + n \epsilon_n I )^{-1} \hat K_Z ] \big)^2 }.
\end{align}
Here we use $(P_Z^{\epsilon_n} \hat K_{X_{j}}) \circ (P_Z^{\epsilon_n} \hat K_{X_{j+}})$ to represent the $Z$-explained component obtained by separately smoothing the two marginal feature maps. Hence for our $S = \big(\Sigma_{\ddot{X}_{1} X_{1+} \mid Z}, \ldots, \Sigma_{\ddot{X}_{d-1} X_{d} \mid Z} \big)^{\top}$, we use the following practical tuning criterion: 
\begin{align*}
    \mathrm{GCV}_S(\epsilon_n) = \sum \limits_{j=1}^{d-1} \mathrm{GCV}_{X_j X_{j+}} (\epsilon_n).
\end{align*}
Then the regularization parameter $\epsilon_n$ is selected by minimizing $\mathrm{GCV}_S$.

\subsection{Unconditional Case}
In this section, we discuss the coordinate mapping and gamma approximation in the unconditional case. There is no Tikhonov regularization parameter $\epsilon_n$ involved, as we will see immediately in the coordinate mapping. Recall that
\begin{align*}
   \coord{\mathscr{B}_{X_{j}}}{\hat{\Sigma}_{X_{j} X_{j+}}}{\mathscr{B}_{X_{j+}}} = \frac{1}{n} \hat{K}_{X_{j+}}.
\end{align*}
Its Hilbert--Schmidt norm is
\begin{align*}
    \big\|\hat{\Sigma}_{{X}_{j} X_{j+}}\big\|_{\mathrm{HS}}^{2} = & \big\| \hat{K}_{{X}_{j}}^{\frac{1}{2}} \coord{\mathscr{B}_{{X}_{j}}}{\hat{\Sigma}_{{X}_{j} X_{j+}}}{\mathscr{B}_{X_{j+}}} \hat{K}_{X_{j+}}^{-\frac{1}{2}}  \big\|_{F}^{2}
    = \big\| \frac{1}{n} \hat{K}_{{X}_{j}}^{\frac{1}{2}} \hat{K}_{X_{j+}}^{\frac{1}{2}} \big\|_{F}^{2} \\
    = & \frac{1}{n^{2}} \trace \big( \hat{K}_{{X}_{j}}^{\frac{1}{2}} \hat{K}_{X_{j+}}^{\frac{1}{2}} \cdot \hat{K}_{X_{j+}}^{\frac{1}{2}} \hat{K}_{{X}_{j}}^{\frac{1}{2}} \big) = \frac{1}{n^{2}} \trace \big(\hat{K}_{{X}_{j}} \cdot \hat{K}_{X_{j+}}\big).
\end{align*}
Then we get
\begin{align}
    n\left\| \hat S^{\mathrm{uc}} \right\|_{\mathbb H_S^{\mathrm{uc}}}^2 = n \sum \limits_{j=1}^{d-1} \left\|\hat{\Sigma}_{{X}_{j} X_{j+}}\right\|_{\mathrm{HS}}^{2} = \frac{1}{n} \sum \limits_{j=1}^{d-1} \trace \big( \hat{K}_{{X}_j} \cdot \hat{K}_{X_{j+}}\big).
    \label{eqn:unconditional_sample_statistic}
\end{align}
We have shown that the influence function $\psi^{\mathrm{uc}}$ and covariance $\Gamma_S^{\mathrm{uc}}$ are given in $\eqref{eqn:def_psi_uc}$. Their empirical counterparts under the null hypothesis are
\begin{align*}
\hat\psi_{j,i}^{\mathrm{uc}} = & \hat\kappa_{X_j}(\cdot,X_{j,i}) \otimes \hat\kappa_{X_{j+}}(\cdot,X_{j+,i}), \quad
\hat\psi_i^{\mathrm{uc}}
=
\big(
\hat\psi_{1,i}^{\mathrm{uc}},\ldots,
\hat\psi_{d-1,i}^{\mathrm{uc}}
\big)^\top; \\
\hat\Gamma_S^{\mathrm{uc}}
= &
\E_n \big[
\hat\psi^{\mathrm{uc}}\otimes\hat\psi^{\mathrm{uc}}
\big]
=
\frac1n\sum_{i=1}^n
\hat\psi_i^{\mathrm{uc}}\otimes\hat\psi_i^{\mathrm{uc}}.
\end{align*}
Under the null hypothesis, $S^{\mathrm{uc}}=0$, the asymptotic null distribution of the test statistic is
\begin{align*}
    \hat T^{\mathrm{uc}} = n\|\hat S^{\mathrm{uc}}\|_{\mathbb H_S^{\mathrm{uc}}}^2
    \xrightarrow{\mathcal D}
    T^{\mathrm{uc}}
    =
    \sum_{i=1}^{\infty}\lambda_i Z_i^2.
\end{align*}
Again we use a Gamma distribution $Gamma(\alpha,\beta)$ with $\alpha = \frac{\E[T^{\mathrm{uc}}]^2}{\var(T^{\mathrm{uc}})}$, $\beta = \frac{\E[T^{\mathrm{uc}}]}{\var(T^{\mathrm{uc}})}$ to approximate the weighted sum of $\chi^{2}$ distribution
\begin{align*}
    \E[T^{\mathrm{uc}}] = \trace(\Gamma_{S}^{\mathrm{uc}}),\quad
    \var(T^{\mathrm{uc}}) = 2\trace((\Gamma_{S}^{\mathrm{uc}})^{2}).
\end{align*}
\begin{proposition}\label{prop:uncondional_gamma_plugin}
Under the null hypothesis $S^{\mathrm{uc}}=0$, the plug-in estimates of mean $\E[T]$ and variance $\var(T)$ above are
\begin{align*}
\trace(\hat\Gamma_{S}^{\mathrm{uc}}) = & \frac{1}{n}\sum_{j=1}^{d-1}
\trace\big(
\hat K_{X_j}
\circ
\hat K_{X_{j+}}
\big),\\
2\trace((\hat\Gamma_{S}^{\mathrm{uc}})^{2}) = & \frac{2}{n^2}
\sum_{j=1}^{d-1}\sum_{k=1}^{d-1}
\trace\big[
\big(
\hat K_{X_j}
\circ
\hat K_{X_{j+}}
\big)
\big(
\hat K_{X_k}
\circ
\hat K_{X_{k+}}
\big)
\big],
\end{align*}
where $\hat\Gamma_{S}^{\mathrm{uc}}$ is the plug-in estimator of the
covariance operator $\Gamma_{S}^{\mathrm{uc}}$ under null hypothesis.
\end{proposition}
The proof is parallel to Proposition~\ref{prop:gamma_plugin}, and thus omitted. Then similarly to Section~\ref{subsec:gamma}, the asymptotic distribution of $\hat T^{\mathrm{uc}}$ can be approximated by a $Gamma(\hat\alpha,\hat\beta)$ distribution, where $\hat \alpha = \frac{\trace(\hat\Gamma_{S}^{\mathrm{uc}})^2}{2\trace((\hat\Gamma_{S}^{\mathrm{uc}})^{2})}$, $\hat \beta = \frac{\trace(\hat\Gamma_{S}^{\mathrm{uc}})}{2\trace((\hat\Gamma_{S}^{\mathrm{uc}})^{2})}$.

\section{Simulation and Application}\label{sec:simulation}

\subsection{Simulation with Linear and Nonlinear Models}

We evaluate the finite-sample behavior of the proposed multi-way conditional independence test under linear and nonlinear models. Throughout, we take $d=4$ and let $Z\in\mathbb{R}^2$. For each replication, we generate $Z \sim N(0,I_2)$ and $\epsilon \sim N(0,\Psi_\rho)$ independently, where $\epsilon =(\epsilon_1,\ldots,\epsilon_4)^\top$,
\begin{align*}
    (\Psi_\rho)_{jk}=\rho^{|j-k|}, \qquad j,k=1,\ldots,4.
\end{align*}
Here $\rho=0$ corresponds to the conditional independence case, whereas increasing $\rho$ produces stronger conditional dependence. We consider
$\rho\in\{0,0.2,0.4,0.6,0.8\}$, $n\in\{100,500\}$,
with $200$ replications for each combination of model, sample size, and $\rho$.

The five models of the data-generating process are summarized in Table \ref{table:comparison}. Polynomials of $Z$ in Model $3$ and product with $\epsilon$ in Model $5$ are taken component-wise. The coefficient matrices are given in Table~\ref {table:comparison}.
\begin{table}[tbp]
\centering
\caption{Linear and nonlinear simulation models}
\label{table:comparison}

\begin{tabular}{cll}
\hline
Model & Model equation & Coefficient matrices \\
\hline
1 & $X=C_1 Z+\epsilon$ &
$C_1=\begin{pmatrix}
0.80 & -0.60 & 0.05 & 0.70 \\
-0.40 & 0.90 & -0.03 & 0.50
\end{pmatrix}^{\top}$
\\
\hline
2 & $X=C_2\log(1+|Z|)+\epsilon$ &
$C_2=\begin{pmatrix}
0.70 & -0.50 & 0.10 & 0.60 \\
0.30 & 0.80 & -0.05 & -0.40
\end{pmatrix}^{\top}$
\\
\hline
3 & $X=C_3 Z^{3}+C_3'Z^{2}+\epsilon$ &
$\begin{aligned}
C_3 &=
\begin{pmatrix}
0.35 & -0.30 & 0.04 & 0.25 \\
-0.20 & 0.25 & -0.03 & 0.20
\end{pmatrix}^{\top}
\\
C_3' &=
\begin{pmatrix}
0.20 & -0.15 & 0.03 & 0.20 \\
0.10 & 0.25 & 0.02 & -0.10
\end{pmatrix}^{\top}
\end{aligned}$
\\
\hline
4 & $X=C_4\sin(Z)+\epsilon$ &
$C_4=\begin{pmatrix}
0.90 & -0.50 & 0.08 & 0.60 \\
-0.30 & 0.70 & -0.04 & 0.40
\end{pmatrix}^{\top}$
\\
\hline
5 & $X=\exp(C_5 Z)\cdot \epsilon$ &
$C_5=\begin{pmatrix}
0.30 & -0.20 & 0.05 & 0.20 \\
-0.15 & 0.25 & -0.03 & 0.15
\end{pmatrix}^{\top}$
\\
\hline
\end{tabular}
\end{table}
We therefore examine whether the proposed test remains effective when the effect of $Z$ is linear, nonlinear, or enters through the conditional variance rather than only through the conditional mean.

For every generated data set, Gaussian radial basis function kernels are used. Kernel bandwidths are selected by the median pairwise-distance rule separately for each $X_j$, for $Z$, and for each grouped vector $X_{j+}$, where $d=4$. For each replication, $\epsilon_n$ is chosen by minimizing the GCV criterion discussed in Section~\ref{sec:tunning} over $51$ logarithmically equally spaced values in $[10^{-5},10^{-1}]$. After conditional residualization, we compute $n\|\hat S\|_{\mathbb H_S}^2$ and approximate its null distribution by the moment-matched Gamma distribution. In most scenarios, the selected $\epsilon_n$ values are not at the boundary, and the GCV curve is nearly flat when they are close to it. The resulting test rejects at significance level $\alpha=0.05$.

\begin{figure}[htb!]
    \centering
    \includegraphics[width=\textwidth]{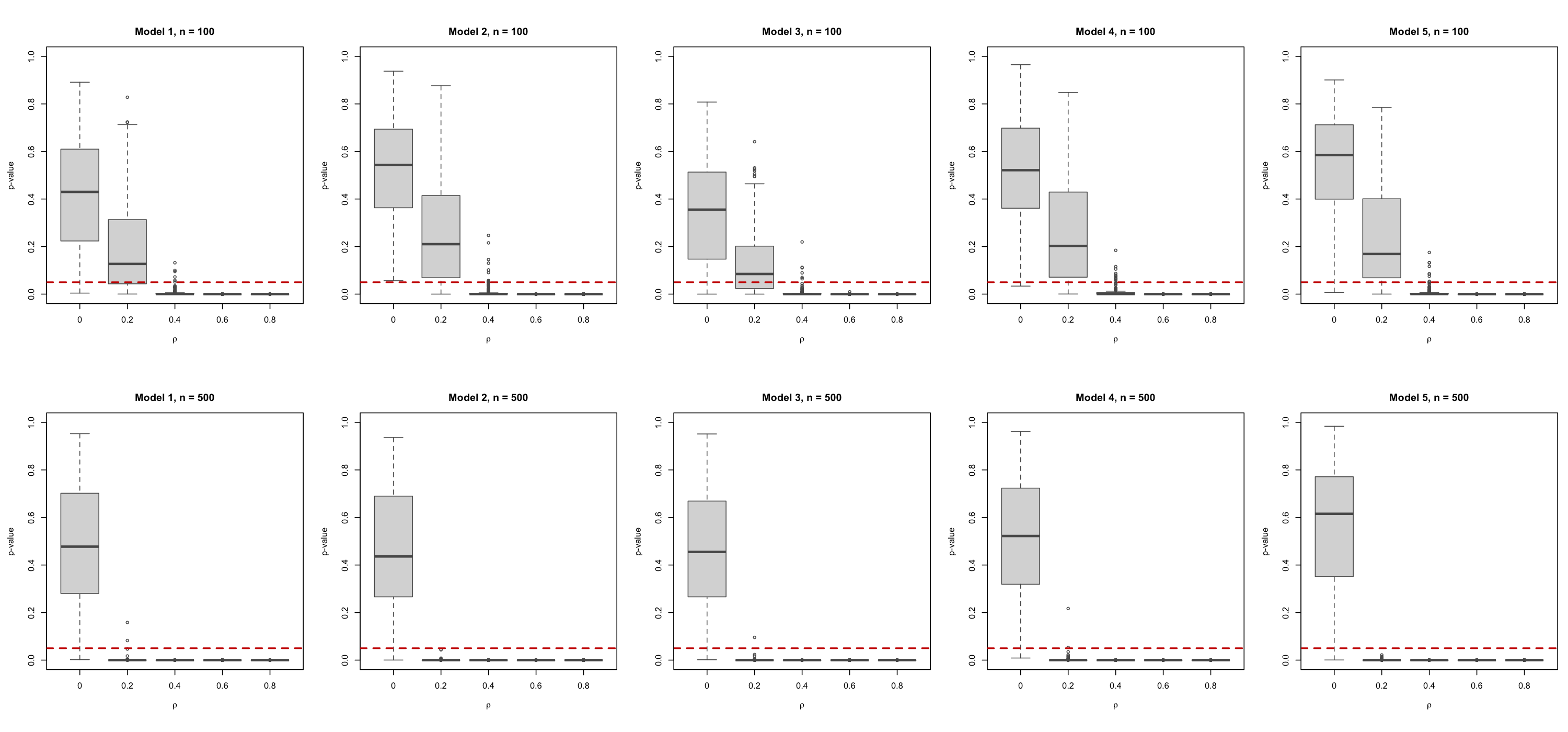}
    \caption{$p$-value distributions for the five simulation models. Columns correspond to Models 1--5 and rows to $n=100$ and $n=500$. Within each panel, boxplots are shown for $\rho\in\{0,0.2,0.4,0.6,0.8\}$. The red dashed line marks the nominal significance level $\alpha=0.05$. Each boxplot is based on 200 Monte Carlo replications.}
    \label{fig:five-model-pvalues}
\end{figure}

Figure~\ref{fig:five-model-pvalues} displays the distributions of the resulting $p$-values. Under the null case $\rho=0$, the $p$-values are broadly spread over $[0,1]$ for all five models and for both sample sizes, with medians near 0.5. This behavior is consistent with the intended null calibration and, importantly, persists across all four nonlinear mechanisms, including the heteroscedastic Model 5. For $n=100$, weak residual dependence may be difficult to detect, whereas the test becomes substantially more reliable as the sample size increases.

\subsection{Application to the OneK1K Single-Cell RNA-seq Data}
\label{sec:onek1k}

Gene co-expression can arise from multiple, potentially overlapping sources. At the molecular level, coordinated gene expression may arise from regulation by common transcription factors \citep{allocco2004quantifying}, coordinated downstream responses to common signaling pathways \citep{oelen2022single}, or shared or co-active cis-regulatory elements \citep{ribeiro2021molecular}. At the cellular level, observed co-expression may reflect heterogeneity in cell identity and cell state \citep{kotliar2019identifying}. At the
population level, genetic background, age, sex, and external exposures, and disease conditions can generate context- and cell-type-specific expression
variation \citep{piasecka2018distinctive,oelen2022single,
yazar2022single, su2023cell}. Moreover, technical features of single-cell RNA sequencing, including sequencing-depth variation and measurement error, may create distortion \citep{su2023cell}. Since these mechanisms are complex and only partially observed, latent factors are therefore commonly used to account for otherwise unexplained shared variation \citep{argelaguet2020mofa,xue2023pitfalls}.

The proposed test provides a practical tool for evaluating a hypothesized explanation: By conditioning on the biological, technical, or even latent covariates, it assesses whether any multi-way dependence among the gene-expression modules
remains unexplained. In this paper, we only consider a very simple case. We compare two conditioning sets:
\begin{align*}
    Z_1 = & \{\text{batch}, \log(1+\texttt{nCount\_RNA}), \log(1+\texttt{nFeature\_RNA})\}; \\
    Z_2 = & Z_1\cup\{\text{cell type}\}.
\end{align*}
Thus, $Z_1$ adjusts for major technical variation, while $Z_2$ additionally removes variation associated with cell-type heterogeneity.

We analyze the OneK1K data, a large population-scale single-cell
RNA-sequencing study of peripheral blood mononuclear cells
\citep{yazar2022single}. The data release used in the present analysis
contains cells from $980$ donors classified into $14$ annotated cell
types, together with cell-level technical measurements and donor-level covariates.

For each cell $i$ and each predefined gene set $\mathcal G_j$, we define
the corresponding module score as
\begin{align*}
    X_{ij}
    =
    \frac{1}{|\mathcal G_j|}
    \sum_{g\in\mathcal G_j}
    \frac{Y_{ig}-\bar Y_g}{s_g},
    \qquad j=1,2,3,
\end{align*}
where $Y_{ig}$ is the normalized expression of gene $g$ in cell $i$,
and $\bar Y_g$ and $s_g$ are, respectively, the sample mean and sample
standard deviation of that gene across the analyzed cells. Genes that
were not detected or had zero sample variance were excluded before the
scores were constructed.

We focus on the following three modules:
\begin{align*}
    X_1 &: \text{classical-monocyte identity},\\
    \mathcal G_1
        &= \{\textit{LYZ},\textit{S100A8},\textit{S100A9},
        \textit{CTSS},\textit{LST1},\textit{VCAN}\};\\[2mm]
    X_2 &: \text{inflammatory NF-$\kappa$B signaling},\\
    \mathcal G_2
        &= \{\textit{NFKBIA},\textit{RELB},\textit{TNFAIP3},
        \textit{CXCL8},\textit{CXCL2},\textit{PTGS2},
        \textit{IER3},\textit{BCL3}\};\\[2mm]
    X_3 &: \text{oxidative-stress response},\\
    \mathcal G_3
        &= \{\textit{HMOX1},\textit{SOD2},\textit{PRDX1},
        \textit{TXN},\textit{NQO1}\}.
\end{align*}
The first score represents classical-monocyte identity, whereas the other two capture inflammatory and oxidative-stress states that may vary within cell types. Classical monocytes exhibit prominent innate immune and inflammatory programs, for which NF-$\kappa$B is a key transcriptional regulator \citep{wong2011gene,mussbacher2023nfkb}. The inflammatory and oxidative-stress modules may interact because ROS can modulate NF-$\kappa$B activation, while NF-$\kappa$B regulates ROS production and antioxidant defenses \citep{morgan2011crosstalk}. For each $n\in\{2000,4000,6000\}$, cells are sampled stratified by cell type and batch.

\begin{table}[htbp!]
    \centering
    \caption{P-values for the multiway conditional independence test of the
    classical-monocyte, inflammatory NF-$\kappa$B, and oxidative-stress
    modules.}
    \label{tab:onek1k-results}
    \begin{tabular}{ccc}
        \hline
        Sample size & $Z_1$ & $Z_2$ \\
        \hline
        $n=2000$ & $0.0120$ & $0.1390$ \\
        $n=4000$ & $3.37\times10^{-11}$ & $0.5168$ \\
        $n=6000$ & $6.88\times10^{-15}$ & $0.7410$ \\
        \hline
    \end{tabular}
\end{table}

These results suggest that, for this triplet, the observed multi-way dependence is attributable to differences among cell types rather than to a stable pattern of coordinated within-cell-type variation. After additionally conditioning on cell type, the dependence is no longer significant. Consistent with the simulation results, the distinction between the two conditioning schemes becomes clearer as the sample size increases.

\section{Appendix}

\subsection{Proofs for Section~\ref{sec:asy}}
\begin{proof}[Proof of Theorem \ref{thm:rate_reg_operator}]
Recall we have the conditional mean-zero property of residual terms in \eqref{eqn:mean_zero_reg_op}:
\begin{align*}
\E\left[ \xi_{X_J\mid Z}(\cdot,X_J;Z)\mid Z \right] = 0 \ \text{almost surely}, \quad \E\left[ \xi_{\ddot X_J\mid Z}(\cdot,\ddot X_J;Z)\mid Z \right] = 0 \ \text{almost surely}.
\end{align*}
Consequently, $\E\left[ \xi_{X_J\mid Z}(\cdot,X_J;Z) \otimes \tilde\kappa_Z(\cdot,Z) \right] =0$. Recall that $\tilde\kappa_{X_J}(\cdot,X_J) = B_{X_J\mid Z}\tilde\kappa_Z(\cdot,Z) + \xi_{X_J\mid Z}(\cdot,X_J;Z)$. It follows from the definitions of the sample-level covariance
operators that 
\begin{align*}
\hat\Sigma_{X_JZ}
=  B_{X_J\mid Z}\hat\Sigma_{ZZ} 
+ \E_n\left[ \xi_{X_J\mid Z}(\cdot,X_J;Z) \otimes \tilde\kappa_Z(\cdot,Z) \right]
- \E_n\left[ \xi_{X_J\mid Z}(\cdot,X_J;Z) \right] \otimes \E_n\left[ \tilde\kappa_Z(\cdot,Z) \right].
\end{align*}
By the resolvent identity
\begin{align}\label{eq:resolvent}
    (\hat{\Sigma}_{ZZ}+\epsilon_n I)^{-1} - (\Sigma_{ZZ}+\epsilon_n I)^{-1}
    = (\Sigma_{ZZ}+\epsilon_n I)^{-1} (\Sigma_{ZZ}-\hat{\Sigma}_{ZZ}) (\hat{\Sigma}_{ZZ}+\epsilon_n I)^{-1},
\end{align}
we have
\begin{align}
\hat B_{X_J\mid Z}-B_{X_J\mid Z}
= & -\epsilon_n B_{X_J\mid Z} (\hat\Sigma_{ZZ}+\epsilon_nI)^{-1} 
+ \E_n\left[ \xi_{X_J\mid Z}(\cdot,X_J;Z) \otimes \tilde\kappa_Z(\cdot,Z) \right]
(\hat\Sigma_{ZZ}+\epsilon_nI)^{-1}  \notag\\
& - \left( \E_n\left[ \xi_{X_J\mid Z}(\cdot,X_J;Z) \right] \otimes \E_n\left[ \tilde\kappa_Z(\cdot,Z) \right] \right) (\hat\Sigma_{ZZ}+\epsilon_nI)^{-1} \notag \\
= &
-\epsilon_n B_{X_J\mid Z}
(\Sigma_{ZZ}+\epsilon_nI)^{-1}
\notag
-
\epsilon_n B_{X_J\mid Z}
(\Sigma_{ZZ}+\epsilon_nI)^{-1}
(\Sigma_{ZZ}-\hat\Sigma_{ZZ})
(\hat\Sigma_{ZZ}+\epsilon_nI)^{-1}
\notag\\
&+
\E_n\left[
\xi_{X_J\mid Z}(\cdot,X_J;Z)
\otimes
\tilde\kappa_Z(\cdot,Z)
\right]
(\Sigma_{ZZ}+\epsilon_nI)^{-1}
\notag\\
&+
\E_n\left[
\xi_{X_J\mid Z}(\cdot,X_J;Z)
\otimes
\tilde\kappa_Z(\cdot,Z)
\right]
(\Sigma_{ZZ}+\epsilon_nI)^{-1}
(\Sigma_{ZZ}-\hat\Sigma_{ZZ})
(\hat\Sigma_{ZZ}+\epsilon_nI)^{-1}
\notag\\
&-
\left(
\E_n\left[
\xi_{X_J\mid Z}(\cdot,X_J;Z)
\right]
\otimes
\E_n\left[
\tilde\kappa_Z(\cdot,Z)
\right]
\right)
(\hat\Sigma_{ZZ}+\epsilon_nI)^{-1}.
\label{eqn:BX-HS-decomposition}
\end{align}
We control the five terms in \eqref{eqn:BX-HS-decomposition} separately.
For the first term, by Assumption \ref{asm:inverse_beta},
\begin{align*}
    B_{X_J\mid Z} = \Sigma_{X_JZ}\Sigma_{ZZ}^{\dagger} = S_{X_JZ}\Sigma_{ZZ}^{1+\beta}\Sigma_{ZZ}^{\dagger},
\end{align*}
where the operator product on the right-hand side is understood through
its bounded extension. Since $S_{X_J Z}$ is bounded,
\begin{align*}
\left\| B_{X_J\mid Z} (\Sigma_{ZZ}+\epsilon_nI)^{-1} \right\|_{\mathrm{op}}
\leq & \|S_{X_JZ}\|_{\mathrm{op}} \left\| \Sigma_{ZZ}^{1+\beta} \Sigma_{ZZ}^\dagger (\Sigma_{ZZ}+\epsilon_nI)^{-1}  \right\|_{\mathrm{op}}.
\end{align*}
Let $\{e_k\}_k$ be the orthonormal eigen functions of $\Sigma_{ZZ}$, $\{\gamma_k\}_k$ be the corresponding eigen values. Then
    \begin{align*}
        \left\| \Sigma_{ZZ}^{1+\beta} \Sigma_{ZZ}^\dagger (\Sigma_{ZZ}+\epsilon_nI)^{-1}  \right\|_{\mathrm{op}}
        =  \sup_k \left\langle \Sigma_{ZZ}^{1+\beta} \Sigma_{ZZ}^\dagger (\Sigma_{ZZ}+\epsilon_nI)^{-1} e_k, e_k \right \rangle_{\mathcal{H}_Z} 
        =  \sup_k \frac{\gamma_k^\beta}{\gamma_k + \epsilon_n}.
    \end{align*}
    If $0 < \beta < 1$, $\frac{\gamma_k^\beta}{\gamma_k + \epsilon_n} \lesssim \epsilon_n^{\beta - 1}$; If $\beta \ge 1$, $\frac{\gamma_k^\beta}{\gamma_k + \epsilon_n} \leq \gamma_k^{\beta - 1} \leq \norm{\Sigma_{ZZ}}_\mathrm{op}^{\beta - 1} < \infty$.  This holds for any $k$.
This gives
\begin{align}\label{eq:beta_inverse}
        \left\| \Sigma_{ZZ}^{1+\beta} \Sigma_{ZZ}^\dagger (\Sigma_{ZZ}+\epsilon_nI)^{-1}  \right\|_{\mathrm{op}} \lesssim \epsilon_n^{\beta \wedge 1 - 1}.
    \end{align}
Hence,
\begin{align}
&\left\|
\epsilon_n B_{X_J\mid Z}
(\Sigma_{ZZ}+\epsilon_nI)^{-1}
\right\|_{\mathrm{op}}
\lesssim
\epsilon_n^{\beta\wedge1}.
\label{rate:BX-HS-term-1}
\end{align}
For the second term of \eqref{eqn:BX-HS-decomposition}, we have
\begin{align}\label{eq:CLT_Sigma_Z}
\norm{\hat\Sigma_{ZZ}-\Sigma_{ZZ}}_{\mathrm{op}} 
\leq & \norm{\hat\Sigma_{ZZ}-\Sigma_{ZZ}}_{\mathrm{HS}} \notag\\
\leq & \norm{\E_n[\tilde \kappa_Z(\cdot,Z) \otimes \tilde \kappa_Z(\cdot,Z)] - \Sigma_{ZZ}}_{\mathrm{HS}} +  \norm{\E_n[\tilde \kappa_Z(\cdot,Z)] \otimes \E_n[\tilde \kappa_Z(\cdot,Z)]}_{\mathrm{HS}} \notag \\
= & \norm{\E_n[\tilde \kappa_Z(\cdot,Z) \otimes \tilde \kappa_Z(\cdot,Z)] - \Sigma_{ZZ}}_{\mathrm{HS}} +  \norm{\E_n[\tilde \kappa_Z(\cdot,Z)]}_{\mathcal{H}_Z} \norm{\E_n[\tilde \kappa_Z(\cdot,Z)]}_{\mathcal{H}_Z} \notag \\
= & O_P(n^{-\frac12}).
\end{align}
And by the spectral theorem, we have
\begin{align}\label{eq:inverse_hat_Sigma_Z}
    \norm{(\hat\Sigma_{ZZ}+\epsilon_nI)^{-1}}_{\mathrm{op}} \leq \epsilon_n^{-1}.
\end{align}
Combining \eqref{rate:BX-HS-term-1}, \eqref{eq:CLT_Sigma_Z} and \eqref{eq:inverse_hat_Sigma_Z} gives
\begin{align}
&\norm{ \epsilon_n B_{X_J\mid Z} (\Sigma_{ZZ}+\epsilon_nI)^{-1} (\Sigma_{ZZ}-\hat\Sigma_{ZZ}) (\hat\Sigma_{ZZ}+\epsilon_nI)^{-1} }_{\mathrm{op}} \notag\\
\leq &  \norm{ \epsilon_n B_{X_J\mid Z} (\Sigma_{ZZ}+\epsilon_nI)^{-1} }_{\mathrm{op}}
\norm{\Sigma_{ZZ}-\hat\Sigma_{ZZ}}_{\mathrm{op}}
\norm{(\hat\Sigma_{ZZ}+\epsilon_nI)^{-1}}_{\mathrm{op}} \notag\\
\leq & O_P(\epsilon_n^{\beta \wedge 1}) O_P(n^{-\frac12}) \epsilon_n^{-1} \notag\\
= & O_P\left( n^{-\frac12}\epsilon_n^{\beta\wedge1-1} \right).
\label{rate:BX-HS-term-2}
\end{align}
We next consider the third term of \eqref{eqn:BX-HS-decomposition}. The boundedness of the kernels and $B_{X_J\mid Z}$ implies the boundedness of the residual term $\xi_{X_J\mid Z}(\cdot,X_J;Z)$. By \eqref{eqn:mean_zero_reg_op} and independence across observations, the cross terms in the second moment vanish. The boundedness of the residuals and the spectral decomposition of $\Sigma_{ZZ}$ therefore give
\begin{align}
&\E \norm{ \E_n\left[ \xi_{X_J\mid Z}(\cdot,X_J;Z) \otimes \tilde\kappa_Z(\cdot,Z) \right] (\Sigma_{ZZ}+\epsilon_nI)^{-1} }_{\mathrm{HS}}^2 \notag \\
= & \frac1n \E\left[ \norm{ \xi_{X_J\mid Z}(\cdot,X_J;Z) }_{\mathcal H_{X_J}}^2
\norm{ (\Sigma_{ZZ}+\epsilon_nI)^{-1} \tilde\kappa_Z(\cdot,Z) }_{\mathcal H_Z}^2 \right] \notag\\
\lesssim & \frac1n \E\left[ \norm{ (\Sigma_{ZZ}+\epsilon_nI)^{-1} \tilde\kappa_Z(\cdot,Z) }_{\mathcal H_Z}^2 \right] \notag\\
= & \frac1n \sum_{k=1}^{\infty} \frac{\gamma_k}{(\gamma_k+\epsilon_n)^2} \notag
\end{align}
It remains to estimate the eigenvalue sum. Using $\gamma_k\asymp k^{-\alpha}$ with $\alpha>1$, we split the sum at $\epsilon_n^{-1/\alpha}$ and apply standard bounds for sums of powers:
\begin{align}
n^{-1}\sum_{k=1}^{\infty}
\frac{\gamma_k}{(\gamma_k+\epsilon_n)^2}
&\asymp
n^{-1}\left\{
\sum_{k\leq\epsilon_n^{-1/\alpha}}\gamma_k^{-1}
+
\epsilon_n^{-2}
\sum_{k>\epsilon_n^{-1/\alpha}}\gamma_k
\right\}\notag\\
&\asymp
n^{-1}\left\{
\epsilon_n^{-(\alpha+1)/\alpha}
+
\epsilon_n^{-2}\epsilon_n^{(\alpha-1)/\alpha}
\right\}\notag\\
&\asymp
n^{-1}\epsilon_n^{-(\alpha+1)/\alpha}.
\label{eqn:gamma_2_rate_alpha}
\end{align}
By Markov's inequality, we have
\begin{align}
\norm{ \E_n\left[ \xi_{X_J\mid Z}(\cdot,X_J;Z) \otimes \tilde\kappa_Z(\cdot,Z) \right] (\Sigma_{ZZ}+\epsilon_nI)^{-1} }_{\mathrm{HS}}
= O_P\left( n^{-\frac12} \epsilon_n^{-\frac{\alpha+1}{2\alpha}} \right).
\label{rate:BX-HS-term-3}
\end{align}
It also helps with the fourth term of \eqref{eqn:BX-HS-decomposition} as shown below:
\begin{align}
&\norm{ \E_n\left[ \xi_{X_J\mid Z}(\cdot,X_J;Z) \otimes \tilde\kappa_Z(\cdot,Z) \right]
(\Sigma_{ZZ}+\epsilon_nI)^{-1} (\Sigma_{ZZ}-\hat\Sigma_{ZZ}) (\hat\Sigma_{ZZ}+\epsilon_nI)^{-1} }_{\mathrm{HS}} \notag\\
\leq & \norm{ \E_n\left[ \xi_{X_J\mid Z}(\cdot,X_J;Z) \otimes \tilde\kappa_Z(\cdot,Z)
\right] (\Sigma_{ZZ}+\epsilon_nI)^{-1} }_{\mathrm{HS}}
\norm{\Sigma_{ZZ}-\hat\Sigma_{ZZ}}_{\mathrm{op}}
\norm{(\hat\Sigma_{ZZ}+\epsilon_nI)^{-1}}_{\mathrm{op}} \notag\\
\leq & O_P\left( n^{-\frac12} \epsilon_n^{-\frac{\alpha+1}{2\alpha}} \right) O_P(n^{-\frac12}) \epsilon_n^{-1} \notag \\
\leq & O_P\left( n^{-1} \epsilon_n^{-\frac{3\alpha+1}{2\alpha}} \right).
\label{rate:BX-HS-term-4}
\end{align}
Finally, for the fifth term of \eqref{eqn:BX-HS-decomposition}, by the Hilbert-space central limit theorem,
\begin{align*}
\norm{ \E_n\left[ \xi_{X_J\mid Z}(\cdot,X_J;Z)
\right] }_{\mathcal H_{X_J}}
=  O_P(n^{- \frac12}), \qquad
\norm{ \E_n\left[ \tilde\kappa_Z(\cdot,Z) \right] }_{\mathcal H_Z}
=  O_P(n^{-\frac12}).
\end{align*}
Applying the above results and \eqref{eq:inverse_hat_Sigma_Z}, we obtain
\begin{align}
&\norm{ \left( \E_n\left[ \xi_{X_J\mid Z}(\cdot,X_J;Z) \right] \otimes 
\E_n\left[ \tilde\kappa_Z(\cdot,Z) \right] \right) (\hat\Sigma_{ZZ}+\epsilon_nI)^{-1} }_{\mathrm{HS}} \notag\\
\leq & \epsilon_n^{-1} \norm{ \E_n\left[ \xi_{X_J\mid Z}(\cdot,X_J;Z) \right] }_{\mathcal H_{X_J}}
\norm{ \E_n\left[ \tilde\kappa_Z(\cdot,Z) \right] }_{\mathcal H_Z} \notag\\
= & O_P(n^{-1}\epsilon_n^{-1}).
\label{rate:BX-HS-term-5}
\end{align}
Since $\alpha>1$ and $\epsilon_n\to0$,
\begin{align*}
n^{-1}\epsilon_n^{-1} = o\left( n^{-1} \epsilon_n^{-\frac{3\alpha+1}{2\alpha}} \right).
\end{align*}
Combining \eqref{rate:BX-HS-term-1}, \eqref{rate:BX-HS-term-2}, \eqref{rate:BX-HS-term-3}, \eqref{rate:BX-HS-term-4} and \eqref{rate:BX-HS-term-5} gives
\begin{align*}
\norm{ \hat B_{X_J\mid Z}-B_{X_J\mid Z} }_{\mathrm{op}}
= O_P\left( n^{-\frac12}\epsilon_n^{\beta\wedge1-1}
+ \epsilon_n^{\beta\wedge1}
+ n^{-1}\epsilon_n^{-\frac{3\alpha+1}{2\alpha}}
+ n^{-\frac12}\epsilon_n^{-\frac{\alpha+1}{2\alpha}}
\right),
\end{align*}
which proves assertion 1.

For assertion $2$, recall that $\hat{\kappa}_Z(\cdot,Z) = \tilde{\kappa}_Z(\cdot,Z) -\left(\hat{\mu}_Z-\mu_Z\right)$ and $\E_n\left[\hat{\kappa}_Z(\cdot,Z)\right]=0$. Those facts give us
\begin{align*}
    \hat{\Sigma}_{\ddot X_J Z}
    = \E_n\left[ \hat{\kappa}_{\ddot X_J}(\cdot,\ddot X_J) \otimes \hat{\kappa}_Z(\cdot,Z) \right]
    = \E_n\left[ \tilde{\kappa}_{\ddot X_J}(\cdot,\ddot X_J) \otimes \hat{\kappa}_Z(\cdot,Z) \right]
\end{align*}
and
\begin{align*}
    \hat \Sigma_{ZZ}
    = \E_n\left[ \hat{\kappa}_Z(\cdot,Z) \otimes \hat{\kappa}_Z(\cdot,Z) \right]
    = \E_n\left[ \tilde{\kappa}_Z(\cdot,Z) \otimes \hat{\kappa}_Z(\cdot,Z) \right] 
\end{align*}
immediately. By the definition of the residual term,
\begin{align*}
    \tilde{\kappa}_{\ddot X_J}(\cdot,\ddot X_J)
    = B_{\ddot X_J| Z} \tilde{\kappa}_Z(\cdot,Z) + \xi_{\ddot X_J| Z}(\cdot,\ddot X_J;Z).
\end{align*}
Substituting this identity gives
\begin{align*}
    \hat{\Sigma}_{\ddot X_J Z}
    = &
    B_{\ddot X_J| Z}
    \E_n\left[ \tilde{\kappa}_Z(\cdot,Z) \otimes \hat{\kappa}_Z(\cdot,Z) \right]
    + \E_n\left[ \xi_{\ddot X_J| Z}(\cdot,\ddot X_J;Z) \otimes \hat{\kappa}_Z(\cdot,Z) \right] \\
    = &  B_{\ddot X_J| Z} \hat \Sigma_{ZZ}
    + \E_n\left[ \xi_{\ddot X_J| Z}(\cdot,\ddot X_J;Z) \otimes \hat{\kappa}_Z(\cdot,Z) \right].
\end{align*}
Subtracting $T_{\hat{\mu}_{X_J}}\hat{\Sigma}_{ZZ}$ from both sides gives
\begin{align*}
    \hat{\Sigma}_{\ddot X_J Z} -T_{\hat{\mu}_{X_J}}\hat{\Sigma}_{ZZ}
    = \left( B_{\ddot X_J| Z}-T_{\mu_{X_J}} \right)\hat{\Sigma}_{ZZ} -T_{\hat{\mu}_{X_J} - \mu_{X_J}} \hat{\Sigma}_{ZZ} +\E_n\left[
        \xi_{\ddot X_J| Z}(\cdot,\ddot X_J;Z)
        \otimes\hat{\kappa}_Z(\cdot,Z)
    \right].
\end{align*}
Since $\operatorname{ran}(\hat{\Sigma}_{ZZ}) \subset \overline{\operatorname{ran}(\Sigma_{ZZ})}$ almost surely, we have
$\Sigma_{ZZ}\Sigma_{ZZ}^{\dagger} \hat{\Sigma}_{ZZ} = \hat{\Sigma}_{ZZ}$ almost surely.
Here, $\Sigma_{ZZ}\Sigma_{ZZ}^{\dagger}$, as mentioned above, denoting the unique bounded extension of $\Sigma_{ZZ}\Sigma_{ZZ}^{\dagger}$, is indeed the orthogonal projection operator onto $\overline{\operatorname{ran}(\Sigma_{ZZ})}$.

Consequently,
\begin{align*}
    \left( B_{\ddot X_J| Z} - T_{\mu_{X_J}} \right)\hat{\Sigma}_{ZZ}
    = \left( B_{\ddot X_J| Z} -T_{\mu_{X_J}} \Sigma_{ZZ}\Sigma_{ZZ}^{\dagger} \right)\hat{\Sigma}_{ZZ}.
\end{align*}
Therefore, we have almost surely that
\begin{align}\label{eq:est_decomp}
    \hat{\Sigma}_{\ddot X_J Z}
    -T_{\hat{\mu}_{X_J}}\hat{\Sigma}_{ZZ} 
    =& 
    \left(
    B_{\ddot X_J| Z} -T_{\mu_{X_J}} \Sigma_{ZZ}\Sigma_{ZZ}^{\dagger} \right)\hat{\Sigma}_{ZZ}
    +\E_n\left[ \xi_{\ddot X_J| Z}(\cdot,\ddot X_J;Z) \otimes \tilde{\kappa}_Z(\cdot,Z) \right]\notag\\
    &
    -\E_n\left[ \xi_{\ddot X_J| Z}(\cdot,\ddot X_J;Z) \right] \otimes \left(\hat{\mu}_Z-\mu_Z\right)
    -T_{\hat{\mu}_{X_J}-\mu_{X_J}} \hat{\Sigma}_{ZZ}.
\end{align}
We first rewrite the quantity of interest as
\begin{align*}
    & \left(\hat{B}_{\ddot X_{J}\mid Z} - T_{\hat \mu_{X_J}} \hat \Sigma_{ZZ} (\hat{\Sigma}_{ZZ} + \epsilon_n I)^{-1}\right) - \left(B_{\ddot X_{J}\mid Z} - T_{\mu_{X_J}} \Sigma_{ZZ} \Sigma_{ZZ}^\dagger \right) \\
    = & \left( \hat \Sigma_{\ddot X_J Z} - T_{\hat \mu_{X_J}} \hat \Sigma_{ZZ} \right) (\hat{\Sigma}_{ZZ}+\epsilon_n I )^{-1} - \left(B_{\ddot X_{J}\mid Z} - T_{\mu_{X_J}} \Sigma_{ZZ} \Sigma_{ZZ}^\dagger \right).
\end{align*}
Substituting \eqref{eq:est_decomp} into the right-hand side gives
\begin{align*}
    & \left( B_{\ddot X_J\mid Z} -T_{\mu_{X_J}} \Sigma_{ZZ}\Sigma_{ZZ}^{\dagger} \right)\hat{\Sigma}_{ZZ} (\hat{\Sigma}_{ZZ}+\epsilon_n I )^{-1}\\
    & +\E_n\left[ \xi_{\ddot X_J\mid Z}(\cdot,\ddot X_J;Z) \otimes \tilde{\kappa}_Z(\cdot,Z) \right] (\hat{\Sigma}_{ZZ}+\epsilon_n I )^{-1}\\
    & - \left(\E_n\left[ \xi_{\ddot X_J\mid Z}(\cdot,\ddot X_J;Z) \right] \otimes \left(\hat{\mu}_Z-\mu_Z\right)\right)(\hat{\Sigma}_{ZZ}+\epsilon_n I )^{-1}\\
    & -T_{\hat{\mu}_{X_J}-\mu_{X_J}} \hat{\Sigma}_{ZZ} (\hat{\Sigma}_{ZZ}+\epsilon_n I )^{-1}
    - \left(B_{\ddot X_{J}\mid Z} - T_{\mu_{X_J}} \Sigma_{ZZ} \Sigma_{ZZ}^\dagger \right).
\end{align*}
Combining the first and last terms, this expression reduces to
\begin{align*}
    & -\epsilon_n \left( B_{\ddot X_J\mid Z} - T_{\mu_{X_J}} \Sigma_{ZZ}\Sigma_{ZZ}^{\dagger} \right) (\hat{\Sigma}_{ZZ}+\epsilon_n I )^{-1}\\
    & + \E_n\left[ \xi_{\ddot X_J\mid Z}(\cdot,\ddot X_J;Z) \otimes \tilde{\kappa}_Z(\cdot,Z) \right] (\hat{\Sigma}_{ZZ}+\epsilon_n I )^{-1}\\
    & - \left( \E_n\left[\xi_{\ddot X_J\mid Z}(\cdot,\ddot X_J;Z)\right] \otimes (\hat{\mu}_Z-\mu_Z) \right)(\hat{\Sigma}_{ZZ}+\epsilon_n I)^{-1}\\
    & - T_{\hat{\mu}_{X_J}-\mu_{X_J}} \hat{\Sigma}_{ZZ} (\hat{\Sigma}_{ZZ}+\epsilon_n I )^{-1}.
\end{align*}
Finally, applying the resolvent identity \eqref{eq:resolvent} to the first two terms yields
\begin{align}
    & \left(\hat{B}_{\ddot X_{J}\mid Z} - T_{\hat \mu_{X_J}} \hat \Sigma_{ZZ} (\hat{\Sigma}_{ZZ} + \epsilon_n I)^{-1}\right) - \left(B_{\ddot X_{J}\mid Z} - T_{\mu_{X_J}} \Sigma_{ZZ} \Sigma_{ZZ}^\dagger \right) \notag\\
    = & - \epsilon_n \left( B_{\ddot X_J\mid Z} - T_{\mu_{X_J}} \Sigma_{ZZ}\Sigma_{ZZ}^{\dagger} \right) (\Sigma_{ZZ}+\epsilon_n I )^{-1} \notag\\
    & + \E_n\left[ \xi_{\ddot X_J\mid Z}(\cdot,\ddot X_J;Z) \otimes \tilde{\kappa}_Z(\cdot,Z) \right]\left(\Sigma_{ZZ}+\epsilon_n I\right)^{-1} \notag\\
    & -\left( \E_n\left[ \xi_{\ddot X_J\mid Z}(\cdot,\ddot X_J;Z) \right] \otimes (\hat{\mu}_Z-\mu_Z) \right) (\hat{\Sigma}_{ZZ}+\epsilon_n I)^{-1} \notag\\
    & - T_{\hat{\mu}_{X_J}-\mu_{X_J}} \hat{\Sigma}_{ZZ} (\hat{\Sigma}_{ZZ}+\epsilon_n I)^{-1} \notag\\
    & + \epsilon_n \left( B_{\ddot X_J\mid Z} - T_{\mu_{X_J}} \Sigma_{ZZ}\Sigma_{ZZ}^{\dagger} \right)
    (\Sigma_{ZZ}+\epsilon_n I)^{-1} (\hat{\Sigma}_{ZZ}-\Sigma_{ZZ}) (\hat{\Sigma}_{ZZ}+\epsilon_n I )^{-1} \notag\\
    & - \E_n\left[ \xi_{\ddot X_J\mid Z}(\cdot,\ddot X_J;Z) \otimes \tilde{\kappa}_Z(\cdot,Z) \right]
    \left(\Sigma_{ZZ}+\epsilon_n I\right)^{-1} (\hat{\Sigma}_{ZZ}-\Sigma_{ZZ}) (\hat{\Sigma}_{ZZ}+\epsilon_n I )^{-1}.
    \label{eq:B-ddot-decomp}
\end{align}
By Assumption \ref{asm:inverse_beta}, we have
\begin{align*}
    B_{\ddot X_J| Z} -T_{\mu_{X_J}} \Sigma_{ZZ}\Sigma_{ZZ}^{\dagger} = \left( R_{\mu_Z} S_{X_J Z} + S_{\ddot X_J Z} \right) \Sigma_{ZZ}^{1+\beta} \Sigma_{ZZ}^\dagger
\end{align*}
Then
\begin{align*}
    & \norm{\left( B_{\ddot X_J| Z} -T_{\mu_{X_J}} \Sigma_{ZZ}\Sigma_{ZZ}^{\dagger}  \right) \left(\Sigma_{ZZ}+\epsilon_n I\right)^{-1}}_\mathrm{op}  \\
    \leq & \norm{R_{\mu_Z} S_{X_J Z} + S_{\ddot X_J Z}}_\mathrm{op} \norm{\Sigma_{ZZ}^{1+\beta} \Sigma_{ZZ}^\dagger \left(\Sigma_{ZZ}+\epsilon_n I\right)^{-1} }_\mathrm{op}  \\
    \leq & \left( \norm{\mu_Z}_{\mathcal{H}_Z} \norm{S_{X_J Z}}_\mathrm{op} + \norm{S_{\ddot X_J Z}}_\mathrm{op} \right) \norm{\Sigma_{ZZ}^{1+\beta} \Sigma_{ZZ}^\dagger \left(\Sigma_{ZZ}+\epsilon_n I\right)^{-1} }_\mathrm{op} \\
    \lesssim & \epsilon_n^{(\beta \wedge 1) - 1},
\end{align*}
where the last line holds because $R_{\mu_Z}$, $S_{X_J Z}$ and $S_{\ddot X_J Z}$ are all bounded and \eqref{eq:beta_inverse} holds.
Hence, the first term in \eqref{eq:B-ddot-decomp} is $O_P(\epsilon_n^{\beta \wedge 1})$, which follows from
\begin{align}
    \norm{\epsilon_n \left( B_{\ddot X_J| Z} -T_{\mu_{X_J}} \Sigma_{ZZ}\Sigma_{ZZ}^{\dagger}  \right) \left(\Sigma_{ZZ}+\epsilon_n I\right)^{-1}}_\mathrm{op} 
    \lesssim \epsilon_n \epsilon_n^{(\beta \wedge 1) -1} \lesssim \epsilon_n^{\beta \wedge 1} \label{rate:reg_term_1}
\end{align}
By \eqref{eqn:mean_zero_reg_op}, we have $\E\left[ \xi_{\ddot X_J\mid Z}(\cdot,\ddot X_J;Z) \otimes\tilde\kappa_Z(\cdot,Z) \right]
= 0$. Moreover, the boundedness of the kernels and of $B_{\ddot X_J\mid Z}$ implies the boundedness of $\xi_{\ddot X_J\mid Z}(\cdot,\ddot X_J;Z)$ under the Hilbert norm.
We further use the independence across observations and obtain
\begin{align*}
&\E \norm{ \E_n\left[ \xi_{\ddot X_J\mid Z}(\cdot,\ddot X_J;Z) \otimes \tilde\kappa_Z(\cdot,Z) \right] (\Sigma_{ZZ}+\epsilon_nI)^{-1} }_{\mathrm{HS}}^2\\
= & \frac1n \E\left[ \norm{ \xi_{\ddot X_J\mid Z}(\cdot,\ddot X_J;Z) }_{\mathcal H_{\ddot X_J}}^2
\norm{ (\Sigma_{ZZ}+\epsilon_nI)^{-1} \tilde\kappa_Z(\cdot,Z) }_{\mathcal H_Z}^2 \right]\\
\lesssim & \frac1n \E\left[ \norm{ (\Sigma_{ZZ}+\epsilon_nI)^{-1} \tilde\kappa_Z(\cdot,Z) }_{\mathcal H_Z}^2 \right] \\
= & \frac1n \sum_{k=1}^{\infty} \frac{\gamma_k}{(\gamma_k+\epsilon_n)^2}\\
\asymp & n^{-1} \epsilon_n^{-\frac{\alpha+1}{\alpha}},
\end{align*}
Hence, by Markov's inequality, we have the convergence rate for the second term  in \eqref{eq:B-ddot-decomp}
\begin{align}
    \norm{ \E_n\left[
\xi_{\ddot X_J\mid Z}(\cdot,\ddot X_J;Z)
\otimes\tilde\kappa_Z(\cdot,Z)
\right]
(\Sigma_{ZZ}+\epsilon_n I)^{-1}
}_\mathrm{op} = O_P(n^{-\frac{1}{2}}\epsilon_n^{-\frac{\alpha+1}{2 \alpha}})    \label{rate:reg_term_2}
\end{align}
And also for the sixth term in \eqref{eq:B-ddot-decomp}
\begin{align}
    & \norm{\E_n\left[ \xi_{\ddot X_J| Z}(\cdot,\ddot X_J;Z) \otimes \tilde{\kappa}_Z(\cdot,Z) \right] 
     \left(\Sigma_{ZZ}+\epsilon_n I \right)^{-1} (\hat{\Sigma}_{ZZ}-\Sigma_{ZZ}) (\hat{\Sigma}_{ZZ}+\epsilon_n I )^{-1}}_\mathrm{op} \notag \\
     = & O_P(n^{-\frac{1}{2}}\epsilon_n^{-\frac{\alpha+1}{2 \alpha}}) O_P(n^{-\frac{1}{2}}) O_P(\epsilon_n^{-1})
     =  O_P(n^{-1}\epsilon_n^{-\frac{3\alpha+1}{2 \alpha}}).     \label{rate:reg_term_6}
\end{align}

For the third term, since $(\hat{\Sigma}_{ZZ}+\epsilon_n I)^{-1}$ is self-adjoint, we have
\begin{align}
&\norm{ \left( \E_n\left[ \xi_{\ddot X_J\mid Z}(\cdot,\ddot X_J;Z) \right] \otimes (\hat{\mu}_Z-\mu_Z) \right) (\hat{\Sigma}_{ZZ}+\epsilon_n I)^{-1} }_\mathrm{op}
\notag\\
= & \norm{ \E_n\left[ \xi_{\ddot X_J\mid Z}(\cdot,\ddot X_J;Z) \right] }_{\mathcal H_{\ddot X_J}} \norm{ (\hat{\Sigma}_{ZZ}+\epsilon_n I)^{-1} (\hat{\mu}_Z-\mu_Z) }_{\mathcal H_Z} \notag\\
\leq & O_P(n^{-\frac12}) \epsilon_n^{-1} O_P(n^{-\frac12}) \notag\\
= & O_P(n^{-1}\epsilon_n^{-1})
\label{rate:reg_term_3}
\end{align}
which is absorbed by the sixth term because $O_P(n^{-1}\epsilon_n^{-1}) = o_P(n^{-1}\epsilon_n^{-\frac{3\alpha+1}{2\alpha}})$. The inequality holds because
\begin{align*}
    \norm{ \E_n\left[ \xi_{\ddot X_J\mid Z}(\cdot,\ddot X_J;Z) \right] }_{\mathcal H_{\ddot X_J}} = O_P(n^{-\frac12})
    \quad {and} \quad
    \norm{\hat{\mu}_Z-\mu_Z }_{\mathcal H_Z} = O_P(n^{-\frac12})
\end{align*}
by Hilbert space CLT; and $\norm{\left(\hat \Sigma_{ZZ} + \epsilon_n I\right)^{-1}}_{\mathrm{op}} \leq \epsilon_n^{-1}$ by spectrum calculation.

For the fourth term, by the definition of $T_{\hat{\mu}_{X_J}-\mu_{X_J}}$ and $\norm{
\hat{\Sigma}_{ZZ}
(\hat{\Sigma}_{ZZ}+\epsilon_n I)^{-1}
}_\mathrm{op} \leq 1$, we have
\begin{align}
& \norm{ T_{\hat{\mu}_{X_J}-\mu_{X_J}} \hat{\Sigma}_{ZZ}
(\hat{\Sigma}_{ZZ}+\epsilon_n I)^{-1} }_\mathrm{op} \notag
\\
\leq & \norm{ T_{\hat{\mu}_{X_J}-\mu_{X_J}} }_\mathrm{op} \norm{
\hat{\Sigma}_{ZZ} (\hat{\Sigma}_{ZZ}+\epsilon_n I)^{-1} }_\mathrm{op} \notag
\\
\leq & \norm{ \hat{\mu}_{X_J}-\mu_{X_J}}_{\mathcal H_{X_J}} \notag\\
= & O_P(n^{-\frac{1}{2}})   \label{rate:reg_term_4}
\end{align}

For the fifth term, 
\begin{align}
& \norm{ \epsilon_n \left( B_{\ddot X_J\mid Z} - T_{\mu_{X_J}} \Sigma_{ZZ}\Sigma_{ZZ}^{\dagger} \right) (\Sigma_{ZZ}+\epsilon_n I)^{-1} (\hat{\Sigma}_{ZZ}-\Sigma_{ZZ}) (\hat{\Sigma}_{ZZ}+\epsilon_n I)^{-1}}_\mathrm{op}   \notag
\\
\leq & \epsilon_n \norm{ \left( B_{\ddot X_J\mid Z} - T_{\mu_{X_J}} \Sigma_{ZZ}\Sigma_{ZZ}^{\dagger} \right) (\Sigma_{ZZ}+\epsilon_n I)^{-1}}_\mathrm{op} 
\norm{ \hat{\Sigma}_{ZZ}-\Sigma_{ZZ} }_\mathrm{op}
\norm{ (\hat{\Sigma}_{ZZ}+\epsilon_n I)^{-1} }_\mathrm{op}  \notag 
\\
\lesssim & \epsilon_n \epsilon_n^{(\beta \wedge 1) - 1} O_P(n^{- \frac{1}{2}}) \epsilon_n^{-1} \notag
\\
= & O_P(n^{- \frac{1}{2}} \epsilon_n^{(\beta \wedge 1) - 1})  \label{rate:reg_term_5}
\end{align}
Therefore, we combine previous results \eqref{rate:reg_term_1}--\eqref{rate:reg_term_5} and conclude that
\begin{align*}
    & \norm{\left(\hat{B}_{\ddot X_{J}|Z} - T_{\hat \mu_{X_J}} \hat \Sigma_{ZZ} (\hat{\Sigma}_{ZZ} + \epsilon_n I)^{-1}\right) - \left(B_{\ddot X_{J}|Z} - T_{\mu_{X_J}} \Sigma_{ZZ} \Sigma_{ZZ}^\dagger \right)}_\mathrm{op} \\
    = & O_P \left(n^{-\frac{1}{2}} \epsilon_{n}^{\beta \wedge 1 - 1} + \epsilon_{n}^{\beta \wedge 1} + n^{-1} \epsilon_{n}^{-\frac{3\alpha+1}{2\alpha}} + n^{-\frac{1}{2}} \epsilon_{n}^{-\frac{1+\alpha}{2\alpha}} + n^{-\frac{1}{2}} \right),
\end{align*}
which proves assertion $2$.
\end{proof}
\begin{proof}[Proof of Lemma~\ref{lem:smoothed_projection_error}]
By the definition of $R_n$, adding and subtracting $T_{\mu_{X_J}} \hat\Sigma_{ZZ} (\hat\Sigma_{ZZ}+\epsilon_nI)^{-1}$ we obtain
\begin{align*}
    R_n\Sigma_{ZZ}^{\theta}
    = T_{\hat\mu_{X_J}-\mu_{X_J}} \hat\Sigma_{ZZ} (\hat\Sigma_{ZZ}+\epsilon_nI)^{-1} \Sigma_{ZZ}^{\theta}
    + T_{\mu_{X_J}} \left[ \hat\Sigma_{ZZ} (\hat\Sigma_{ZZ}+\epsilon_nI)^{-1} - \Sigma_{ZZ}\Sigma_{ZZ}^{\dagger} \right] \Sigma_{ZZ}^{\theta}.
\end{align*}

For the first term
\begin{align}
    \left\|
        T_{\hat\mu_{X_J}-\mu_{X_J}}
        \hat\Sigma_{ZZ} (\hat\Sigma_{ZZ}+\epsilon_nI)^{-1}
        \Sigma_{ZZ}^{\theta}
    \right\|_{\mathrm{op}}
    \leq \norm{\hat\mu_{X_J}-\mu_{X_J}}_{\mathcal{H}_{X_J}} \norm{\hat\Sigma_{ZZ} (\hat\Sigma_{ZZ}+\epsilon_nI)^{-1}}_\mathrm{op} \norm{\Sigma_{ZZ}^{\theta}}_\mathrm{op} = O_P(n^{-1/2})
    \label{eqn:Rn_mean_term}
\end{align}

For the second term, since $\theta>0$, we have $\Sigma_{ZZ}\Sigma_{ZZ}^{\dagger} \Sigma_{ZZ}^{\theta} = \Sigma_{ZZ}^{\theta}$. Then we obtain
\begin{align}
    & \left[ \hat\Sigma_{ZZ} (\hat\Sigma_{ZZ}+\epsilon_nI)^{-1}
        - \Sigma_{ZZ}\Sigma_{ZZ}^{\dagger} \right] \Sigma_{ZZ}^{\theta} \notag \\
    = & \left[ \hat\Sigma_{ZZ} (\hat\Sigma_{ZZ}+\epsilon_nI)^{-1}
        - I \right] \Sigma_{ZZ}^{\theta}
        + \left[ I - \Sigma_{ZZ}\Sigma_{ZZ}^{\dagger} \right] \Sigma_{ZZ}^{\theta} \notag \notag \\
    = & - \epsilon_n (\hat\Sigma_{ZZ}+\epsilon_nI)^{-1} \Sigma_{ZZ}^{\theta}
    \label{eqn:Rn_projection_identity}
\end{align}

By the resolvent identity,
\begin{align*}
    (\hat\Sigma_{ZZ}+\epsilon_nI)^{-1}
    \Sigma_{ZZ}^{\theta}
    = (\Sigma_{ZZ}+\epsilon_nI)^{-1} \Sigma_{ZZ}^{\theta}
    + (\hat\Sigma_{ZZ}+\epsilon_nI)^{-1} (\Sigma_{ZZ}-\hat\Sigma_{ZZ}) (\Sigma_{ZZ}+\epsilon_nI)^{-1} \Sigma_{ZZ}^{\theta}.
\end{align*}
The spectral theorem gives $\norm{(\Sigma_{ZZ}+\epsilon_nI)^{-1} \Sigma_{ZZ}^{\theta}}_{\mathrm{op}}
= O(\epsilon_n^{\theta\wedge1-1})$. Combined with \eqref{eq:CLT_Sigma_Z}
\begin{align}
    &\epsilon_n \left\| (\hat\Sigma_{ZZ}+\epsilon_nI)^{-1} \Sigma_{ZZ}^{\theta} \right\|_{\mathrm{op}}  \notag \\
    \leq & \epsilon_n \norm{(\Sigma_{ZZ}+\epsilon_nI)^{-1} \Sigma_{ZZ}^{\theta}}_{\mathrm{op}}
    + \epsilon_n \norm{(\hat\Sigma_{ZZ}+\epsilon_nI)^{-1}}_\mathrm{op}
    \norm{(\Sigma_{ZZ}-\hat\Sigma_{ZZ})}_\mathrm{op}
    \norm{(\Sigma_{ZZ}+\epsilon_nI)^{-1} \Sigma_{ZZ}^{\theta}}_{\mathrm{op}}   \notag \\
    \leq & \epsilon_n O(\epsilon_n^{\theta\wedge1 - 1}) + \epsilon_n \epsilon_n^{-1} O_P(n^{-1/2}) O(\epsilon_n^{\theta\wedge1 - 1})
    \leq O_P(\epsilon_n^{\theta \wedge 1} + n^{-1/2} \epsilon_n^{\theta\wedge1 - 1} ). \label{eqn:Rn_2nd_term}
\end{align}
Since
$\|T_{\mu_{X_J}}\|_{\mathrm{op}}
=\|\mu_{X_J}\|_{\mathcal H_{X_J}}<\infty$, equations
\eqref{eqn:Rn_mean_term}--\eqref{eqn:Rn_2nd_term} imply
\begin{align*}
    \|R_n\Sigma_{ZZ}^{\theta}\|_{\mathrm{op}}
    = O_P( n^{-1/2} + n^{-1/2}\epsilon_n^{\theta\wedge1-1} + \epsilon_n^{\theta\wedge1} ).
\end{align*}
Finally, since $\epsilon_n\to0$ and
$\theta\wedge1-1\leq0$, we have $n^{-1/2}= O(n^{-1/2}\epsilon_n^{\theta\wedge1-1})$. Therefore,
\begin{align*}
    \|R_n\Sigma_{ZZ}^{\theta}\|_{\mathrm{op}}
    = O_P( n^{-1/2}\epsilon_n^{\theta\wedge1-1} + \epsilon_n^{\theta\wedge1}),
\end{align*}
which proves \eqref{eqn:smoothed_Rn_rate}.
\end{proof}
\begin{proof}[Proof of Proposition~\ref{prop:remainder_op1}]
Fix $j\in\{1,\ldots,d-1\}$, and let $R_n$ denote the operator in \eqref{eqn:def_Rn} with $J=\{j\}$.
    Since $\Sigma_{ZZ}$ is a positive self-adjoint compact operator, its fractional power $\Sigma_{ZZ}^{1-\theta}$ admits the spectral representation with eigenvalues $\{\gamma_k^{1-\theta}\}_k$. If we choose $\theta$ such that
\[
\frac{\alpha+1}{2\alpha}
<
\theta
<
1-\frac{1}{2\alpha}.
\]
Such a $\theta$ exists because $\alpha>2$. Moreover, the upper
bound on $\theta$, together with Assumption
\ref{asm:Sig_ZZ_eigen}, implies that
\begin{align*}
\norm{\Sigma_{ZZ}^{1-\theta}}_{\mathrm{HS}}^2
=
\sum_{k=1}^{\infty}\gamma_k^{2(1-\theta)}
\lesssim
\sum_{k=1}^{\infty}k^{-2\alpha(1-\theta)}
<\infty.
\end{align*}
It then follows from Lemma~\ref{lem:smoothed_projection_error} and Proposition~\ref{prop:optimal_rate_reg_operator} that
\begin{align*}
& \sqrt n \norm{ R_n\Sigma_{ZZ} \left( \hat B_{X_{j+}\mid Z} - B_{X_{j+}\mid Z} \right)^* }_{\mathrm{HS}}
\\
\leq & \sqrt n \norm{R_n\Sigma_{ZZ}^{\theta}}_{\mathrm{op}} 
\norm{\Sigma_{ZZ}^{1-\theta}}_{\mathrm{HS}}
\norm{\hat B_{X_{j+}\mid Z}
- B_{X_{j+}\mid Z} }_{\mathrm{op}}
\\
\leq &
O_P\left( \epsilon_n^{\theta-1} n^{-\alpha(\beta\wedge1)/(2\alpha(\beta\wedge1)+\alpha+1)}
+ \sqrt n\,\epsilon_n^\theta n^{-\alpha(\beta\wedge1) /(2\alpha(\beta\wedge1)+\alpha+1)} \right)
\\
\leq &
O_P\left( n^{ -\alpha(\beta\wedge1+\theta-1)/(2\alpha(\beta\wedge1)+\alpha+1)}
+ n^{ 1/2- \alpha(\beta\wedge1+\theta)/(2\alpha(\beta\wedge1)+\alpha+1)}\right) \\
= & o_P(1).
\end{align*}
Indeed, the exponent in the first term is negative because
\begin{align*}
\beta\wedge1+\theta-1
> \frac{\alpha+1}{2\alpha}
+ \frac{\alpha+1}{2\alpha}
-1
= \frac1\alpha
>0,
\end{align*}
whereas the exponent in the second term is negative because
\begin{align*}
\frac12-
\frac{\alpha(\beta\wedge1+\theta)}
{2\alpha(\beta\wedge1)+\alpha+1}
<0
\quad\Longleftrightarrow\quad
\theta>\frac{\alpha+1}{2\alpha}
\end{align*}
which proves \eqref{eqn:mixed_projection_remainder}.
\end{proof}

\begin{proof}[Proof of Lemma~\ref{lem:asymptotic_normality}]
Fix $j\in\{1,\ldots,d-1\}$. Throughout this proof, $R_n$ denotes the operator in \eqref{eqn:def_Rn} with $J=\{j\}$. Two intermediate operators are needed here. Let
\begin{align*}
    \tilde{\Sigma}^{1}_{\ddot{X}_{j} X_{j+}\mid Z}
    = & \E_n\left[
        \xi_{\ddot{X}_{j}\mid Z}(\cdot,\ddot{X}_{j};Z)
        \otimes \xi_{X_{j+}\mid Z}(\cdot,X_{j+};Z)
    \right], \\
    \tilde{\Sigma}^{2}_{\ddot{X}_{j} X_{j+}\mid Z}
    = & \E_n\left[
        \left(\tilde{\kappa}_{\ddot{X}_{j}}(\cdot,\ddot{X}_{j})
        -\hat{B}_{\ddot{X}_{j}\mid Z}\tilde{\kappa}_{Z}(\cdot,Z)\right)
        \otimes
        \left(\tilde{\kappa}_{X_{j+}}(\cdot,X_{j+})
        -\hat{B}_{X_{j+}\mid Z}\tilde{\kappa}_{Z}(\cdot,Z)\right)
    \right].
\end{align*}
Here $\hat{\Sigma}_{\ddot{X}_{j} X_{j+}\mid Z}$ is the empirical cross-covariance of the fitted residuals, obtained from $\tilde{\Sigma}^{2}_{\ddot{X}_{j} X_{j+}\mid Z}$ by replacing the population-centered kernels $\tilde{\kappa}$ with the sample-centered kernels $\hat{\kappa}$. We first show that $\hat{\Sigma}_{\ddot{X}_{j} X_{j+}\mid Z}$ and $\tilde{\Sigma}^{1}_{\ddot{X}_{j} X_{j+}\mid Z}$ differ by $o_P(n^{-1/2})$ in Hilbert--Schmidt norm. We split it into two parts.

Let $A=\hat{B}_{\ddot{X}_{j}\mid Z}-B_{\ddot{X}_{j}\mid Z}-R_n$ and $C=\hat{B}_{X_{j+}\mid Z}-B_{X_{j+}\mid Z}$. By Theorem~\ref{thm:rate_reg_operator} and Proposition~\ref{prop:optimal_rate_reg_operator},
\begin{align}\label{eqn:operator-rates-final}
    \norm{A}_{\mathrm{op}}+\norm{C}_{\mathrm{op}}
    = & O_P\left(n^{-\alpha(\beta\wedge1)
        /(2\alpha(\beta\wedge1)+\alpha+1)}\right)
    = o_P(n^{-1/4}).
\end{align}
Since the kernels are bounded and both $\hat{\Sigma}_{ZZ}(\hat{\Sigma}_{ZZ}+\epsilon_n I)^{-1}$ and $\Sigma_{ZZ}\Sigma_{ZZ}^{\dagger}$ have operator norm at most one,
\begin{align*}
    \norm{R_n}_{\mathrm{op}}
    \leq & \norm{\hat{\mu}_{X_j}}_{\mathcal H_{X_j}}
        +\norm{\mu_{X_j}}_{\mathcal H_{X_j}}
    = O_P(1).
\end{align*}
Together with the boundedness of the population regression operators, this gives
\begin{align*}
    \norm{\hat{B}_{\ddot{X}_{j}\mid Z}}_{\mathrm{op}}
    +\norm{\hat{B}_{X_{j+}\mid Z}}_{\mathrm{op}}=O_P(1).
\end{align*}

We first consider the difference between $\hat{\Sigma}_{\ddot{X}_{j} X_{j+}\mid Z}$ and $\tilde{\Sigma}^{2}_{\ddot{X}_{j} X_{j+}\mid Z}$. Since the kernels are bounded, $\norm{\hat{\mu}_X-\mu_X}_{\mathcal H_X}=O_P(n^{-1/2})$. Notice that $\hat{\kappa}_X(\cdot,X) =\tilde{\kappa}_X(\cdot,X)-(\hat{\mu}_X-\mu_X)$.
Then by definition
\begin{align*}
    \hat{\Sigma}_{\ddot{X}_{j} X_{j+}\mid Z}
    -\tilde{\Sigma}^{2}_{\ddot{X}_{j} X_{j+}\mid Z}
    = & -\Big[
        (\hat{\mu}_{\ddot{X}_{j}}-\mu_{\ddot{X}_{j}})
        -\hat{B}_{\ddot{X}_{j}\mid Z}(\hat{\mu}_Z-\mu_Z)
    \Big]
    \otimes \Big[
        (\hat{\mu}_{X_{j+}}-\mu_{X_{j+}})
        -\hat{B}_{X_{j+}\mid Z}(\hat{\mu}_Z-\mu_Z)
    \Big].
\end{align*}
Both factors are $O_P(n^{-1/2})$ in their respective reproducing kernel Hilbert space norms. Hence
\begin{align*}
    \sqrt n\norm{
        \hat{\Sigma}_{\ddot{X}_{j} X_{j+}\mid Z}
        -\tilde{\Sigma}^{2}_{\ddot{X}_{j} X_{j+}\mid Z}
    }_{\mathrm{HS}}
    =O_P(n^{-1/2})=o_P(1).
\end{align*}

We next consider the difference between $\tilde{\Sigma}^{2}_{\ddot{X}_{j} X_{j+}\mid Z}$ and $\tilde{\Sigma}^{1}_{\ddot{X}_{j} X_{j+}\mid Z}$. The boundedness of the kernels and the central limit theorem in Hilbert space give
\begin{align*}
    \norm{\E_n\left[
        \tilde{\kappa}_Z(\cdot,Z)\otimes\tilde{\kappa}_Z(\cdot,Z)
    \right]-\Sigma_{ZZ}}_{\mathrm{HS}}
    = & O_P(n^{-1/2}), \\
    \norm{\E_n\left[
        \tilde{\kappa}_Z(\cdot,Z)\otimes\tilde{\kappa}_Z(\cdot,Z)
    \right]}_{\mathrm{HS}}
    = & O_P(1).
\end{align*}
We also know that
\begin{align*}
    \E\left[\xi_{\ddot{X}_{j}\mid Z}(\cdot,\ddot{X}_{j};Z)\mid Z\right]=0,
    \qquad
    \E\left[\xi_{X_{j+}\mid Z}(\cdot,X_{j+};Z)\mid Z\right]=0.
\end{align*}
Since the residuals are bounded, another application of the central limit theorem in Hilbert space gives
\begin{align*}
    \norm{\E_n\left[
        \xi_{\ddot{X}_{j}\mid Z}(\cdot,\ddot{X}_{j};Z)
        \otimes\tilde{\kappa}_Z(\cdot,Z)
    \right]}_{\mathrm{HS}}
    = & O_P(n^{-1/2}), \\
    \norm{\E_n\left[
        \tilde{\kappa}_Z(\cdot,Z)
        \otimes\xi_{X_{j+}\mid Z}(\cdot,X_{j+};Z)
    \right]}_{\mathrm{HS}}
    = & O_P(n^{-1/2}).
\end{align*}
Notice that $\hat{B}_{\ddot{X}_{j}\mid Z}-B_{\ddot{X}_{j}\mid Z}=A+R_n$. Expanding the difference, we get
\begin{align*}
    \tilde{\Sigma}^{2}_{\ddot{X}_{j} X_{j+}\mid Z}
    -\tilde{\Sigma}^{1}_{\ddot{X}_{j} X_{j+}\mid Z}
    = & A\E_n\left[
        \tilde{\kappa}_Z(\cdot,Z)\otimes\tilde{\kappa}_Z(\cdot,Z)
    \right]C^* \\
    & -A\E_n\left[
        \tilde{\kappa}_Z(\cdot,Z)
        \otimes\xi_{X_{j+}\mid Z}(\cdot,X_{j+};Z)
    \right] \\
    & -\E_n\left[
        \xi_{\ddot{X}_{j}\mid Z}(\cdot,\ddot{X}_{j};Z)
        \otimes\tilde{\kappa}_Z(\cdot,Z)
    \right]C^* \\
    & +R_n\E_n\left[
        \tilde{\kappa}_Z(\cdot,Z)\otimes\tilde{\kappa}_Z(\cdot,Z)
    \right]C^* \\
    & -R_n\E_n\left[
        \tilde{\kappa}_Z(\cdot,Z)
        \otimes\xi_{X_{j+}\mid Z}(\cdot,X_{j+};Z)
    \right].
\end{align*}
By \eqref{eqn:operator-rates-final} and the preceding bounds, the first three terms are $o_P(n^{-1/2})$ in Hilbert--Schmidt norm. Now we only need to control the two terms involving $R_n$.

By Proposition~\ref{prop:remainder_op1}, $\sqrt n\norm{R_n\Sigma_{ZZ}C^*}_{\mathrm{HS}}=o_P(1)$. Therefore,
\begin{align}\label{eqn:R_n_ZZ_C}
    & \sqrt n\norm{R_n\E_n\left[
        \tilde{\kappa}_Z(\cdot,Z)\otimes\tilde{\kappa}_Z(\cdot,Z)
    \right]C^*}_{\mathrm{HS}} \notag\\
    \leq & \sqrt n\norm{R_n\Sigma_{ZZ}C^*}_{\mathrm{HS}}
    +\norm{R_n}_{\mathrm{op}}\sqrt n
    \norm{\E_n\left[
        \tilde{\kappa}_Z(\cdot,Z)\otimes\tilde{\kappa}_Z(\cdot,Z)
    \right]-\Sigma_{ZZ}}_{\mathrm{HS}}\norm{C}_{\mathrm{op}} \notag\\
    = & o_P(1)+O_P(1)O_P(1)o_P(1)=o_P(1).
\end{align}

For the remaining term, choose some constant $\theta$ such that $0<\theta<\frac{\alpha-1}{2\alpha}$. Let $\{e_k\}_{k\geq1}$ be orthonormal eigenfunctions of $\Sigma_{ZZ}$ corresponding to its positive eigenvalues $\{\gamma_k\}_{k\geq1}$. Since $\tilde{\kappa}_Z(\cdot,Z)\in\ker(\Sigma_{ZZ})^\perp$ almost surely, Parseval's identity and Tonelli's theorem give
\begin{align*}
    \E\norm{(\Sigma_{ZZ}^{\dagger})^\theta
        \tilde{\kappa}_Z(\cdot,Z)}_{\mathcal H_Z}^2
    = & \sum_{k=1}^{\infty}\gamma_k^{-2\theta}
        \E\left[\left|
            \left\langle\tilde{\kappa}_Z(\cdot,Z),e_k\right\rangle_{\mathcal H_Z}
        \right|^2\right] \\
    = & \sum_{k=1}^{\infty}\gamma_k^{1-2\theta}
    \lesssim \sum_{k=1}^{\infty}k^{-\alpha(1-2\theta)}
    <\infty.
\end{align*}
In particular, $(\Sigma_{ZZ}^{\dagger})^\theta\tilde{\kappa}_Z(\cdot,Z)$ is well defined almost surely. Notice that $\Sigma_{ZZ}^{\theta}(\Sigma_{ZZ}^{\dagger})^\theta\tilde{\kappa}_Z(\cdot,Z)=\tilde{\kappa}_Z(\cdot,Z)$ almost surely. Then
\begin{align}\label{eqn:split_R_n_theta}
    & \sqrt n\norm{R_n\E_n\left[
        \tilde{\kappa}_Z(\cdot,Z)
        \otimes\xi_{X_{j+}\mid Z}(\cdot,X_{j+};Z)
    \right]}_{\mathrm{HS}} \notag\\
    \leq & \norm{R_n\Sigma_{ZZ}^{\theta}}_{\mathrm{op}}
    \sqrt n\norm{\E_n\left[
        (\Sigma_{ZZ}^{\dagger})^\theta\tilde{\kappa}_Z(\cdot,Z)
        \otimes\xi_{X_{j+}\mid Z}(\cdot,X_{j+};Z)
    \right]}_{\mathrm{HS}}.
\end{align}
Under the choice $\epsilon_n=n^{-\delta}$, where $\delta=\frac{\alpha}{2\alpha(\beta\wedge1)+\alpha+1}<\frac12$, Lemma~\ref{lem:smoothed_projection_error} gives
\begin{align*}
    \norm{R_n\Sigma_{ZZ}^{\theta}}_{\mathrm{op}}
    = & O_P\left(n^{-1/2}\epsilon_n^{\theta-1}
        +\epsilon_n^\theta\right) \\
    = & O_P\left(n^{-1/2+\delta(1-\theta)}
        +n^{-\delta\theta}\right)=o_P(1).
\end{align*}
Furthermore, the conditional mean-zero property of the residual gives
\begin{align*}
    \E\left[
        (\Sigma_{ZZ}^{\dagger})^\theta\tilde{\kappa}_Z(\cdot,Z)
        \otimes\xi_{X_{j+}\mid Z}(\cdot,X_{j+};Z)
    \right]=0.
\end{align*}
Since $\xi_{X_{j+}\mid Z}$ is bounded, we also have
\begin{align*}
    \E\norm{
        (\Sigma_{ZZ}^{\dagger})^\theta\tilde{\kappa}_Z(\cdot,Z)
        \otimes\xi_{X_{j+}\mid Z}(\cdot,X_{j+};Z)
    }_{\mathrm{HS}}^2
    \lesssim & \E\norm{
        (\Sigma_{ZZ}^{\dagger})^\theta\tilde{\kappa}_Z(\cdot,Z)
    }_{\mathcal H_Z}^2<\infty.
\end{align*}
It follows from independence across observations that
\begin{align*}
    & n\E\norm{\E_n\left[
        (\Sigma_{ZZ}^{\dagger})^\theta\tilde{\kappa}_Z(\cdot,Z)
        \otimes\xi_{X_{j+}\mid Z}(\cdot,X_{j+};Z)
    \right]}_{\mathrm{HS}}^2 \\
    = & \E\norm{
        (\Sigma_{ZZ}^{\dagger})^\theta\tilde{\kappa}_Z(\cdot,Z)
        \otimes\xi_{X_{j+}\mid Z}(\cdot,X_{j+};Z)
    }_{\mathrm{HS}}^2<\infty.
\end{align*}
Then by Chebyshev's inequality,
\begin{align*}
    \sqrt n\norm{\E_n\left[
        (\Sigma_{ZZ}^{\dagger})^\theta\tilde{\kappa}_Z(\cdot,Z)
        \otimes\xi_{X_{j+}\mid Z}(\cdot,X_{j+};Z)
    \right]}_{\mathrm{HS}}=O_P(1).
\end{align*}
Now we come back to \eqref{eqn:split_R_n_theta} and obtain
\begin{align}\label{eqn:R_n_Z_residual}
    \sqrt n\norm{R_n\E_n\left[
        \tilde{\kappa}_Z(\cdot,Z)
        \otimes\xi_{X_{j+}\mid Z}(\cdot,X_{j+};Z)
    \right]}_{\mathrm{HS}}=o_P(1).
\end{align}
Combining \eqref{eqn:R_n_ZZ_C} and \eqref{eqn:R_n_Z_residual} with the bounds for the first three terms proves
\begin{align*}
    \sqrt n\norm{
        \tilde{\Sigma}^{2}_{\ddot{X}_{j} X_{j+}\mid Z}
        -\tilde{\Sigma}^{1}_{\ddot{X}_{j} X_{j+}\mid Z}
    }_{\mathrm{HS}}=o_P(1).
\end{align*}
Together with the first part, this gives
\begin{align*}
    \sqrt n\norm{
        \hat{\Sigma}_{\ddot{X}_{j} X_{j+}\mid Z}
        -\tilde{\Sigma}^{1}_{\ddot{X}_{j} X_{j+}\mid Z}
    }_{\mathrm{HS}}=o_P(1).
\end{align*}

Recall that the influence function is
\begin{align}\label{eqn:influence_func}
    \psi(\ddot{X}_{j},X_{j+};Z)
    = & \xi_{\ddot{X}_{j}\mid Z}(\cdot,\ddot{X}_{j};Z)
        \otimes\xi_{X_{j+}\mid Z}(\cdot,X_{j+};Z)
        -\Sigma_{\ddot{X}_{j} X_{j+}\mid Z}.
\end{align}
By the definition of $\tilde{\Sigma}^{1}_{\ddot{X}_{j} X_{j+}\mid Z}$,
\begin{align*}
    \sqrt n\left(
        \hat{\Sigma}_{\ddot{X}_{j} X_{j+}\mid Z}
        -\Sigma_{\ddot{X}_{j} X_{j+}\mid Z}
    \right)
    = & \sqrt n\left(
        \tilde{\Sigma}^{1}_{\ddot{X}_{j} X_{j+}\mid Z}
        -\Sigma_{\ddot{X}_{j} X_{j+}\mid Z}
    \right)+o_P(1) \\
    = & \frac{1}{\sqrt n}\sum_{i=1}^{n}
        \psi(\ddot{X}_{j,i},X_{j+,i};Z_i)+o_P(1),
\end{align*}
where the remainders are measured in Hilbert--Schmidt norm. Since the kernels and regression operators are bounded,
\begin{align*}
    \E\norm{
        \xi_{\ddot{X}_{j}\mid Z}(\cdot,\ddot{X}_{j};Z)
        \otimes\xi_{X_{j+}\mid Z}(\cdot,X_{j+};Z)
    }_{\mathrm{HS}}^2
    = & \E\left[
        \norm{\xi_{\ddot{X}_{j}\mid Z}(\cdot,\ddot{X}_{j};Z)}_{\mathcal H_{\ddot{X}_{j}}}^2
        \norm{\xi_{X_{j+}\mid Z}(\cdot,X_{j+};Z)}_{\mathcal H_{X_{j+}}}^2
    \right]<\infty.
\end{align*}
By the residual representation of $\Sigma_{\ddot{X}_{j} X_{j+}\mid Z}$, $\E[\psi(\ddot{X}_{j},X_{j+};Z)]=0$. The preceding bound also gives $\E\norm{\psi(\ddot{X}_{j},X_{j+};Z)}_{\mathrm{HS}}^2<\infty$. We can apply the central limit theorem in Hilbert space and Slutsky's theorem to get
\begin{align*}
    \sqrt n\left(
        \hat{\Sigma}_{\ddot{X}_{j} X_{j+}\mid Z}
        -\Sigma_{\ddot{X}_{j} X_{j+}\mid Z}
    \right)
    \xrightarrow{\mathcal D}
    N\left(0,\Gamma_{\ddot{X}_{j} X_{j+}\mid Z}\right).
\end{align*}

\end{proof}
\begin{proof}[Proof of Theorem~\ref{thm:S_asymptotic}]
For each $j=1,\ldots,d-1$, the proof of Lemma~\ref{lem:asymptotic_normality} gives
\begin{align*}
&\sqrt n\left(
\hat\Sigma_{\ddot X_jX_{j+}\mid Z} - \Sigma_{\ddot X_jX_{j+}\mid Z} \right)
=
\frac{1}{\sqrt n}\sum_{i=1}^n\psi_{j,i} + o_P(1),
\end{align*}
where
\begin{align*}
\psi_{j,i}
=
\xi_{\ddot X_j\mid Z} (\cdot,\ddot X_{j,i};Z_i) \otimes \xi_{X_{j+}\mid Z}(\cdot,X_{j+,i};Z_i)
- \Sigma_{\ddot X_jX_{j+}\mid Z}.
\end{align*}
Let $\psi_i = (\psi_{1,i},\ldots,\psi_{d-1,i})^\top$. Then $\psi_i$, as a random element of $\mathbb H_S$, is the joint influence function of $\hat S$. That is,
\begin{align}
\sqrt n(\hat S-S)
=
\frac{1}{\sqrt n}\sum_{i=1}^n\psi_i
+o_P(1).
\label{eqn:joint_asymptotic_linearity}
\end{align}
By the residual representation of the conditional
cross-covariance operators, $\E[\psi_i]=0$.
Moreover, $\Gamma_{S\mid Z}=\E[\psi_i\otimes\psi_i]$ is trace-class. Therefore, the Hilbert space CLT gives
\begin{align*}
\frac{1}{\sqrt n}\sum_{i=1}^n\psi_i
\xrightarrow{\mathcal D}
N(0,\Gamma_{S\mid Z}),
\end{align*}
For any $j,k\in\{1,\ldots,d-1\}$, the $(j,k)$ block of this covariance
operator is
\begin{align*}
\left(\Gamma_{S\mid Z}\right)_{(j,k)} = \E\left[\psi_{j,i}\otimes\psi_{k,i}\right].
\end{align*}
Combining this Hilbert-space central limit theorem with
\eqref{eqn:joint_asymptotic_linearity} and Slutsky's theorem proves \eqref{eqn:test_asym}.
\end{proof}

\begin{proof}[of Theorem~\ref{thm:local_alternative}]
We first verify that the asymptotic linear representation established in the proof of Lemma \ref{lem:asymptotic_normality} remains valid along $\{P_n\}$. Assumption \ref{asm:uniform_local} makes every bound
used there uniform in $n$. Therefore, uniformly over all relevant subsets $J$,
\begin{align}
\norm{\hat\mu_{X_J,n}-\mu_{X_J,n}}_{\mathcal H_{X_J}}
= & O_{P_n}(n^{-1/2}), \quad
\norm{\hat\mu_{\ddot X_J, n}-\mu_{\ddot X_J,n}}_{\mathcal H_{\ddot X_J}}
= O_{P_n}(n^{-1/2}), \notag\\
\norm{\hat\mu_{Z,n}-\mu_{Z,n}}_{\mathcal H_Z} \quad
= & O_{P_n}(n^{-1/2}),
\label{eqn:local_mean_rates}
\end{align}
and
\begin{align}
\norm{\hat\Sigma_{X_JZ, n}-\Sigma_{X_JZ,n}}_{\mathrm{HS}}
= & O_{P_n}(n^{-1/2}), \quad
\norm{\hat\Sigma_{\ddot X_JZ, n}-\Sigma_{\ddot X_JZ,n}}_{\mathrm{HS}} 
= O_{P_n}(n^{-1/2}), \notag \\
\norm{\hat\Sigma_{ZZ,n}-\Sigma_{ZZ,n}}_{\mathrm{HS}} \ \
= & O_{P_n}(n^{-1/2}).
\label{eqn:local_covariance_rates}
\end{align}

The uniform source bounds and the uniform decay rate of eigenvalues imply, by the proofs of Theorem~\ref{thm:rate_reg_operator}, Corollary~\ref{prop:optimal_rate_reg_operator}, and Lemma~\ref{lem:smoothed_projection_error}, that their conclusions hold their conclusions hold with $O_P$ and $o_P$ replaced by $O_{P_n}$ and $o_{P_n}$, respectively. Here $R_n$ is defined as in
\eqref{eqn:def_Rn}, with the population quantities replaced by their
$P_n$ versions; its dependence on the subset $J$ is again suppressed. In particular,
\begin{align}
& \norm{\hat B_{X_J\mid Z,n}-B_{X_J\mid Z,n}}_{\mathrm{op}}
=O_{P_n}\left(
n^{-\alpha(\beta\wedge1)
/(2\alpha(\beta\wedge1)+\alpha+1)}
\right)
=o_{P_n}(n^{-1/4}),
\label{eqn:local_B_rate}\\
&\big\| \hat B_{\ddot X_J\mid Z,n} - B_{\ddot X_J\mid Z,n} - R_n \big\|_{\mathrm{op}}
= o_{P_n}(n^{-1/4}),
\label{eqn:local_corrected_B_rate}\\
&\norm{R_n\Sigma_{ZZ,n}^{\theta}}_{\mathrm{op}}
=O_{P_n}\left(
n^{-1/2}\epsilon_n^{\theta\wedge1-1}
+\epsilon_n^{\theta\wedge1}
\right).
\label{eqn:local_R_rate}
\end{align}
{
The uniform eigenvalue bounds imply
\begin{align}
\sup_{n\geq1} \sum_{k=1}^{\infty}\gamma_{k,n}^{1-2\theta}
<\infty & \quad\text{if }
0 < \theta < \frac{\alpha-1}{2\alpha},
\label{eqn:local_inverse_moment}\\
\sup_{n\geq1} \sum_{k=1}^{\infty}\gamma_{k,n}^{2(1-\theta)}
<\infty & \quad\text{if }
\theta<1-\frac{1}{2\alpha}.
\label{eqn:local_smoothing_moment}
\end{align}
We now verify explicitly that the remainder terms in the asymptotic expansion are uniformly negligible. Fix $j\in\{1,\ldots,d-1\}$. In the remainder calculations below, $R_n$ denotes the operator in \eqref{eqn:def_Rn} with $J=\{j\}$ and population quantities evaluated under $P_n$. First, fix $0<\theta<\frac{\alpha-1}{2\alpha}$. By \eqref{eqn:local_inverse_moment},
\begin{align*}
\sup_{n\geq1} \E_{P_n}\left\| (\Sigma_{ZZ,n}^{\dagger})^\theta \tilde\kappa_Z(\cdot,Z) \right\|_{\mathcal H_Z}^2
=
\sup_{n\geq1} \sum_{k=1}^{\infty}\gamma_{k,n}^{1-2\theta}
<\infty.
\end{align*}
The conditional mean-zero property and the uniform boundedness of the
residuals therefore give
\begin{align}
\sqrt n \norm{ \E_n\left[ (\Sigma_{ZZ,n}^{\dagger})^\theta \tilde\kappa_Z(\cdot,Z) \otimes \xi_{X_{j+}\mid Z,n}(\cdot,X_{j+};Z) \right] }_{\mathrm{HS}}
= O_{P_n}(1).
\label{eqn:local_smoothed_cross_rate}
\end{align}
Moreover, \eqref{eqn:local_R_rate} gives
\begin{align*}
\norm{ R_n\Sigma_{ZZ,n}^{\theta} }_{\mathrm{op}}
= O_{P_n}\left( n^{-1/2}\epsilon_n^{\theta-1} +\epsilon_n^\theta \right)
= o_{P_n}(1).
\end{align*}
Consequently,
\begin{align}
& \sqrt n \norm{ \E_n\left[ R_n\tilde\kappa_Z(\cdot,Z) \otimes \xi_{X_{j+}\mid Z,n}(\cdot,X_{j+};Z) \right] }_{\mathrm{HS}} \notag\\
\leq & \norm{ R_n\Sigma_{ZZ,n}^{\theta} }_{\mathrm{op}}
\sqrt n \norm{ \E_n\left[ (\Sigma_{ZZ,n}^{\dagger})^\theta \tilde\kappa_Z(\cdot,Z) \otimes \xi_{X_{j+}\mid Z,n}(\cdot,X_{j+};Z) \right] }_{\mathrm{HS}} \notag\\
= & o_{P_n}(1).
\label{eqn:local_R_cross_remainder}
\end{align}

For the ``quadratic'' remainder, choose a possibly different value
of $\theta$ such that
\begin{align*}
\frac{\alpha+1}{2\alpha} < \theta < 1-\frac{1}{2\alpha}.
\end{align*}
Such a choice is possible because $\alpha>2$. By
\eqref{eqn:local_smoothing_moment},
\begin{align*}
\sup_{n\geq1} \norm{ \Sigma_{ZZ,n}^{1-\theta} }_{\mathrm{HS}}^2 
= \sup_{n\geq1} \sum_{k=1}^{\infty}\gamma_{k,n}^{2(1-\theta)}
<\infty.
\end{align*}
The argument used in Proposition~\ref{prop:remainder_op1}, together
with \eqref{eqn:local_B_rate} and \eqref{eqn:local_R_rate}, then gives
\begin{align}
\sqrt n \norm{ R_n\Sigma_{ZZ,n} 
\left( \hat B_{X_{j+}\mid Z,n} - B_{X_{j+}\mid Z,n}\right)^*
}_{\mathrm{HS}}
= o_{P_n}(1).
\label{eqn:local_mixed_remainder}
\end{align}
We now list the remaining terms in the algebraic expansion used in the
proof of Theorem~\ref{lem:asymptotic_normality}. The sample-centering
terms satisfy
\begin{align}
\sqrt n
\norm{ \hat\Sigma_{\ddot X_{j}X_{j+}\mid Z,n} - \tilde\Sigma^{\,2}_{\ddot X_{j}X_{j+}\mid Z,n} }_{\mathrm{HS}}
= o_{P_n}(1).
\label{eqn:local_centering_remainder}
\end{align}
The linear terms involving the corrected regression-operator error
satisfy
\begin{align}
\sqrt n
\norm{ \left(\hat B_{\ddot X_{j}\mid Z,n} - B_{\ddot X_{j}\mid Z,n} - R_n \right)
\E_n\left[ \tilde\kappa_Z(\cdot,Z) \otimes \xi_{X_{j+}\mid Z,n}(\cdot,X_{j+};Z) \right]
}_{\mathrm{HS}}
= o_{P_n}(1),
\label{eqn:local_corrected_linear_remainder}
\end{align}
because the operator error is \(o_{P_n}(n^{-1/4})\), while all the mean-zero empirical average is \(O_{P_n}(n^{-1/2})\) by Hilber space CLT. Similarly,
\begin{align}
&\sqrt n
\norm{ \E_n\left[ \xi_{\ddot X_{j}\mid Z,n}(\cdot,\ddot X_{j};Z) \otimes \tilde\kappa_Z(\cdot,Z) \right]
\left( \hat B_{X_{j+}\mid Z,n} - B_{X_{j+}\mid Z,n} \right)^* }_{\mathrm{HS}}
= o_{P_n}(1).
\label{eqn:local_linear_remainder}
\end{align}
Finally, the quadratic operator-estimation term satisfies
\begin{align}
\sqrt n
\norm{ \left( \hat B_{\ddot X_{j}\mid Z,n} - B_{\ddot X_{j}\mid Z,n} \right)
\E_n\left[ \tilde\kappa_Z(\cdot,Z) \otimes \tilde\kappa_Z(\cdot,Z) \right]
\left( \hat B_{X_{j+}\mid Z,n} - B_{X_{j+}\mid Z,n} \right)^* }_{\mathrm{HS}}
= o_{P_n}(1).
\label{eqn:local_quadratic_remainder}
\end{align}
Indeed, after writing
\begin{align*}
\hat B_{\ddot X_{j}\mid Z,n} - B_{\ddot X_{j}\mid Z,n}
= \left( \hat B_{\ddot X_{j}\mid Z,n} - B_{\ddot X_{j}\mid Z,n} - R_n \right) + R_n
\end{align*}
and
\begin{align*}
\E_n[ \tilde\kappa_Z\otimes\tilde\kappa_Z]
= \Sigma_{ZZ,n} + \left\{ \E_n[ \tilde\kappa_Z\otimes\tilde\kappa_Z] - \Sigma_{ZZ,n} \right\},
\end{align*}
the terms not containing \(R_n\Sigma_{ZZ,n}\) are
\(o_{P_n}(1)\) by
\eqref{eqn:local_B_rate}--\eqref{eqn:local_corrected_B_rate} and
\eqref{eqn:local_covariance_rates}, whereas the remaining term is
\(o_{P_n}(1)\) by \eqref{eqn:local_mixed_remainder}.

Combining \eqref{eqn:local_R_cross_remainder} and \eqref{eqn:local_centering_remainder}--\eqref{eqn:local_quadratic_remainder}, we obtain, for every $j=1,\ldots,d-1$,
\begin{align*}
\sqrt n\left( \hat\Sigma_{\ddot X_{j}X_{j+}\mid Z,n} - \Sigma_{\ddot X_{j}X_{j+}\mid Z,n} \right)
=
\frac1{\sqrt n} \sum_{i=1}^n \psi_{j,n,i} + o_{P_n}(1).
\end{align*}
Here $\psi_{j,n,i}$ denotes the influence function $\psi_{j,n}$ evaluated at the $i$th observation, and $\psi_{n,i}=(\psi_{1,n,i},\ldots,\psi_{d-1,n,i})^\top$.
Combining the $d-1$ components gives the vector-form influence function
\begin{align}
\sqrt n(\hat S_n-S_n)
=
\frac1{\sqrt n}\sum_{i=1}^n\psi_{n,i} + o_{P_n}(1).
\label{eqn:local_asymptotic_linearity}
\end{align}}
The uniform kernel and source bounds imply for any $n \geq 1$
\begin{align*}
& P_n\left(\norm{\psi_n}_{\mathbb H_S}\leq C\right) \\
\geq & P_n\left(\sum_{j=1}^{d-1} \norm{\psi_{j,n}}_{\mathrm{HS}}^2 \leq C^2 \right)  \\
\geq & P_n \left(  \sum_{j=1}^{d-1}
\left(\norm{ \xi_{\ddot X_j\mid Z,n}}_{\mathcal H_{\ddot X_j}}
\norm{\xi_{X_{j+}\mid Z,n}
}_{\mathcal H_{X_{j+}}}
+ 
\norm{ \Sigma_{\ddot X_jX_{j+}\mid Z,n} }_{\mathrm{HS}}\right)^2 \leq C^2 \right) \\
=& 1
\end{align*}
for some constant $C<\infty$ independent of $n$. Hence almost surely
\begin{align*}
\max_{1\leq i\leq n} \norm{\frac{\psi_{n,i}}{\sqrt n}}_{\mathbb H_S}
\leq \frac{C}{\sqrt n}
\longrightarrow 0.
\end{align*}
Then the Lindeberg condition holds. Together with
\eqref{eqn:local_covariance_convergence}, the Hilbert-space CLT for triangular arrays gives
\begin{align*}
\frac{1}{\sqrt n}\sum_{i=1}^n\psi_{n,i} \xrightarrow{\mathcal D} G,
\quad \text{where } \ G\sim N(0,\Gamma_{S\mid Z,0}).
\end{align*}
Combining this with \eqref{eqn:local_asymptotic_linearity} proves \eqref{eqn:local_clt}.

Under \eqref{eqn:local_alternative},
\begin{align*}
\sqrt n\hat S_n
= \sqrt n(\hat S_n - S_n) + S_h
\xrightarrow{\mathcal D}
G+S_h.
\end{align*}
The continuous mapping theorem proves \eqref{eqn:local_norm_limit}.
Finally, the Karhunen--Lo\`eve expansion gives
\begin{align*}
G=\sum_{k=1}^{\infty}\sqrt{\lambda_k}Z_kv_k.
\end{align*}
Since $S_h\in\ker(\Gamma_{S\mid Z,0})^\perp$, Parseval's identity therefore yields
\begin{align*}
\norm{G+S_h}_{\mathbb H_S}^2
=\sum_{k=1}^{\infty}
\left(\sqrt{\lambda_k}Z_k+\sigma_k\right)^2,
\end{align*}
which proves \eqref{eqn:local_weighted_chisq}.
\end{proof}

\subsection{Proofs for Section \ref{sec:unc}}
\begin{proof}[Proof of Theorem~\ref{thm:unconditional}]
    Assertion~$1$ follows immediately from Theorem~\ref{thm:cross_cor_op}. Now we prove assertion~$2$. We need to show for any $j \in \{1,\ldots,d-1\}$, 
    \begin{align}\label{eqn:uncond_Sigma}
        \sqrt n \left(\hat \Sigma_{X_{j} X_{j+}} - \Sigma_{X_{j} X_{j+}} \right) \xrightarrow{\mathcal D} N\left(0, \Gamma_{X_{j} X_{j+}}\right)
    \end{align}
    where $\Gamma_{X_{j} X_{j+}} = \E\left[\psi_j^{\mathrm{uc}} \otimes \psi_j^{\mathrm{uc}}\right]$. Recall that
    \begin{align*}
        \hat \Sigma_{X_{j} X_{j+}} 
        = \E_n\left[ \tilde \kappa_{X_j} \otimes \tilde \kappa_{X_{j+}} \right]
        - \E_n \left[ \tilde \kappa_{X_j}\right] \otimes \E_n\left[\tilde \kappa_{X_{j+}} \right]
    \end{align*}
    Thus
    \begin{align}
        & \sqrt n \left(\hat \Sigma_{X_{j} X_{j+}} - \Sigma_{X_{j} X_{j+}} \right) \notag \\
        = & \sqrt n  \left( \E_n\left[ \tilde \kappa_{X_j} \otimes \tilde \kappa_{X_{j+}} \right] - \E\left[ \tilde \kappa_{X_j} \otimes \tilde \kappa_{X_{j+}} \right] \right)
        - \sqrt n \E_n \left[ \tilde \kappa_{X_j}\right] \otimes \E_n\left[\tilde \kappa_{X_{j+}} \right]. \label{eqn:uncond_decompose}
    \end{align}
    Since all the kernels are bounded, by Hilbert space CLT,
    \begin{align}\label{eqn:uncond_CLT}
        \sqrt n  \left( \E_n\left[ \tilde \kappa_{X_j} \otimes \tilde \kappa_{X_{j+}} \right] - \E\left[ \tilde \kappa_{X_j} \otimes \tilde \kappa_{X_{j+}} \right] \right) \xrightarrow{\mathcal D} N\left(0, \Gamma_{X_{j} X_{j+}}\right).
    \end{align}
    The centered kernels have mean zero, then again by Hilbert space CLT (or Chebyshev inequality)
    \begin{align}\label{eqn:uncond_op1}
        \sqrt n \E_n \left[ \tilde \kappa_{X_j}\right] \otimes \E_n\left[\tilde \kappa_{X_{j+}} \right] = n^{1/2} O_P(n^{-1/2})O_P(n^{-1/2}) = o_P(1).
    \end{align}
    Combining \eqref{eqn:uncond_decompose}--\eqref{eqn:uncond_op1} gives the desired results \eqref{eqn:uncond_Sigma}. 
    For assertion~$2$, we can use the fact obtained above
    \begin{align*}
        \sqrt n\left(
        \hat\Sigma_{X_jX_{j+}}-\Sigma_{X_jX_{j+}}
        \right)
        =
        \frac1{\sqrt n}\sum_{i=1}^n\psi_{j,i}^{\mathrm{uc}}
        +o_P(1).
    \end{align*}
    Let
    $\psi_i^{\mathrm{uc}}
    =(\psi_{1,i}^{\mathrm{uc}},\ldots,
    \psi_{d-1,i}^{\mathrm{uc}})^\top$.
    Combining $d-1$ components, we get
    \begin{align*}
        \sqrt n(\hat S^{\mathrm{uc}}-S^{\mathrm{uc}})
        =
        \frac1{\sqrt n}\sum_{i=1}^n\psi_i^{\mathrm{uc}}
        +o_P(1).
    \end{align*}
The random elements $\psi_i^{\mathrm{uc}}$ are i.i.d., with $\E[\psi_i^{\mathrm{uc}}]=0$. Boundedness of kernels implies
$\E\|\psi_i^{\mathrm{uc}}\|_{\mathbb H_S}^2<\infty$. Therefore, the Hilbert space CLT and Slutsky's theorem give
\begin{align*}
    \sqrt n(\hat S^{\mathrm{uc}}-S^{\mathrm{uc}})
    \xrightarrow{\mathcal D}
    N(0,\Gamma_S^{\mathrm{uc}}),
    \quad \text{ where }
    \Gamma_S^{\mathrm{uc}}
    =
    \E[\psi_i^{\mathrm{uc}}\otimes\psi_i^{\mathrm{uc}}],
\end{align*}
which proves assertion~2. Assertions~$3$ just follow as corollaries.
\end{proof}


\begin{proof}[Proof of Theorem~\ref{thm:unconditional_local_alternative}]
For each relevant subset $J$, boundedness of kernels and independence imply
\begin{align*}
\E_{P_n}\left\|
\hat\mu_{X_J,n}-\mu_{X_J,n}
\right\|_{\mathcal H_{X_J}}^2
= &
\frac1n
\E_{P_n}\left\|
\tilde\kappa_{X_J,n}(\cdot,X_J)
\right\|_{\mathcal H_{X_J}}^2\\
= &
\frac1n\left(
\E_{P_n}[\kappa_{X_J}(X_J,X_J)]
-\|\mu_{X_J,n}\|_{\mathcal H_{X_J}}^2
\right)\\
\leq &
\frac1n\sup_{x_J}\kappa_{X_J}(x_J,x_J).
\end{align*}
Thus
\begin{align*}
\left\|
\hat\mu_{X_J,n}-\mu_{X_J,n}
\right\|_{\mathcal H_{X_J}}
=
O_{P_n}(n^{-1/2}).
\end{align*}
Fix $j\in\{1,\ldots,d-1\}$.
By simple calculation,
\begin{align*}
\hat\Sigma_{X_jX_{j+},n}
=
\E_n\left[
\tilde\kappa_{X_j,n}(\cdot,X_j)
\otimes
\tilde\kappa_{X_{j+},n}(\cdot,X_{j+})
\right]
-
(\hat\mu_{X_j,n}-\mu_{X_j,n})
\otimes
(\hat\mu_{X_{j+},n}-\mu_{X_{j+},n}).
\end{align*}
Let $\psi_{j,n,i}^{\mathrm{uc}}$ denote
$\psi_{j,n}^{\mathrm{uc}}$ evaluated at observation $i$. Then
\begin{align*}
\sqrt n\left(
\hat\Sigma_{X_jX_{j+},n}-\Sigma_{X_jX_{j+},n}
\right)
=
\frac1{\sqrt n}\sum_{i=1}^n\psi_{j,n,i}^{\mathrm{uc}}
-
\sqrt n\,
(\hat\mu_{X_j,n}-\mu_{X_j,n})
\otimes
(\hat\mu_{X_{j+},n}-\mu_{X_{j+},n}).
\end{align*}
It follows that
\begin{align*}
&\sqrt n\left\|
(\hat\mu_{X_j,n}-\mu_{X_j,n})
\otimes
(\hat\mu_{X_{j+},n}-\mu_{X_{j+},n})
\right\|_{\mathrm{HS}}\\
= &
\sqrt n\,
\|\hat\mu_{X_j,n}-\mu_{X_j,n}\|_{\mathcal H_{X_j}}
\|\hat\mu_{X_{j+},n}-\mu_{X_{j+},n}\|_{\mathcal H_{X_{j+}}}\\
= &
\ O_{P_n}(n^{-1/2})=o_{P_n}(1).
\end{align*}
Combining the $d-1$ components gives the vector-form influence function
\begin{align}
\sqrt n\left(
\hat S_n^{\mathrm{uc}}-S_n^{\mathrm{uc}}
\right)
=
\frac1{\sqrt n}\sum_{i=1}^n\psi_{n,i}^{\mathrm{uc}}
+o_{P_n}(1)
\label{eqn:uc_local_asymptotic_linearity}
\end{align}
where
$\psi_{n,i}^{\mathrm{uc}}
=(\psi_{1,n,i}^{\mathrm{uc}},\ldots,
\psi_{d-1,n,i}^{\mathrm{uc}})^\top$.

Recall that the residual term has mean-zero property, i.e.,
$\E_{P_n}[\psi_n^{\mathrm{uc}}]=0$.
The boundedness of kernels then implies for every $n\geq1$,
\begin{align*}
&P_n\left(
\norm{\psi_n^{\mathrm{uc}}}_{\mathbb H_S}\leq C
\right)\\
={}&P_n\left(
\sum_{j=1}^{d-1}
\norm{\psi_{j,n}^{\mathrm{uc}}}_{\mathrm{HS}}^2
\leq C^2
\right)\\
\geq{}&
P_n\left(
\sum_{j=1}^{d-1}
\left(
\norm{
\tilde\kappa_{X_j,n}(\cdot,X_j)
}_{\mathcal H_{X_j}}
\norm{
\tilde\kappa_{X_{j+},n}(\cdot,X_{j+})
}_{\mathcal H_{X_{j+}}}
+
\norm{\Sigma_{X_jX_{j+},n}}_{\mathrm{HS}}
\right)^2
\leq C^2
\right)\\
={}&1
\end{align*}
for some constant $C<\infty$ independent of $n$.
Hence almost surely,
\begin{align*}
\max_{1\leq i\leq n}
\norm{
\frac{\psi_{n,i}^{\mathrm{uc}}}{\sqrt n}
}_{\mathbb H_S}
\leq\frac{C}{\sqrt n}
\longrightarrow0.
\end{align*}
Thus the Lindeberg condition holds. Together with \eqref{eqn:uc_local_covariance_convergence}, the Hilbert-space CLT for triangular arrays gives
\begin{align*}
\frac1{\sqrt n}\sum_{i=1}^n\psi_{n,i}^{\mathrm{uc}}
\xrightarrow{\mathcal D}G,
\quad\text{where }\;
G\sim N\left(0,\Gamma_{S,0}^{\mathrm{uc}}\right).
\end{align*}
Combining this with
\eqref{eqn:uc_local_asymptotic_linearity}
proves \eqref{eqn:uc_local_clt}. Under \eqref{eqn:uc_local_alternative},
\begin{align*}
\sqrt n\,\hat S_n^{\mathrm{uc}}
=
\sqrt n\left(
\hat S_n^{\mathrm{uc}}-S_n^{\mathrm{uc}}
\right)
+S_h^{\mathrm{uc}}
\xrightarrow{\mathcal D}
G+S_h^{\mathrm{uc}}.
\end{align*}
The continuous mapping theorem proves
\eqref{eqn:uc_local_norm_limit}.
Finally, the Karhunen--Lo\`eve expansion gives
\begin{align*}
G=\sum_{k=1}^\infty \sqrt{\lambda_k}Z_kv_k.
\end{align*}
Since
$S_h^{\mathrm{uc}}\in
\ker(\Gamma_{S,0}^{\mathrm{uc}})^\perp$,
Parseval's identity therefore yields
\begin{align*}
\norm{G+S_h^{\mathrm{uc}}}_{\mathbb H_S}^2
=
\sum_{k=1}^\infty
\left(\sqrt{\lambda_k}Z_k+\sigma_k\right)^2,
\end{align*}
which proves \eqref{eqn:uc_local_weighted_chisq}.
\end{proof}

\subsection{Proofs for Section \ref{sec:num}}
\begin{proof}[Proof of Lemma~\ref{lem:gamma_consistent}]
Let $\hat\psi_i$ and $\psi_i$ be the estimated and population influence functions. We first show that $\E_n \norm{R_n \hat \kappa_Z}_{\mathcal H_{\ddot X_j}}^2 = o_P(1)$.
From the boundedness of kernels, it's easy to derive that $\norm{R_n}$ is bounded uniformly over $n$. Also by Lemma~\ref{lem:smoothed_projection_error} with $\theta=1$,
\begin{align*}
    \norm{R_n\Sigma_{ZZ}}_{\mathrm{op}}
    =
    O_P(n^{-1/2}+\epsilon_n)
    =o_P(1).
\end{align*}
Let $\{\lambda_\ell\}$ and $\{e_\ell\}$ be positive eigenvalues and corresponding orthonormal eigenvectors of $\Sigma_{ZZ}$. Since $\Sigma_{ZZ}$ is a trace-class operator, $\sum_\ell\lambda_\ell<\infty$. Then, for some constant $C>0$, we have
\begin{align*}
    \norm{R_n\Sigma_{ZZ}^{\frac 12}}_{\mathrm{HS}}^2
    = &
    \sum_\ell
    \lambda_\ell
    \norm{R_ne_\ell}_{\mathcal H_{\ddot X_J}}^2\\
    = &
    \sum_{\ell=1}^{m}
    \lambda_\ell^{-1}
    \norm{R_n\Sigma_{ZZ}e_\ell}_{\mathcal H_{\ddot X_J}}^2
    +
    \sum_{\ell>m}
    \lambda_\ell
    \norm{R_ne_\ell}_{\mathcal H_{\ddot X_J}}^2\\
    \le &
    \norm{R_n\Sigma_{ZZ}}_{\mathrm{op}}^2
    \sum_{\ell=1}^{m}\lambda_\ell^{-1}
    +
    C\sum_{\ell>m}\lambda_\ell.
\end{align*}
Note that the above argument holds for all positive integers $m$. Thus, given any $\eta>0$, we fix a sufficiently large $m$ such that $C\sum_{\ell>m}\lambda_\ell < \eta/2$, which implies
\begin{align*}
\Pr\left( \norm{R_n\Sigma_{ZZ}^{1/2}}_{\mathrm{HS}}^2>\eta \right)
\le \Pr\left( \norm{R_n\Sigma_{ZZ}}_{\mathrm{op}}^2 \sum_{\ell=1}^{m}\lambda_\ell^{-1} > \frac{\eta}{2} \right)
\to 0,
\end{align*}
where the limit holds because $\norm{R_n\Sigma_{ZZ}}_{\mathrm{op}} = o_P(1)$. Thus,
$\norm{R_n\Sigma_{ZZ}^{1/2}}_{\mathrm{HS}}^2=o_P(1)$. By the Banach-space strong law, we get
\begin{align*}
&\|\hat\Sigma_{ZZ}-\Sigma_{ZZ}\|_{\mathrm{tr}} \\
\le &
\left\|
\frac1n\sum_{i=1}^{n}
(k_Z(\cdot,Z_i)-\mu_Z)\otimes(k_Z(\cdot,Z_i)-\mu_Z)
-\Sigma_{ZZ}
\right\|_{\mathrm{tr}}
+
\left\|
\frac1n\sum_{i=1}^{n}k_Z(\cdot,Z_i)-\mu_Z
\right\|_{\mathcal H_Z}^{2} \\
= & o_P(1).
\end{align*}
Then
\begin{align}
    \E_n\norm{R_n \hat \kappa_Z}_{\mathcal H_{\ddot X_j}}^2 
    = & \trace\left(R_n \hat \Sigma_{ZZ} R_n^* \right) \notag \\
    = & \trace\left(R_n (\hat \Sigma_{ZZ} - \Sigma_{ZZ}) R_n^* \right) + \trace\left(R_n \Sigma_{ZZ} R_n^* \right) \notag \\
    \leq & \norm{R_n (\hat \Sigma_{ZZ} - \Sigma_{ZZ}) R_n^*}_{\mathrm{tr}} + \norm{R_n\Sigma_{ZZ}^{\frac 12}}_{\mathrm{HS}}^2 \notag \\
    \leq & \norm{R_n}_{\mathrm{op}}^2 \norm{\hat \Sigma_{ZZ} - \Sigma_{ZZ}}_{\mathrm{tr}} + \norm{R_n\Sigma_{ZZ}^{\frac 12}}_{\mathrm{HS}}^2 \notag \\
    = & o_P(1). \label{eqn:R_n_hat_kappa_Z}
\end{align}
The definition of residuals gives the decomposition
\begin{align*}
    \hat\xi_{\ddot X_j\mid Z}(\cdot,\ddot X_{j,i};Z_i)
    -\xi_{\ddot X_j\mid Z}(\cdot,\ddot X_{j,i};Z_i)
    =
    -(\hat\mu_{\ddot X_{j,i}}-\mu_{\ddot X_{j,i}})
    -(\hat B_{\ddot X_j\mid Z}-B_{\ddot X_j\mid Z})
    \hat\kappa_Z(\cdot,Z_i)
    +B_{\ddot X_j\mid Z}(\hat\mu_Z-\mu_Z).
\end{align*}
Applying Cauchy--Schwartz inequality, we have
\begin{align}
    &
    \E_n\norm{
    \hat\xi_{\ddot X_j\mid Z}
    -\xi_{\ddot X_j\mid Z}
    }_{\mathcal H_{\ddot X_j}}^2 \notag \\
    \le & 4\norm{\hat\mu_{\ddot X_j}-\mu_{\ddot X_j}}_{\mathcal H_{\ddot X_j}}^2
    + 4\norm{\hat B_{\ddot X_j\mid Z} - B_{\ddot X_j\mid Z}-R_n}_{\mathrm{op}}^2 \E_n\norm{\hat\kappa_Z}_{\mathcal H_Z}^2
    + 4\E_n\norm{R_n\hat\kappa_Z}_{\mathcal H_{\ddot X_j}}^2
    + 4\norm{B_{\ddot X_j\mid Z}}_{\mathrm{op}}^2\norm{\hat\mu_Z-\mu_Z}_{\mathcal H_Z}^2 \notag \\
    = & o_P(1). \label{eqn:E_n_residual_j_2}
\end{align}
Similarly, the residual decomposition for $X_{j+}\mid Z$ gives
\begin{align}
    &
    \E_n\norm{\hat\xi_{X_{j+}\mid Z} - \xi_{X_{j+}\mid Z}}_{\mathcal H_{X_{j+}}}^2 \notag \\
    \le & 3\norm{\hat\mu_{X_{j+}}-\mu_{X_{j+}}
    }_{\mathcal H_{X_{j+}}}^2
    + 3\norm{\hat B_{X_{j+}\mid Z} - B_{X_{j+}\mid Z}}_{\mathrm{op}}^2 \E_n\norm{\hat\kappa_Z}_{\mathcal H_Z}^2
    + 3\norm{B_{X_{j+}\mid Z}}_{\mathrm{op}}^2 \norm{\hat\mu_Z-\mu_Z}_{\mathcal H_Z}^2 \notag \\
    = & o_P(1). \label{eqn:E_n_residual_j+_2}
\end{align}
We also have universal boundedness of the residual terms, both the estimated and the population ones, which is an immediate result from the boundedness of kernels and regression operators.
\begin{align}
    & \max_{1\le i\le n} \norm{\xi_{\ddot X_j\mid Z}(\cdot,\ddot X_{j,i};Z_i)}_{\mathcal H_{\ddot X_j}} =O_P(1), \quad
    \max_{1\le i\le n} \norm{\hat\xi_{X_{j+}\mid Z}(\cdot,X_{j+,i};Z_i)}_{\mathcal H_{X_{j+}}} =O_P(1). \label{eqn:max_residual}
\end{align}
Combining \eqref{eqn:R_n_hat_kappa_Z}--\eqref{eqn:max_residual} we get
\begin{align*}
    &
    \E_n\norm{
    \hat\xi_{\ddot X_j\mid Z}\otimes\hat\xi_{X_{j+}\mid Z}
    -
    \xi_{\ddot X_j\mid Z}\otimes\xi_{X_{j+}\mid Z}
    }_{\mathrm{HS}}^2\\
    \le &
    2\max_{1\le i\le n}
    \norm{
    \hat\xi_{X_{j+}\mid Z}(\cdot,X_{j+,i};Z_i)
    }_{\mathcal H_{X_{j+}}}^2
    \E_n\norm{
    \hat\xi_{\ddot X_j\mid Z}
    -\xi_{\ddot X_j\mid Z}
    }_{\mathcal H_{\ddot X_j}}^2 \\
    & +
    2\max_{1\le i\le n}
    \norm{
    \xi_{\ddot X_j\mid Z}(\cdot,\ddot X_{j,i};Z_i)
    }_{\mathcal H_{\ddot X_j}}^2
    \E_n\norm{
    \hat\xi_{X_{j+}\mid Z}
    -\xi_{X_{j+}\mid Z}
    }_{\mathcal H_{X_{j+}}}^2\\
    =& o_P(1).
\end{align*}
Combining the results above yields
\begin{align*}
    \E_n\norm{\hat\psi_j-\psi_j}_{\mathrm{HS}}^2
    \le &
    2\E_n\norm{\hat\xi_{\ddot X_j\mid Z}\otimes\hat\xi_{X_{j+}\mid Z} - \xi_{\ddot X_j\mid Z}\otimes\xi_{X_{j+}\mid Z}}_{\mathrm{HS}}^2
    +
    2\norm{\hat\Sigma_{\ddot X_jX_{j+}\mid Z} - \Sigma_{\ddot X_jX_{j+}\mid Z}}_{\mathrm{HS}}^2\\
    = &o_P(1).
\end{align*}
Thus
\begin{align*}
    \E_n\norm{\hat\psi-\psi}_{\mathbb H_S}^2
    =
    \sum_{j=1}^{d-1} \E_n\norm{\hat\psi_j-\psi_j}_{\mathrm{HS}}^2
    =o_P(1).
\end{align*}
We next establish consistency of the covariance estimator. By Cauchy--Schwarz inequality,
\begin{align*}
\norm{
\hat\Gamma_{S\mid Z}
-\E_n[\psi\otimes\psi]
}_{\mathrm{tr}}
= & \norm{\frac1n \sum_{i=1}^n \left(\hat \psi_i\otimes \hat \psi_i - \psi_i \otimes \psi_i \right)}_{\mathrm{tr}}\\
\leq & \frac1n \sum_{i=1}^n \norm{\left(\hat \psi_i\otimes \hat \psi_i - \psi_i \otimes \psi_i \right)}_{\mathrm{tr}} \\
\le & \frac1n \sum_{i=1}^n \left(2 \norm{\hat \psi_i - \psi_i}_{\mathbb H_S} \norm{\psi_i}_{\mathbb H_S} +  \norm{\hat \psi_i - \psi_i}_{\mathbb H_S}^2 \right) \\
\le &
2\left(
\E_n\norm{\hat\psi-\psi}_{\mathbb H_S}^{2}
\right)^{1/2}
\left(
\E_n\norm{\psi}_{\mathbb H_S}^{2}
\right)^{1/2}
+
\E_n\norm{\hat\psi-\psi}_{\mathbb H_S}^{2}\\
= & o_P(1).
\end{align*}
Combining with the fact that $\Gamma_{S\mid Z}$ is a trace-class operator, the law of large numbers in the separable Banach space of trace-class operators gives
\begin{align*}
\norm{
\E_n[\psi\otimes\psi]-\Gamma_{S\mid Z}
}_{\mathrm{tr}}
=o_P(1).
\end{align*}
Thus
\begin{align*}
\norm{\hat\Gamma_{S\mid Z}-\Gamma_{S\mid Z}}_{\mathrm{tr}}
\leq \norm{\hat\Gamma_{S\mid Z}-\E_n[\psi\otimes\psi]}_{\mathrm{tr}} + \norm{\E_n[\psi\otimes\psi]-\Gamma_{S\mid Z}}_{\mathrm{tr}} = o_P(1),
\end{align*}
which completes the proof.
\end{proof}

\begin{proof}[Proof of Proposition~\ref{prop:gamma_plugin}]
We first derive the coordinate mapping for the numerical influence function.
\begin{align*}
   \coord{\mathscr{B}_{X_{J}}}{\hat{B}_{X_{J}\mid Z}}{\mathscr{B}_{Z}}
    = & \hat{K}_{Z} \left( \hat{K}_{Z} + n \epsilon_{n} I \right)^{-1}
     \qquad
    \left[\hat \kappa_{X_ J} (\cdot, X_{J,i})\right]_{\mathscr{B}_{X_J}} = e_{i}
    \qquad
    \left[\hat \kappa_{Z} (\cdot, Z_ i)\right]_{\mathscr{B}_{Z}} = e_{i} \\
    \left[ \hat \xi_{X_{J}\mid Z} (\cdot, X_{J,i};Z_i) \right]_{\mathscr{B}_{X_J}}
    = & \left[\hat \kappa_{X_ J} (\cdot, X_{J,i}) - \hat{B}_{X_{J}\mid Z} \hat \kappa_{Z} (\cdot, Z_ i) \right]_{\mathscr{B}_{X_J}} = \left( I - \hat{K}_{Z} \left( \hat{K}_{Z} + n \epsilon_{n} I \right)^{-1} \right)  e_{i} = R_{Z} e_{i}
\end{align*}
For the Gamma approximation under the null hypothesis, we use $\Sigma_{\ddot X_j X_{j+}\mid Z}=0$ when constructing the empirical influence function. Accordingly, throughout this proof, $\hat\psi_{j,i}$ denotes the tensor product of the estimated residual feature maps, without subtracting $\hat\Sigma_{\ddot X_j X_{j+}\mid Z}$. Its coordinate representation is
\begin{align*}
   \coord{\mathscr{B}_{\ddot X_{j}}}{ \hat \psi (\ddot{X}_{j,i}, X_{j+,i}; Z_i) }{\mathscr{B}_{X_{j+}}}
    = &\coord{\mathscr{B}_{\ddot X_j}}{ \hat\xi_{\ddot X_{j}\mid Z} (\cdot, \ddot X_{j,i};Z_i) \otimes \hat \xi_{X_{j+}\mid Z} (\cdot, X_{j+,i};Z_i) }{\mathscr{B}_{X_{j+}}} \\
    = & \left[ \hat\xi_{\ddot X_{j}\mid Z} (\cdot, \ddot X_{j,i};Z_i) \right]_{\mathscr{B}_{\ddot X_j}}
    \left[ \hat\xi_{X_{j+}\mid Z} (\cdot, X_{j+,i};Z_i) \right]_{\mathscr{B}_{X_{j+}}}^{\top} \hat{K}_{X_{j+}} \\
    = & \big(R_{Z} e_{i}\big) \big(R_{Z} e_{i}\big)^\top \hat{K}_{X_{j+}}
\end{align*}

Consider the Gram matrix of the tensor product space $\mathcal{H}_{\ddot X_{j}} \otimes \mathcal{H}_{X_{j+}}$, the spanning system is $\mathscr{B}_{\ddot X_{j} X_{j+}} = \left\{\hat \kappa_{\ddot X_{j} X_{j+}}(\cdot, \ddot X_{j, i}, \star, X_{j+,\ell}) = \hat \kappa_{\ddot X_{j}}(\cdot, \ddot X_{j, i}) \ \hat \kappa_{X_{j+}}(\star, X_{j+,\ell}), i, \ell = 1, \ldots, n \right\} = \mathscr{B}_{\ddot X_{j}} \otimes \mathscr{B}_{X_{j+}}$. If we reconsider the influence function under the tensor product space, we have
\begin{align*}
    \left[ \hat \psi_{j,i} \right]_{\mathscr{B}_{\ddot X_{j} X_{j+}}} = \left[ \hat \psi (\ddot{X}_{j,i}, X_{j+,i}; Z_i) \right]_{\mathscr{B}_{\ddot X_{j} X_{j+}}} =
     R_{Z} e_{i} \otimes R_{Z} e_i 
\end{align*}
Thus
\begin{align*}
    \coord{\mathscr{B}_{\ddot X_{j} X_{j+}}}{\hat \psi_{j,i} \otimes \hat \psi_{k,i}}{\mathscr{B}_{\ddot X_{k} X_{k+}}} 
    = & \left[\hat \psi_{j,i}\right]_{\mathscr{B}_{\ddot X_{j} X_{j+}}} \left[\hat \psi_{k,i}\right]_{\mathscr{B}_{\ddot X_{k} X_{k+}}}^{\top} \hat K_{\ddot X_{k} X_{k+}} \\
    = & (R_{Z} e_{i} \otimes R_{Z} e_i) (R_{Z} e_{i} \otimes R_{Z} e_i)^\top \hat K_{\ddot X_{k} X_{k+}}
\end{align*}
We next compute the trace of $\hat\Gamma_{S\mid Z}$. By its coordinate representation,
\begin{align*}
    \sum_{j=1}^{d-1} \trace\left((\hat\Gamma_{S\mid Z})_{(j,j)}\right)
    = & \sum_{j=1}^{d-1} \trace\left(
    \coord{\mathscr{B}_{\ddot X_{j} X_{j+}}}{
    \frac{1}{n} \sum_{i=1}^{n} \hat\psi_{j,i} \otimes \hat\psi_{j,i}
    }{\mathscr{B}_{\ddot X_{j} X_{j+}}}
    \right) \\
    = & \sum_{j=1}^{d-1} \trace\left(
    \Big( \frac{1}{n} \sum_{i=1}^{n}
    (R_{Z} e_{i} \otimes R_{Z} e_i)
    (R_{Z} e_{i} \otimes R_{Z} e_i)^{\top} \Big)
    \Big( \hat K_{\ddot X_j X_{j+}} \Big)
    \right).
\end{align*}
Using the Kronecker product identities and the symmetry of $R_Z$, it becomes
\begin{align*}
    & \sum_{j=1}^{d-1} \trace \Big(
    \Big( \frac{1}{n} \sum_{i=1}^{n}
    ((R_{Z} e_{i}) (R_{Z} e_i)^\top)
    \otimes ((R_{Z} e_{i}) (R_{Z} e_i)^\top) \Big)
    \Big( \hat K_{\ddot X_j} \otimes \hat K_{X_{j+}} \Big)
    \Big) \\
    = & \sum_{j=1}^{d-1} \frac{1}{n} \sum_{i=1}^{n}
    \trace \left(
    \Big( (R_{Z} e_{i}) (R_{Z} e_i)^\top \hat K_{\ddot X_j} \Big)
    \otimes
    \Big( (R_{Z} e_{i}) (R_{Z} e_i)^\top \hat K_{X_{j+}} \Big)
    \right) \\
    = & \sum_{j=1}^{d-1} \frac{1}{n} \sum_{i=1}^{n}
    \trace \Big(
    (R_{Z} e_{i}) (R_{Z} e_i)^\top \hat K_{\ddot X_j}
    \Big)
    \trace \Big(
    (R_{Z} e_{i}) (R_{Z} e_i)^\top \hat K_{X_{j+}}
    \Big) \\
    = & \sum_{j=1}^{d-1} \frac{1}{n} \sum_{i=1}^{n}
    \left( e_i^\top R_Z \hat{K}_{\ddot X_j} R_Z e_i \right)
    \left( e_i^\top R_Z \hat{K}_{X_{j+}} R_Z e_i \right).
\end{align*}
The two quadratic forms are the $i$th diagonal entries of $\hat K_{\ddot X_j\mid Z}$ and $\hat K_{X_{j+}\mid Z}$, respectively.
Therefore,
\begin{align*}
    \sum_{j=1}^{d-1} \trace\left((\hat\Gamma_{S\mid Z})_{(j,j)}\right)
    = & \sum_{j=1}^{d-1} \frac{1}{n}
    \trace\left(
    \hat{K}_{\ddot X_j \mid Z} \circ \hat{K}_{X_{j+}\mid Z}
    \right).
\end{align*}
We next compute the trace of the second moment $\hat\Gamma_{S\mid Z}^{2}$.
For fixed $j,k=1,\ldots,d-1$, the coordinate representation gives
\begin{align*}
    & \trace\left(
    (\hat\Gamma_{S\mid Z})_{(j,k)}
    (\hat\Gamma_{S\mid Z})_{(k,j)}
    \right) \\
    = & \trace\Big( \coord{\mathscr{B}_{\ddot X_{j} X_{j+}}}{\frac{1}{n} \sum_{i=1}^{n} \hat\psi_{j,i} \otimes \hat\psi_{k,i}}{\mathscr{B}_{\ddot X_{k} X_{k+}}}
    \coord{\mathscr{B}_{\ddot X_{k} X_{k+}}}{\frac{1}{n} \sum_{i'=1}^{n} \hat\psi_{k,i'} \otimes \hat\psi_{j,i'}}{\mathscr{B}_{\ddot X_{j} X_{j+}}} \Big).
\end{align*}

Substituting the matrix representations and using the cyclic property
of the trace, we obtain
\begin{align*}
    & \trace\left(
    (\hat\Gamma_{S\mid Z})_{(j,k)}
    (\hat\Gamma_{S\mid Z})_{(k,j)}
    \right) \\
    = & \trace\Big(
    \big( \frac{1}{n} \sum_{i=1}^{n}
    (R_{Z} e_{i} \otimes R_{Z} e_i)
    (R_{Z} e_{i} \otimes R_{Z} e_i)^{\top} \big)
    \big( \hat K_{\ddot X_j X_{j+}} \big) \\
    & \qquad \times
    \big( \frac{1}{n} \sum_{i'=1}^{n}
    (R_{Z} e_{i'} \otimes R_{Z} e_{i'})
    (R_{Z} e_{i'} \otimes R_{Z} e_{i'})^{\top} \big)
    \big( \hat K_{\ddot X_k X_{k+}} \big)
    \Bigg).
\end{align*}

Expanding the two sums and applying the Kronecker product identities, it becomes
\begin{align*}
    & \frac{1}{n^2} \sum_{i=1}^{n} \sum_{i'=1}^{n}
    \trace \Big(
    (R_{Z} e_{i}) (R_{Z} e_i)^\top
    \hat K_{\ddot X_j}
    (R_{Z} e_{i'}) (R_{Z} e_{i'})^\top
    \hat K_{\ddot X_k}
    \Big) \\
    & \qquad \times
    \trace \Big(
    (R_{Z} e_{i}) (R_{Z} e_i)^\top
    \hat K_{X_{j+}}
    (R_{Z} e_{i'}) (R_{Z} e_{i'})^\top
    \hat K_{X_{k+}}
    \Big). \\
    = & \frac{1}{n^2} \sum_{i=1}^{n} \sum_{i'=1}^{n}
    \Big(
    e_i^\top R_{Z} \hat K_{\ddot X_j} R_{Z} e_{i'}
    \cdot
    e_i^\top R_{Z} \hat K_{X_{j+}} R_{Z} e_{i'}
    \Big) \\
    & \qquad \times
    \Big(
    e_{i'}^\top R_{Z} \hat K_{\ddot X_k} R_{Z} e_i
    \cdot
    e_{i'}^\top R_{Z} \hat K_{X_{k+}} R_{Z} e_i
    \Big) \\
    = & \frac{1}{n^2} \trace\left(
    \big(
    \hat{K}_{\ddot X_j \mid Z} \circ \hat{K}_{X_{j+}\mid Z}
    \big)
    \cdot
    \big(
    \hat{K}_{\ddot X_k \mid Z} \circ \hat{K}_{X_{k+}\mid Z}
    \big)
    \right).
\end{align*}
Summing over $j$ and $k$, we conclude that
\begin{align*}
    & \sum_{j=1}^{d-1} \sum_{k=1}^{d-1}
    \trace\left(
    (\hat\Gamma_{S\mid Z})_{(j,k)}
    (\hat\Gamma_{S\mid Z})_{(k,j)}
    \right) \\
    = & \frac{1}{n^2}
    \sum_{j=1}^{d-1} \sum_{k=1}^{d-1}
    \trace\left(
    \big(
    \hat{K}_{\ddot X_j \mid Z} \circ \hat{K}_{X_{j+}\mid Z}
    \big)
    \cdot
    \big(
    \hat{K}_{\ddot X_k \mid Z} \circ \hat{K}_{X_{k+}\mid Z}
    \big)
    \right),
\end{align*}
which gives the desired formula.
\end{proof}

\bibliographystyle{asa}
\bibliography{references}

\end{document}